\documentclass{IEEEtran}
\usepackage{amsmath}
\usepackage{bm}
\usepackage{amsthm}
\usepackage{graphicx}
\usepackage{float}
\usepackage{subfigure}
\usepackage{cite}
\usepackage{enumerate}
\usepackage{mathrsfs}
\usepackage{multirow}
\usepackage{amsfonts}
\usepackage{makecell}
\usepackage{url}
\usepackage{array}
\usepackage{CJK}
\usepackage{indentfirst}
\usepackage{algorithm}
\usepackage{algpseudocode}
\usepackage{color, soul}
\usepackage{indentfirst} 
\usepackage[nogroupskip,nonumberlist,nopostdot]{glossaries}
\setglossarystyle{alttree}

\makeglossaries

\newacronym{5G}{5G}{Fifth-generation cellular networks}
\newacronym{IOS}{IOS}{Intelligent omni-surface}
\newacronym{6G}{6G}{Sixth-generation cellular networks}
\newacronym{mmWave}{mmWave}{Millimeter wave}
\newacronym{THz}{THz}{Terahertz}
\newacronym{MIMO}{MIMO}{Multiple-input multiple-output}
\newacronym{RF}{RF}{Radio freqency}
\newacronym{AP}{AP}{Access point}
\newacronym{UAV}{UAV}{Unmanned aerial vehicle}
\newacronym{EM}{EM}{Electromagnetic}
\newacronym{RIS}{RIS}{Reconfigurable intelligent surface}
\newacronym{PIN}{PIN}{Positive intrinsic negative}
\newacronym{IRS}{IRS}{Intelligent reflecting surface}
\newacronym{BS}{BS}{Base station}
\newacronym{CCSA}{CCSA}{China communications standards association}
\newacronym{RRS}{RRS}{Reconfigurable refractive surface}
\newacronym{STAR-RIS}{STAR-RIS}{Simultaneously transmitting and reflecting reconfigurable intelligent surface}
\newacronym{CSI}{CSI}{Channel state information}
\newacronym{MISO}{MISO}{Multiple-input single-output}
\newacronym{NLoS}{NLoS}{Non-line-of-sight}
\newacronym{LoS}{LoS}{Line-of-sight}
\newacronym{SINR}{SINR}{Signal-to-interference-plus-noise ratio}
\newacronym{MMSE}{MMSE}{Minimum mean square error}
\newacronym{CDF}{CDF}{Cumulative distribution function}
\newacronym{FPGA}{FPGA}{Field-programmable gate array}
\newacronym{Tx}{Tx}{Transmitter}
\newacronym{Rx}{Rx}{Receiver}
\newacronym{USRP}{USRP}{Universal software radio peripheral}
\newacronym{LNA}{LNA}{Low noise amplifier}
\newacronym{HPBW}{HPBW}{Half-power beamwidth}
\newacronym{SLL2}{SLL}{Sidelobe level}
\newacronym{NOMA}{NOMA}{Non-orthogonal multiple access}
\newacronym{NGMA}{NGMA}{Next generation multiple access}
\newacronym{OMA}{OMA}{Orthogonal multiple access}
\newacronym{CoMP}{CoMP}{Coordinated multi-point transmission}
\newacronym{QoS}{QoS}{Quality-of-Service}
\newacronym{URLLC}{URLLC}{Ultra-reliable low latency communications}
\newacronym{PLS}{PLS}{Physical layer security}
\newacronym{SIC}{SIC}{Successive interference cancellation}
\newacronym{AN}{AN}{Artifical noise}
\newacronym{AAN}{AAN}{Aerial access network}
\newacronym{ISAC}{ISAC}{Integrated sensing and communication}
\newacronym{SWIPT}{SWIPT}{Simultaneous wireless information and power transfer}
\newacronym{IoT}{IoT}{Internet-of-Things}
\newacronym{VLC}{VLC}{Visible light communication}
\newacronym{NR}{NR}{New radio}
\newacronym{ID}{ID}{Identification}
\newacronym{ITU}{ITU}{International telecommunication union}
\newacronym{3GPP}{3GPP}{3rd generation partnership project}
\newacronym{SDO}{SDO}{Standards developing organization}

\glsaddall
\glsfindwidesttoplevelname

\begin{document}
\title{Intelligent Omni-Surfaces: Simultaneous Refraction and Reflection for Full-dimensional Wireless Communications}

\author{
	{Hongliang Zhang}, \IEEEmembership{Member, IEEE}
	{and Boya Di}, \IEEEmembership{Member, IEEE}.
	
	\thanks{H. Zhang is with Department of Electrical and Computer Engineering, Princeton University, NJ, USA.}
	
	\thanks{B. Di is with School of Electronics, Peking University, Beijing, China.}
}

\maketitle

\begin{abstract}
	The development of metasurfaces has unlocked various use cases in wireless communication networks to improve performance by manipulating the propagation environment. Intelligent omni-surface (IOS),  an innovative technique in this category, is proposed for coverage extension. In contrast to the widely studied reflective metasurfaces, i.e., intelligent reflecting surfaces (IRSs), which can only serve receivers located on the same side of the transmitter, the IOS can achieve full-dimensional wireless communications by enabling the simultaneous reflection and refraction of the surface, and thus users on both sides can be served. In this paper, we provide a comprehensive overview of the state-of-the-art in IOS from the perspective of wireless communications, with the emphasis on their design principles, channel modeling, beamforming design, experimental implementation and measurements, as well as possible applications in future cellular networks. We first describe the basic concepts of metasurfaces, and introduce the corresponding design principles for different types of metasurfaces. Moreover, we elaborate on the reflective-refractive model for each IOS element and the channel model for IOS-aided wireless communication systems. Furthermore, we show how to achieve full-dimensional wireless communications with the IOS for three different scenarios. In particular, we present the implementation of an IOS-aided wireless communication prototype and report its experimental measurement results. Finally, we outline some potential future directions and challenges in this area.
\end{abstract}


\printglossary[title = List of Acronyms]

\section{Introduction}

\subsection{Motivation}

With the development of emerging applications such as Internet-of-Things (IoT), augmented reality, virtual reality, and automated driving, we expect the evolution of cellular networks toward higher data rate, wider coverage, higher energy efficiency, and lower latency \cite{saad2019vision,letaief2019roadmap,zhang20196g}. However, existing techniques in the current fifth-generation cellular networks (5G) are not sufficient to meet these stringent requirements \cite{shafi20175g,zhang2020beyond2,andrews2014will}. There are several major challenges for the future sixth-generation cellular networks (6G):
\begin{itemize}
	\item \emph{Conflicts between low hardware cost and high spatial resolution}: Three techniques have been developed in 5G to provide high spatial resolution for multiple access. The first is to use higher frequency bands such as millimeter wave (mmWave) \cite{roh2014millimeter,rappaport2013millimeter} or terahertz (THz) \cite{jornet2011channel,koenig2013wireless} as narrow beams are easier to generate in these bands. These systems, on the other hand, need dedicated radio frequency (RF) chains to support the transmission over higher frequency bands, whose costs could be up to 10 times of that in current sub-6G bands\footnote{Source: \emph{http://luowave.net/en/}.}.  Massive multiple-input multiple-output (MIMO) systems is the second enabling technique \cite{larsson2014massive}. The use of a large number of antennas, each equipped with a phase shifter, also leads to a high hardware cost for network deployment. Ultra-dense networks are the third solution~\cite{zhang2017hypergraph,kamel2016ultra}. The ultra-dense networking, provided by densely-deployed access points (APs), can provide a high spatial resolution but still requires exceedingly high hardware costs for AP deployment and the coordination among these APs.

	\item \emph{Conflicts between flexible network deployment and low energy consumption}: With reduced cell coverage due to the high operating frequency, data traffic in future networks varies far more dramatically ~\cite{zhang2018load}. As a result, APs placed in a fixed position, such as terrestrial infrastructure, will struggle to deal with the traffic fluctuations because it is not cost-effective to deploy extra APs to support extreme situations. In contrast, though flexible APs, such as unmanned aerial vehicles (UAVs), can be applied to alleviate this issue \cite{mozaffari2019tutorial,zhang2019cooperation}, they require extra energy consumption to support the mobility. 
\end{itemize}
In view of these challenges, it necessitates the development of innovative technologies to achieve sustainable capacity growth of future wireless networks with low cost and energy consumption.

Motivated by these challenges, the recent development of metasurfaces has given rise to a profoundly new paradigm for controlling electromagnetic (EM) waves and achieving smart wireless propagation environments, which is commonly referred to as reconfigurable intelligent surfaces (RISs) \cite{di2019smart,zhang2021reconfigurable} or large intelligent surfaces \cite{liang2019large}. The RIS is an engineered planar surface with multiple nearly passive sub-wavelength scattering elements that can manipulate the signals impinging on it. Such a manipulation of signals can be realized through configuring the states of, for example, positive intrinsic negative (PIN) diodes embedded in the surface \cite{tong2022two}. By tuning the ON/OFF state of each PIN diode, the amplitude and phase of the scattered signals can be programmed as desired, enabling signal re-radiation toward the receiver (Rx) at decreased hardware cost and energy \cite{di2020hybrid}.

Existing research in the wireless communication community focuses on the reflective type of RISs, also known as intelligent reflecting surfaces (IRSs), which only allow the signal reflection toward users located on the same side of the transmitter (Tx) \cite{chen2019intelligent,yu2020robust}. In other words, the users located on the opposite side of the surface cannot be served, limiting the coverage of the surface. To address this issue, an emerging implementation of RISs, which we refer to as \emph{intelligent omni-surfaces (IOSs)}, is proposed \cite{zhang2022intelligent}. Unlike the IRS, which provides the only function of reflection, the IOS allows simultaneous signal reflection and refraction. Specifically, the IOS can simultaneously reflect and refract the signals that impinge on any side of the surface toward users on either side of the surface. The power ratio of the reflected and refracted signals is dependent on and can be optimized by the structure of each IOS element \cite{cai2017high}. With the capability of joint reflection and refraction, an IOS can achieve full-dimensional wireless communications despite users' locations with respect to the surface \cite{zhang2020beyond}. 

\subsection{Use Cases}
\begin{figure*}[!t]
	\centering
	\includegraphics[width=0.95\textwidth]{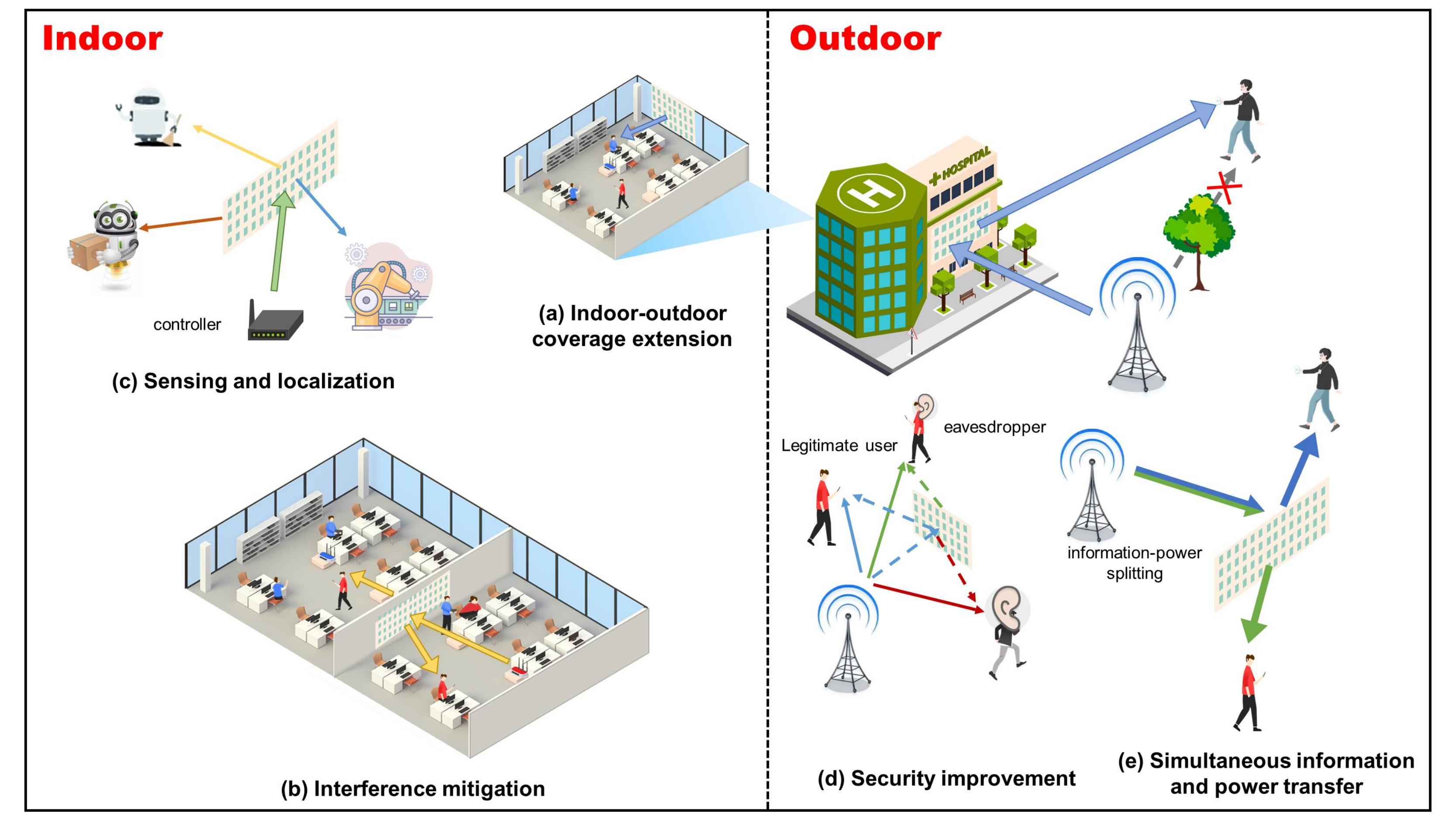}
	\caption{Possible use cases of IOSs in future wireless networks.}
	\label{usecase}
\end{figure*}

In general, an IOS can be placed anywhere the propagation of radio waves should be adjusted to improve the performance of communication systems. Specially, the IOS can be deployed at the billboard and road side units for outdoor scenarios, or be embedded in the walls between two rooms for indoor scenarios~\cite{subrt2012intelligent}. As shown in Fig. \ref{usecase}, potential use cases for IOS-aided wireless communications include:
\begin{itemize}
	\item \emph{Coverage Extension:} Compared to the IRS, the IOS can effectively extend the coverage as users on both sides of the surface can be served simultaneously through its reflection and refraction signals. As the example shown in Fig. \ref{usecase}(a), by replacing the window with an IOS, it can extend the coverage for both indoor and outdoor environments.
	
	\item \emph{Interference Mitigation:} In hotspot areas, especially indoor scenarios, some users are very likely to be located within the coverage of multiple small cells, leading to increasing multi-cell interference \cite{chen2020reconfigurable}. By deploying an IOS between two small cell APs, for example, embedded in the wall, it can help focus the signals toward specified users on one side of the surface while nulling the interference to the other side, as illustrated in Fig. \ref{usecase}(b).
	
	\item \emph{Sensing and Localization:} In future cellular networks, wireless sensing \cite{hu2020reconfigurable} and localization \cite{zhang2021metalocalization} will be critical services. The capability of the IOS to alter the propagation environment can provide more fingerprints about the environment, thus improving the precision. Compared to the IRS, the dual function of reflection and refraction makes the IOS have a larger coverage as shown in Fig. \ref{usecase}(d).
	
	\item \emph{Security Improvement:} With an IOS, the physical layer security (PLS) \cite{mukherjee2014principles} can be further enhanced. For example, even the eavesdroppers are located in both sides of the surface, the IOS can still generate both reflective and refractive signals to confuse them. An example is given in Fig. \ref{usecase}(d).
	
    \item \emph{Simultaneous Wireless Information and Power Transfer:} IOSs also provide an opportunity for simultaneous wireless information and energy transfer (SWIPT), as shown in Fig. \ref{usecase}(e). Different from the traditional SWIPT \cite{krikidis2014simultaneous}, where information and power are splitted at the user, the IOS can split them before arriving the user, which can provide more flexibility.
\end{itemize}

Therefore, IOS is a promising technology that can potentially benefit various vertical industries in 5G/6G ecosystems.

\subsection{Commercialization and Standardization}
The commercialization and standardization of RIS/IOS-related techniques have attracted much attention from the industry, which can be reflected by the following three parts: 

\subsubsection{White papers} 
According to \cite{liu2022path}, the RIS has been considered as a candidate enabling technology for future networks in 17 white papers. For example, Huawei expects that the RISs can work as low-cost passive relays to extend the coverage from outdoor or indoor base stations to users or vehicles, where the direct link between them is not available. In addition to the usage for passive relays, vivo envisages that the RIS can also be applied to the BS and user side, forming a novel transceiver structure with index modulation \cite{vivo}. In \cite{6gflagship}, machine learning-driven RISs and related research problems are discussed. The authors in \cite{bourdoux20206g} explored the possibility of RISs in wireless sensing and localization applications with the capability to reshape and control the electromagnetic response of the propagation environments.  

\subsubsection{Projects}
Funding agencies worldwide are providing more investment in this emerging technology, and a detailed summary can be found in \cite{liu2022path}. These research projects have significantly contributed to the theoretical study and hardware implementation of RISs. 

\subsubsection{Standardization}
\begin{figure}[!t]
	\centering
	\includegraphics[width=0.48\textwidth]{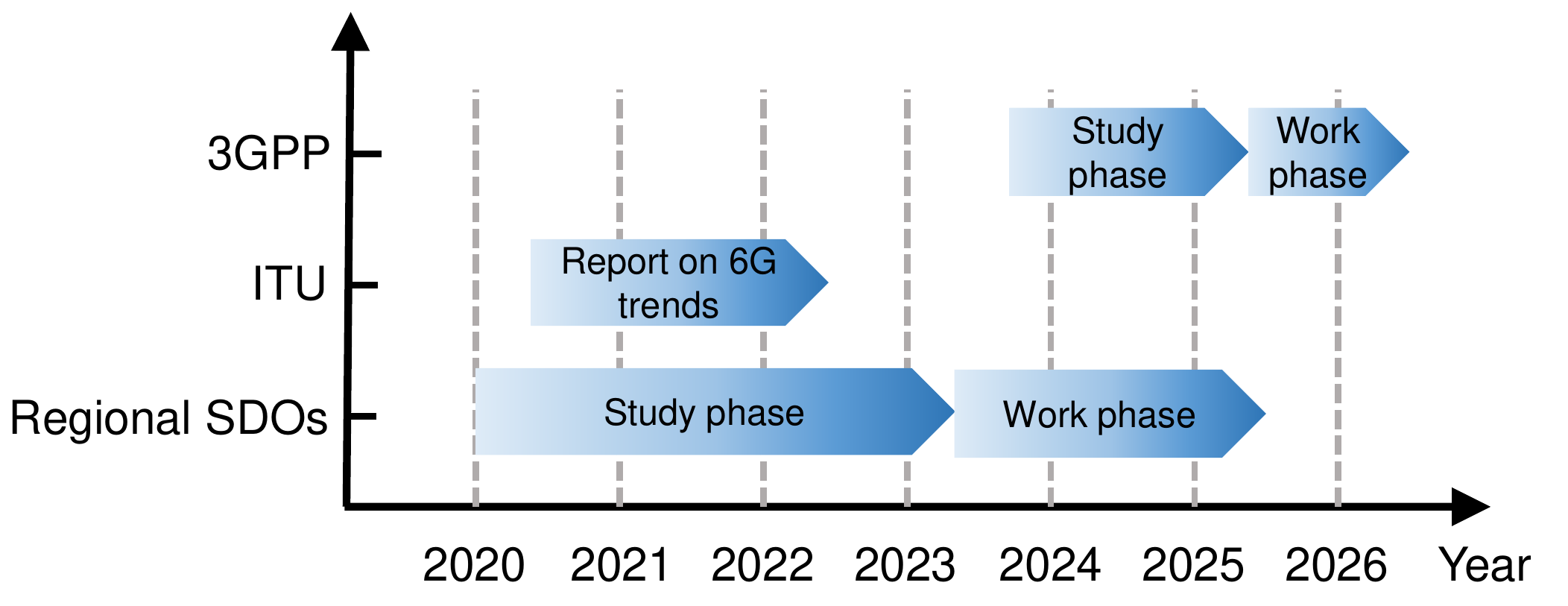}
	\caption{Current status and forecast timeline for the standardization process \cite{jian2022reconfigurable}.}
	\label{timeline}
\end{figure}

With extensive research efforts invested in industry and academia, including theoretical studies and field tests, standardization work has also been kicked off, mainly by some regional standards developing organizations (SDOs). For example, in China, the China Communications Standards Association (CCSA), which is a formal SDO producing normative specifications, has approved the proposal to establish a study item on RIS \cite{meeting}, and aims to conclude in 2022 with a technical report as the major deliverable. In addition, a new industry specification group on RIS was approved by the European Telecommunications Standards Institute (ETSI) in June 2021 \cite{ETSI}, and a stable report is expected to be available by the end of 2022.

In global SDOs, we can also see some attempts at the standardization of RIS. Specifically, during the international telecommunication union (ITU) radiocommunication sector meeting in October 2020, RISs were depicted as critical components for the physical layer of next generation networks, as reported in \cite{meeting2}. In the 3rd Generation Partnership Project (3GPP), proposals related to RISs started to be submitted during the March 2021 meeting \cite{proposal1}, after which more companies are involved to support the RIS to become a key component of future wireless networks \cite{proposal2}.  

In summary, a predicted standardization timeline is depicted in Fig. \ref{timeline} \cite{jian2022reconfigurable}. In the ITU's future standard activities, it is not likely to establish a dedicated focus group or a work item since the ITU focuses on the regulatory aspect of spectrum. However, the RIS could be possible to be studied for the channel modeling. As for 3GPP, its current focus is to start the second stage of 5G by Release 18. In general, each generation of wireless communication standards lasts for ten years. Therefore, it is reasonable to expect that possible discussions on requirements for 6G systems will begin after 2027. In order to standardize the RIS as a new technology, a lot of efforts need to be made. Based on the maturity of this technology, the study phase should be initiated at least three years in advance.

\subsection{Objective and Organization}

\begin{table*}[!t]
	\caption{List of representative survey papers related to RISs}
	\centering
	\renewcommand\arraystretch{1.2}
	\begin{tabular}{|c| m{8cm} |m{3cm}|m{3.5cm}|}
		\Xhline{1pt}
		\textbf{Reference}  & \makecell[c]{\textbf{Focuses}} & \makecell[c]{\textbf{Types of RISs}} & \makecell[c]{\textbf{Hardware}} \\
		\hline
		\cite{di2019smart} & Introduce the concept of RISs to empower smart radio environments, discuss different types of  functionalities for the intelligent surfaces, and overview current research efforts on smart radio environments & Focus on reflective RISs, but other types are also introduced & No hardware implementation\\
		\hline
		\cite{liang2019large} & Discuss the hardware implementations, performance gains, and applications of RISs, emphasizing the similarities and differences compared with backscatter communications and reflective relays.& Reflective RISs & Hardware designs of reflective elements are introduced, but measurement results are not provided\\
		\hline
		\cite{di2020smart} & Give a comprehensive overview of enabling technologies for RIS-empowered wireless networks from a communication-theoretic perspective and discuss the most important research issues to tackle & Focus on reflective RISs, but other types are also introduced & Some hardware structures are introduced, but real-world measurements are not presented\\
		\hline
		\cite{elmossallamy2020reconfigurable} & Present different hardware implementations to achieve smart radio environments, analyze the channel characteristics, and other challenges and opportunities in RIS-aided wireless networks& Reflective RISs & No hardware implementation\\
		\hline
		\cite{wu2021intelligent} & Provide a tutorial of RIS-aided wireless communications to address several main technical challenges from a communication point of view & Reflective RISs & No hardware implementation\\
		\hline
		\cite{bjornson2021reconfigurable} & Overview the RIS fundamentals and up-to-date research efforts from a signal processing standpoint, including communication, localization, and sensing&Reflective RISs & No hardware implementation\\
		\hline
		\cite{yuan2021reconfigurable} & Summarize the state-of-the-art solutions to three physical layer challenges for incorporating RISs into wireless networks, including channel state information acquisition, passive information transfer, and low-complexity phase shift optimization&Reflective RISs & No hardware implementation\\
		\hline
		\cite{zhang2022towards} & Give a tutorial of RIS-aided sensing and localization to address key technical challenges and present the prototype together with experimental results&Focus on reflective RISs, but other types are also introduced & Only provide hardware implementation and measurement results for reflective RISs\\
		\hline
		\cite{liu2021reconfigurable}& Provide a survey on the performance analysis, resource management, as well as machine learning techniques for RIS-aided wireless communications and outline potential applications& Reflective RISs & No hardware implementation\\
		\hline
		\cite{gong2020toward} & Present a literature review on recent applications and design aspects of RISs in wireless networks, and discuss emerging use cases of RISs&Reflective RISs & No hardware implementation\\
		\hline
		\cite{zheng2021survey}& Provide a comprehensive survey on the up-to-date research in RIS-aided wireless communications, with an emphasis on the promising solutions to tackle practical design issues of channel estimation and beamforming design&Reflective RISs & No hardware implementation\\
		\hline
		\cite{aboagye2022ris} & Discuss indoor VLC systems with RIS technology, especially how RISs overcome line-of-sight (LoS) blockage and device orientation issues in VLC systems&Reflective RISs & Some hardware designs for reflective elements are introduced, but no measurement results are reported\\
		\Xhline{1.pt}
	\end{tabular}
	\label{table: literature}
\end{table*}

In the literature, some existing articles \cite{liang2019large,di2019smart,di2020smart,elmossallamy2020reconfigurable,wu2021intelligent,bjornson2021reconfigurable,yuan2021reconfigurable,zhang2022towards,liu2021reconfigurable,gong2020toward,zheng2021survey,aboagye2022ris} have provided comprehensive tutorials or surveys of the state-of-the-art research work related to RISs from different perspectives, but the focus of this paper is quite different. Specifically, the authors in \cite{liang2019large} discussed the applications of RISs as reflectors, especially the comparison with backscatter communications and reflective relays. Papers \cite{di2019smart} and \cite{di2020smart} introduced the concept of RISs, and gave a comprehensive overview of RIS-empowered smart radio environment. The authors in \cite{elmossallamy2020reconfigurable} analyzed the channel characteristics for RIS-aided wireless communications, and \cite{zheng2021survey} surveyed channel estimation schemes for these systems. The authors in \cite{bjornson2021reconfigurable} investigated RIS-aided wireless communications from the perspective of signal processing, and paper \cite{zhang2022towards} gave a tutorial of RIS-aided wireless sensing and localization. The authors in \cite{yuan2021reconfigurable} focused on three physical layer challenges for incorporating RISs into wireless networks, i.e., channel state information acquisition, passive information transfer, and low-complexity phase shift optimization. The authors in \cite{liu2021reconfigurable} discussed the machine learning and resource allocation techniques for RIS-aided wireless communications, while paper \cite{gong2020toward} emphasized the emerging use cases of RISs. The authors in \cite{aboagye2022ris} mainly discussed the applications of RISs in visible light communication (VLC) systems. These articles are summarized in Table \ref{table: literature} for reference.

Different from the above work which focus on the reflective-only surfaces, this paper is the first tutorial on a more generalized type of surfaces, i.e., IOSs, and their applications on wireless communications. Moreover, unlike most of the aforementioned tutorials which do not include the hardware implementation, we are the first one to build an IOS-aided wireless communication prototype and report the measurement results. 

\begin{figure}[!t]
	\centering
	\includegraphics[width=0.48\textwidth]{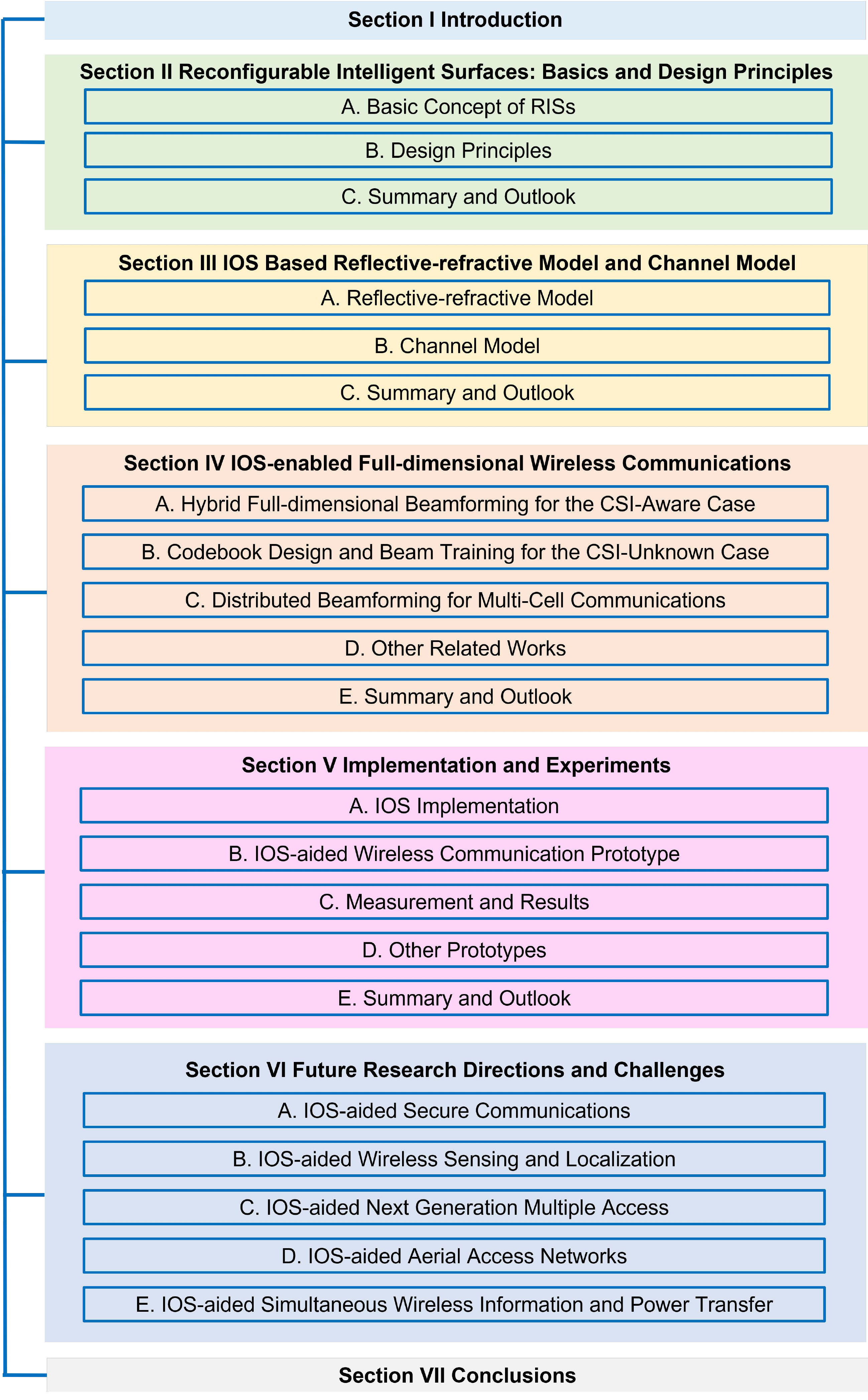}
	\caption{Organization of this paper.}
	\label{org}
\end{figure}

Against the above observations, our main contributions are as follows:
\begin{itemize}
	\item We overview the design principles for different types of RISs. Different from those that focus on the IRS, which can only serve the users located on the same side of the base station (BS), we highlight the IOS-based reflective-refractive models to provide simultaneous coverage for users on both sides of the surface. Additionally, we present the channel models for IOS-aided wireless communications. 
	
	\item We develop three beamforming schemes for IOS-aided full-dimensional wireless communications under different network settings. Moreover, we also review related works in the literature. 
	
	\item We elaborate on how to implement the IOS and build an IOS-aided wireless communication prototype. Based on the prototype, we give some measurement results and discuss the characteristics of the IOS from the perspective of wireless communications.
	
	\item We outline the major potential applications of the IOS in future wireless networks and identify the research challenges aiming to stimulate potential studies.
\end{itemize}

As shown in Fig. \ref{org}, the rest of this paper is organized as follows. Section \ref{basic} introduces the fundamentals of intelligent surfaces, including their basic concepts and design principles. Section \ref{IOS modeling} overviews the reflective-refractive model for the IOS and the corresponding channel model for IOS-aided wireless communications. In Section \ref{fd-IOS}, we present the key techniques for the IOS-aided full-dimensional wireless communications. In Section \ref{implementation}, we introduce the implementation of the IOS and an IOS-aided wireless communication prototype, and show the experimental results obtained from the prototype. In Section \ref{future direction}, we discuss other relevant topics on IOS to broaden its scope. Finally, we conclude this paper in Section \ref{conclusion}.

\section{Reconfigurable Intelligent Surfaces: Basics and Design Principles} \label{basic}
In this section, we start with the basic concepts of RISs in Section \ref{basics}. The general structure of RISs is introduced, based on which the specific structure and design principles of three different variants are shown in Section \ref{principle}, i.e., reflective, refractive, and reflective-refractive types.

\subsection{Basic Concept of RISs}
\label{basics}
The RIS is a two-dimensional meta-surface inlaid with numerous electrically controllable scatters, each of which is an element whose EM response can be reconfigured. When an EM wave projects toward the element, it excites the surface current of this element such that the element emits another EM radiation in the form of reflected and/or refracted signals from the element. By reconfiguring the element structure, the phase and/or amplitude of the reflected/refracted signal can be artificially programmed. Containing numerous elements with an adjacent distance far smaller than a half wavelength, the meta-surface can then reshape the propagation environment by superposing the reshaped EM radiation of all elements such that the incident signal can be reflected and/or refracted by the surface toward desirable directions.

\begin{figure}[!t]
	\centering
	\includegraphics[width=0.45\textwidth]{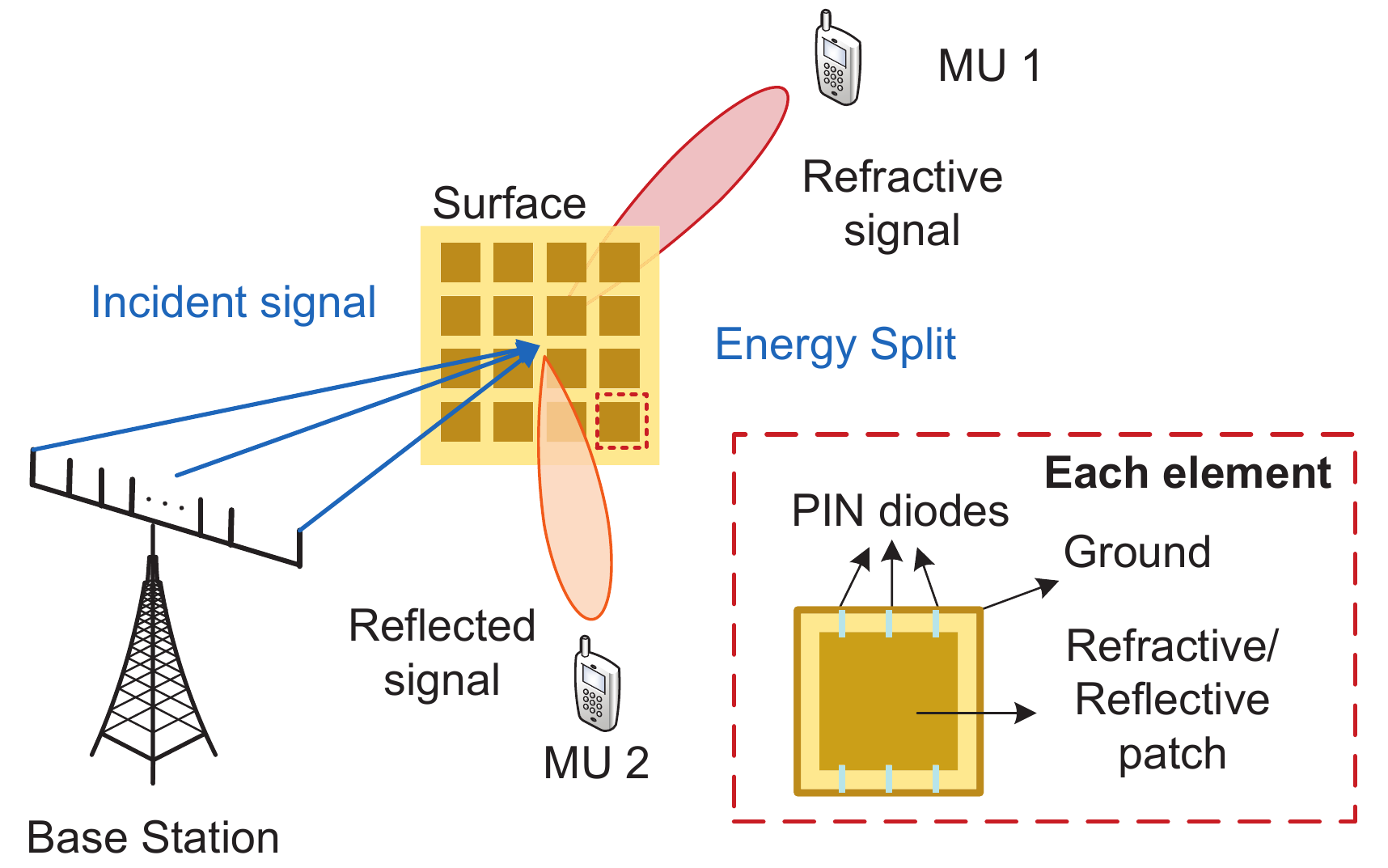}
	\caption{Illustration of reflected and refracted signals in an RIS-assisted communication system.}
	\label{angle}
\end{figure}

As shown in Fig. \ref{angle}, a surface element contains one or more diodes and a metallic patch\footnote{The diode can be either a PIN diode or a varactor, and the shape of this metallic patch can be either rectangular or irregular.}, laid on a dielectric substrate. Take the PIN diode as an example. The PIN diode can be switched between ON and OFF modes by changing its bias voltage. The bias voltage is imposed by the feed line equipped at the bottom of the substrate, which connects to the metallic patch through a via hole. The surface impedance of the element varies with different working modes of the diode, which is the key to reconfigure the EM response of the element. In more detail, for a given EM wave impinging upon the element, the excited surface current varies with its impedance, which is determined by the ON/OFF state of the diode. The EM response, influenced by both the incident EM wave and the surface current, can then be controlled by shifting the element's \emph{state}. 

For an incident signal $x$, the reflected and refracted signals of the $m$-th element of the IOS, i.e., $y_{m}^{(l)}$ and $y_{m}^{(r)}$, are shown below, respectively:
\begin{equation}\label{signal_model}
\begin{split}
& y_{m}^{(l)} = \Gamma_m^{(l)} e^{j\theta_m^{(l)}}x,\\
& y_{m}^{(r)} = \Gamma_m^{(r)} e^{j\theta_m^{(r)}}x,
\end{split}
\end{equation}
where ${\theta}_m^{(l)}$ and ${\theta}_m^{(r)}$ are the phase responses, and $\Gamma_m^{(l)}$ and $\Gamma_m^{(r)}$ are the amplitude responses. In general, the phase and amplitude responses have some correlations as introduced in Section~\ref{principle}. The sum energy of $y_{m}^{(l)}$ and $y_{m}^{(r)}$, i.e., $\left| \Gamma_m^{(l)}\right| ^2 + \left| \Gamma_m^{(r)}\right| ^2$, is smaller than 1 due to the insertion loss. 

In the literature, there exists another type of antenna technique which is typically referred to as holographic MIMO~\cite{pizzo2020spatially}. In holographic MIMO, an infinite number of antenna elements are expected to be placed in a compact space, forming a continuous aperture~\cite{deng2022hdma,deng2021reconfigurable}. As a result, an extremely high spatial resolution can be achieved. In comparison, holographic MIMO and RISs are designed from different perspectives. The holographic MIMO emphasizes small element spacing, whereas the RIS focuses on its own capability of tuning wireless channels. It is worth pointing out that the holographic MIMO can also be implemented by the RIS \cite{wan2021terahertz}.

\subsection{Design Principles}
\label{principle}
Based on the general element structure in Fig. \ref{angle}, by designing its geometric parameters and arranging numerous elements in different manners, three variants of intelligent surfaces are developed, along with specific design principles such that the radiated wave can be either reflective-only, refractive-only, or reflective-refractive. 

\subsubsection{$\left| \Gamma_m^{(r)}\right| = 0$, $\left| \Gamma_m^{(l)}\right| > 0$, Reflective-only} \label{reflect-only}
\begin{figure}[!t]
	\centering
	\includegraphics[width=0.4\textwidth]{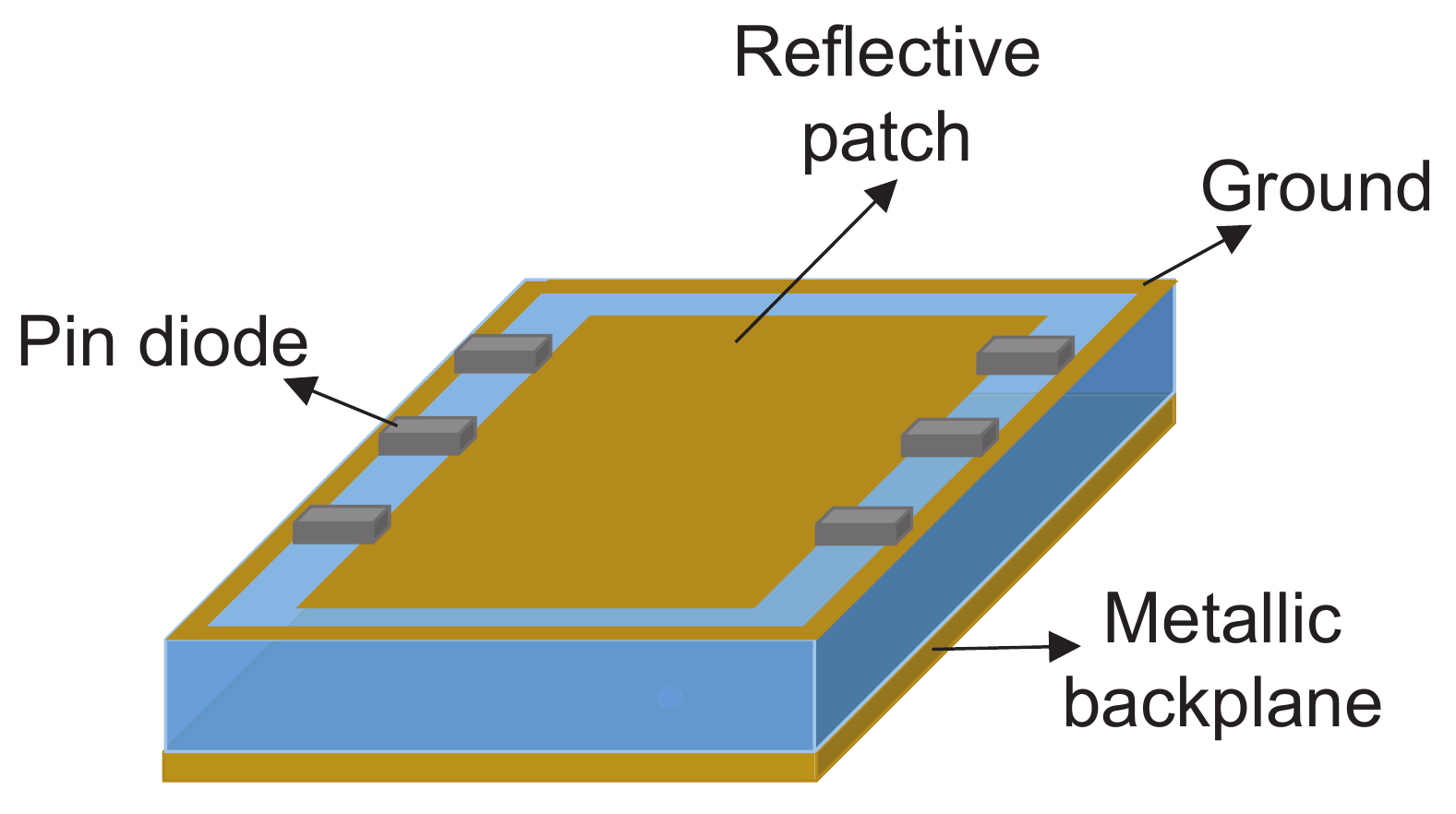}
	\caption{Structure of an IRS element.}
	\label{structure_IRS}
\end{figure}

The surface only reflects the incident signal in the desired direction. Such surfaces are also known as IRSs \cite{yu2021smart}. A typical IRS structure is shown in Fig. \ref{structure_IRS}, consisting of a reflective patch, PIN diodes, a ground, and a metallic backplane. The metallic backplane is used to shield the signal refraction. 

A well-designed IRS is expected to reflect the incident wave in any desired direction. This typically requires the phase response of each element to be reconfigured independently while the amplitude response is fixed, which is compatible with the function of traditional phase arrays. Two design principles of the IRS are then introduced.
\begin{itemize}
	\item The amplitude responses for different element states should be the same;
	
	\item The phase response for different element states should be as different as possible. For example, when each element has two states (tuned by the PIN diode), the phase responses with respect to ON and OFF states should be 0$^\circ$ and 180$^\circ$, respectively.
\end{itemize} 

\subsubsection{$\left| \Gamma_m^{(r)}\right| > 0$, $\left| \Gamma_m^{(l)}\right| = 0$, Refractive-only}
\begin{figure}[!t]
	\centering
	\includegraphics[width=0.4\textwidth]{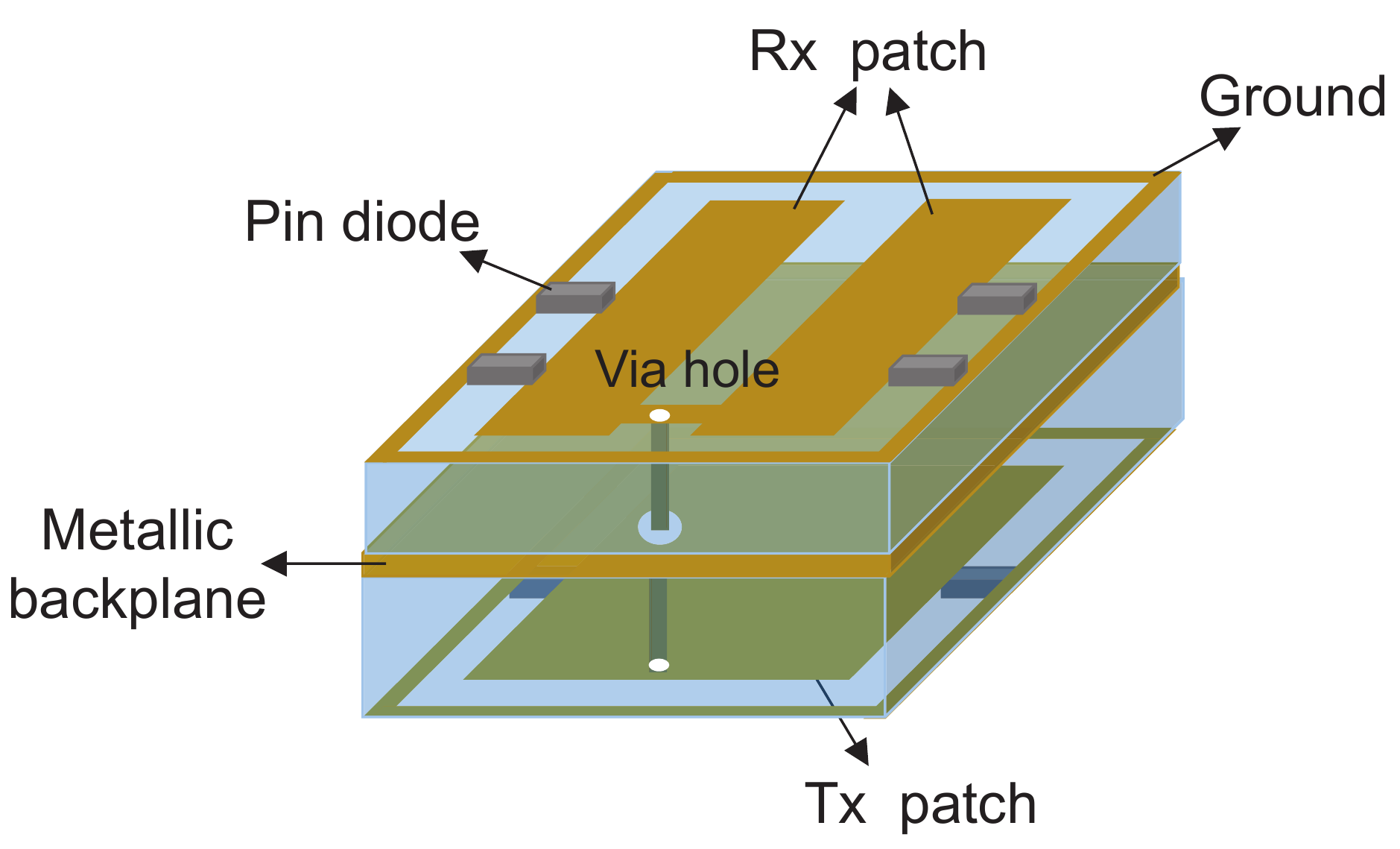}
	\caption{Structure of an RRS element.}
	\label{structure_RRS}
\end{figure}

The surface only refracts the incident signal with no reflection, which is called a reconfigurable refractive surface (RRS) \cite{zeng2022reconfigurable}. A typical RRS structure is shown in Fig. \ref{structure_RRS}. Each RRS is composed of two layers with a metallic backplane in the middle. For each layer, it contains a patch to receive or radiate EM waves, PIN diodes, a ground, and a via hole. The function of the via hole is to transfer the incident energy from one side to the other side with the EM coupling, and the metallic backplane is used to shield the signal reflection. 

Besides the two design principles for the IRS as introduced in Section \ref{reflect-only}, the RRS also requires the reflection-side amplitude response to be close to zero so that the energy of the incident EM wave can be refracted by the surface to the maximum degree.

\subsubsection{$\left| \Gamma_m^{(r)}\right| > 0$, $\left| \Gamma_m^{(l)}\right| > 0$, Reflective-refractive}\label{IOS-epsilon}
The surface divides the energy of an incident signal, and then reflects and refracts it simultaneously, as illustrated in Fig. \ref{angle}. Such surface is also known as IOS \cite{zhang2022intelligent} or simultaneously transmitting and reflecting reconfigurable intelligent surfaces (STAR-RISs) \cite{liu2021star} since the surface can project the incident signal toward all directions in space. Unlike the cases of IRS and RRS, both reflection- and refraction-related amplitude and phase responses of the IOS should be considered. For the case of unified user distribution\footnote{Users are distributed randomly in the area of interest.}, the general designing principles can be given below.
\begin{itemize}
	\item The amplitude responses of the reflected and refracted signals given the same element state should be as close as possible. This enables the maximum sum rate of users on two sides of the surface. 
	
	\item The phase responses of both reflected and refracted signals given different element states should be as different as possible.
\end{itemize}

\begin{figure}[!t]
	\centering
	\includegraphics[width=0.4\textwidth]{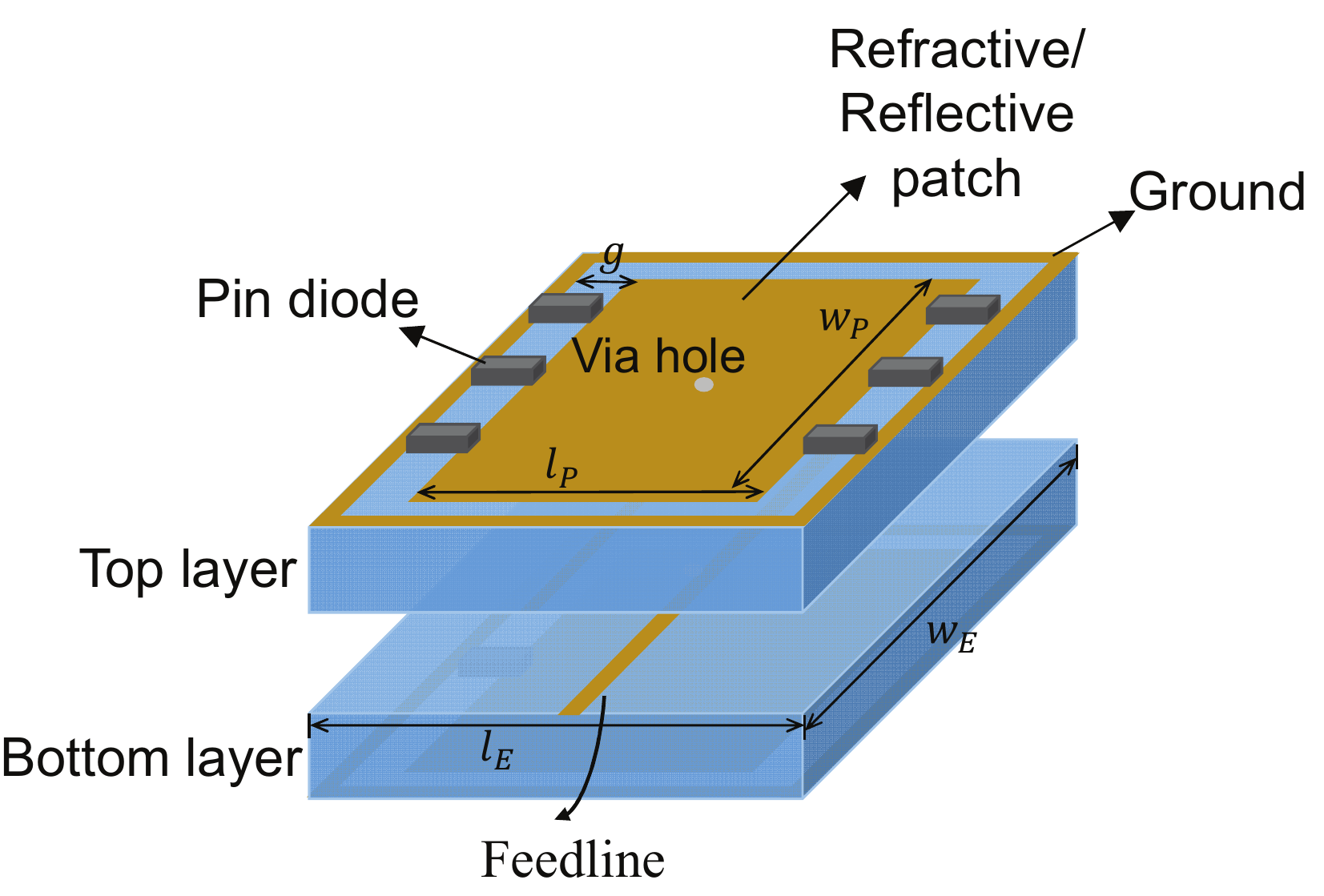}
	\caption{Structure of an IOS element.}
	\label{structure_IOS}
\end{figure}

Satisfying the above principles, an IOS structure widely used in the literature is shown in Fig. \ref{structure_IOS} consisting of two symmetrical layers. Each layer contains a reflective/refractive patch, PIN diodes, a ground, a via hole, and a feedline. The ground and the feedline are connected to the top and bottom of each layer, respectively, applying a varying bias voltage to control the element state. Similar to the RRS element, the via hole transfers the incident energy to the other layer, but without the existence of the metallic backplane, the EM wave can be radiated to both directions, enabling the simultaneous reflection and refraction. We denote two symmetric elements on both layers as one \emph{IOS element}. An IOS element has two symmetric parts, imposing a joint influence on the refraction and reflection responses, i.e., they act as a whole, depicted by the phase responses $\theta_m^{l}$ and $\theta_m^{r}$ in $(\ref{signal_model})$. In practice, $\theta_m^{l}$ and $\theta_m^{r}$ with respect to the same IOS element are usually coupled with each other due to the hardware limitations. By deploying more layers in the metasurface design to adjust its electric and magnetic impedance, independent control over $\theta_m^{l}$ and $\theta_m^{r}$ may be achieved at the cost of extra complexity\cite{xu2017dual}. 

Other IOS structures can also be found in the literature. For example, a sandwich-type IOS structure was proposed in \cite{bao2021programmable}, where the gap between two layers in Fig. \ref{structure_IOS} is replaced by a metallic layer, connecting to the ground. The designed surface can generate an $x$-polarization reflected beam and a $y$-polarization refracted beam separately, but when these two beams are generated simultaneously, the surface suffers inevitable side-lobe effects.

Note that the above design principle works for uniform or unknown user distribution. If users are known to distribute unevenly at two sides of the IOS, the first design principle should be updated such that either $\left| \Gamma_m^{(r)}\right| > \left| \Gamma_m^{(l)}\right|$ or $\left| \Gamma_m^{(r)}\right| < \left| \Gamma_m^{(l)}\right|$ depending on the user distribution, which will be discussed later.

\subsection{Summary and Outlook}
\begin{figure}[!t]
	\centering
	\includegraphics[width=0.45\textwidth]{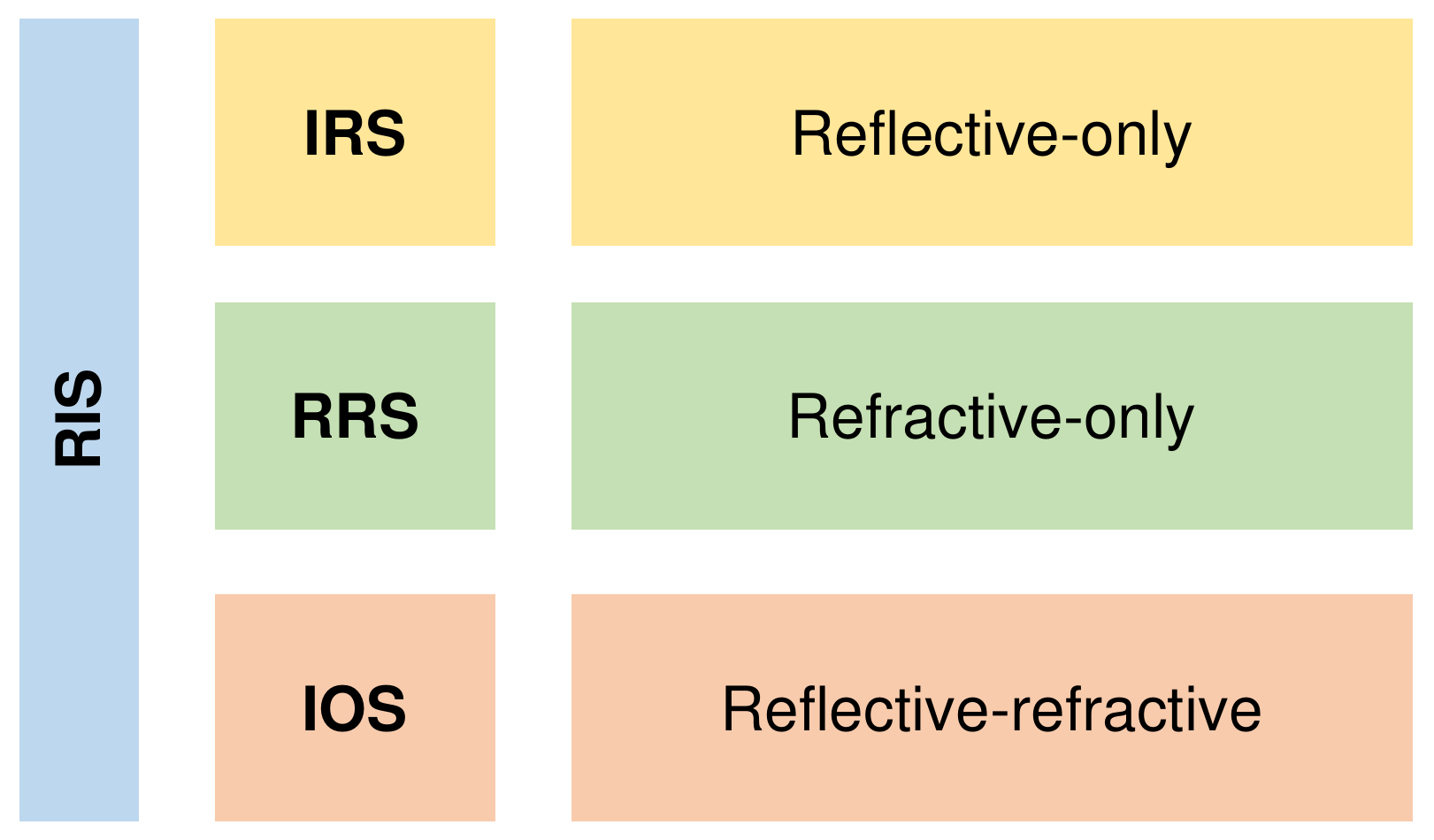}
	\caption{Different types of RISs.}
	\label{type}
\end{figure}

In this section, we have introduced three types of RISs according to their responses to incident signals, as illustrated in Fig. \ref{type}: 
\begin{itemize}
	\item \textbf{IRS}: The incident signals are fully reflected, i.e., only users located on the same side of the Tx with respect to the surface can be served.
	
	\item \textbf{RRS}: The incident signals fully penetrate the surface, i.e., only users located on the opposite side of the Tx with respect to the surface can be served.
	
	\item \textbf{IOS}: One part of the incident signals is reflected and the other is refracted, i.e., users located on both sides of the surface can be served simultaneously.
\end{itemize}

\begin{table*}[!t]
	\renewcommand\arraystretch{1.2}
	\begin{center}
		\caption{Comparison among different surfaces}
		\begin{tabular}{|c|m{4cm} |m{4cm} |m{4cm} |}
			\hline
			{}& \makecell[c]{\textbf{IOS}}& \makecell[c]{\textbf{IRS}}& \makecell[c]{\textbf{RRS}} \\ \hline
			Deployment & Billboard, road side units, or embedded in walls & Ceiling or building surface& Part of the BS\\ \hline
			Structure & Symmetric & Asymmetric & Asymmetric \\ \hline
			Phase Shift & Both reflected and reflected signals, but can be different & Reflected signals & Refracted signals\\ \hline
		\end{tabular}\label{comparison}
	\end{center}
\end{table*}

Based on the above definitions, we can learn that the IOS is a generalized variant of the IRS and RRS. If the amount of reflected signals is equal to 0, an IOS will be reduced to an RRS, while it can also be reduced to an IRS if the amount of refracted signals is equal to 0. Moreover, we can also observe that either IRS or RRS can only serve users located on one side of the surface, limiting the coverage of the surface, which motivates us to propose the IOS to enable full-dimensional wireless communications for coverage extension.

The comparison among IOS, IRS, and RRS is presented in Table \ref{comparison} from the following aspects.
\begin{itemize}
	\item \textbf{Deployment}: Due to its reflective characteristics, the IRS is usually deployed on the ceiling or the building surface to serve as a reflective relay. Since the IOS can reflect and refract signals simultaneously, it can be deployed at the billboard, road side units, or embedded in the walls between two rooms for coverage extension. To exploit the refractive function of RRS, it can be equipped as part of the BS to replace the traditional large-scale phase arrays, enabling novel low-cost and energy-efficient BS antenna arrays.
	
	\item \textbf{Structure}: The element structures of these three types of metasurfaces are different. The IOS element usually has a symmetric structure, while the IRS and RRS elements' structure is asymmetrical. 
	
	\item \textbf{Phase shift}: For the IRS and RRS, each element only has reflected and refracted phase shifts, respectively. In contrast, each IOS element has both reflected and refracted phase shifts simultaneously, but phase shifts for reflection and refraction can be different.
\end{itemize}

It should be noted that the structures given in this section are an example. Exploiting new materials (discussions on various materials will be given in Section \ref{implementation}) or structures to realize a more flexible control of the wireless propagation environment could be a future research direction. 

\section{IOS Based Reflective-refractive Model and Channel Model}
\label{IOS modeling}
In Section \ref{basic}, we have discussed the basic concepts of intelligent surfaces followed by the design principles of different types of surfaces with an emphasis on the IOS. However, the relation between the reflective-refractive coefficients and the IOS structure have not been investigated. Therefore, how the IOSs influence communication performance remains an open problem. In this section, we present a systematic survey of the above issues.
\subsection{Reflective-refractive Model}
\label{modeling}
\begin{figure*}[!t]
	\centering
	\includegraphics[width=0.95\textwidth]{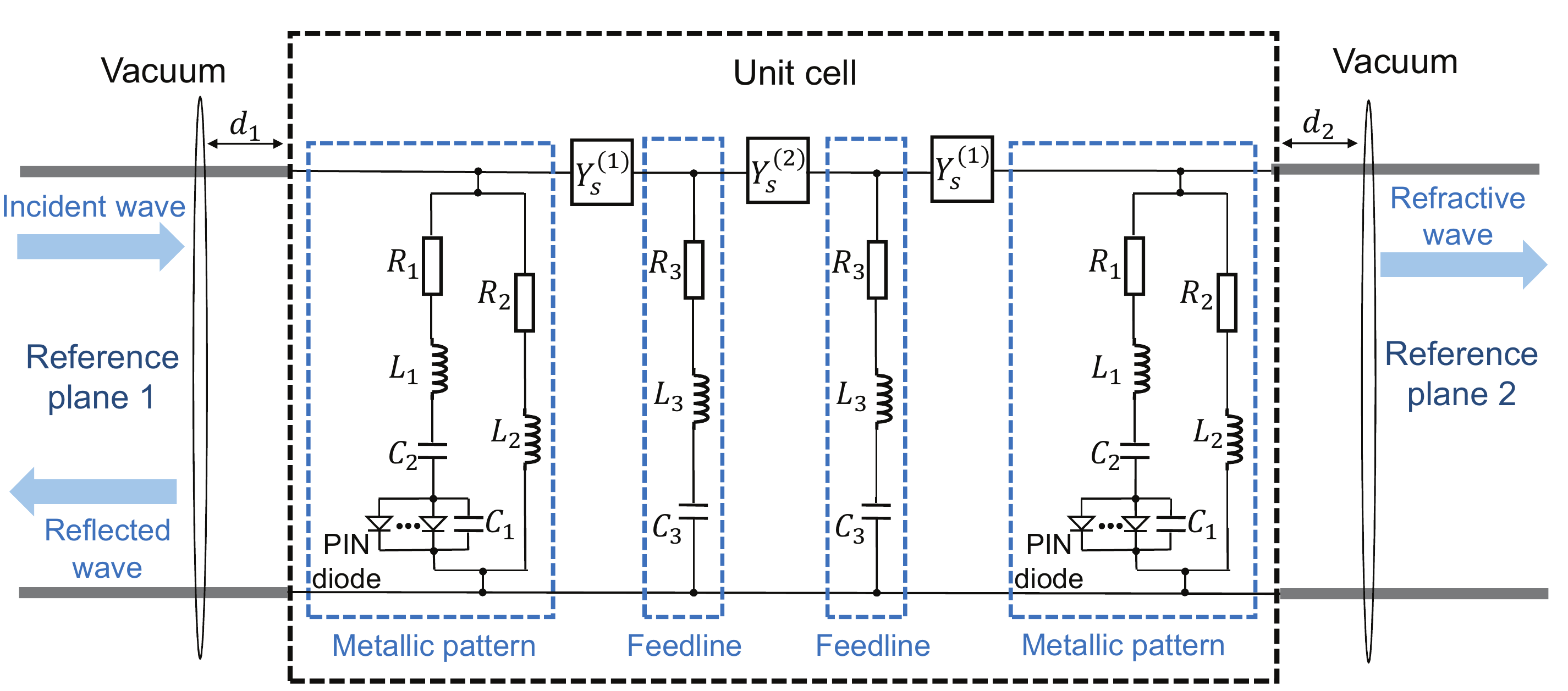}
	\caption{Equivalent circuit model of the IOS element.}
	\label{equivalent_circuit}
\end{figure*}

Existing works have considered the reflective (or refractive) model of IRS (or RRS) by either constructing the equivalence between the metasurface element and the RLC-vibrating circuit or investigating the electric field excited by the surface. In \cite{costa2021circuit}, a circuit based approach was proposed to model the reflection coefficient of the IRS by representing the periodic structure of the IRS as a shunt impedance and the dielectric part as transmission lines. In \cite{danufane2021path}, the electric field was modeled for both the IRS and RRS given different configuration patterns. It also reveals the relevance between the electric field distribution and the incident angles. In \cite{xu2017dual}, a triple-layer dual-mode RRS is designed where the refraction coefficient is fixed as a constant related to the dielectric constant. However, these models cannot be directly used in the IOS as they do not consider the energy split among refraction and reflection signals.

Some early works have considered the modeling of the IOS element. In \cite{xu2021simultaneously,liu2021simultaneously}, a simple load impedance model of the IOS element was presented where the refraction and reflection coefficients are modeled as functions of the electric and magnetic impedances. In \cite{xu2016multifunctional}, a general relation between the refraction and reflection coefficients for a single-mode two-port reciprocal system was developed based on coupled mode theory, which can also be applied to the metasurface design. 

The above works established a link between refraction/reflection coefficients and impedances or electric field distribution. However, to design an IOS with the desired refraction and reflection patterns, it is important to understand how the refraction/reflection coefficients are influenced by the geometric parameters of the IOS element structure. Below we briefly introduce an equivalent circuit model that has recently been developed for IOS, based on which a discussion on the influence of the IOS geometric parameters is delivered.

To model the IOS element as shown in Fig. \ref{structure_IOS}, a two-port microwave network is considered based on the equivalence between the vacuum on both sides of the IOS and the semi-infinite transmission lines\cite{torres2015accurate}. The equivalent circuit model is given in Fig.~\ref{equivalent_circuit} \cite{zeng2022intelligent}. For convenience, we introduce the concept of a metallic pattern, consisting of a metallic patch, a PIN diode, and the substrate. As mentioned in Section \ref{IOS-epsilon}, an IOS element consists of two symmetrical layers, each of which is formed by a metallic pattern and a feedline at the back of this metallic pattern. Thus, the whole circuit model contains four parts, corresponding to two metallic patterns and two feedlines due to the symmetric structure of IOS. 

In more detail, the metallic pattern can be represented by two resistors, two inductors, two capacitors, and a PIN diode. The pairs $\left\lbrace L_1, R_1\right\rbrace$ and $\left\lbrace L_2, R_2\right\rbrace$ refer to the inductance and resistance produced by the metallic patch and the substrate (connected to the ground), respectively. The pair $\left\lbrace C_1, C_2\right\rbrace$ refers to the capacitances with respect to the patch and substrate. We omit the RLC circuit equivalent to the PIN diode since the circuit structure varies with the state of the diode. Similarly, the feedline can be depicted by a series of RLC circuits $\left\lbrace R_3, L_3, C_3\right\rbrace$. The admittances $\left\lbrace Y_s^{(1)}, Y_s^{(2)} \right\rbrace $ depict the substrate-metal coupling and the inter-layer coupling, respectively. 

Given such an equivalent two-port network model, the 2$\times$2 transmission matrix can be given by
\begin{equation}
	\begin{split}
\begin{bmatrix}
	A &B \\C &D
\end{bmatrix}
=&\begin{bmatrix}
	1 &0 \\Y_{U,M} &1
\end{bmatrix}
\begin{bmatrix}
	1 &\frac{1}{Y_s^{(1)}} \\0 &1
\end{bmatrix}
\begin{bmatrix}
	1+\frac{Y_{F}}{Y_s^{(2)}} &\frac{1}{Y_s^{(2)}} \\2Y_F+\frac{Y_F^2}{Y_s^{(2)}} &1+\frac{Y_F}{Y_s^{(2)}}
\end{bmatrix}\\
&\begin{bmatrix}
	1 &\frac{1}{Y_s^{(1)}} \\0 &1
\end{bmatrix}
\begin{bmatrix}
	1 &0 \\Y_{L,M} &1
\end{bmatrix}.
\end{split}
\end{equation}
Note that $Y_F$, $Y_{U,M}$ and $Y_{L,M}$ are parallel admittances with respect to the feedline, the upper- and lower-layer metallic patterns, respectively. Each matrix on the right side of the equation is a scattering matrix of the corresponding network component\cite{costa2014overiew}. The terms $\left\lbrace Y_F, Y_{U,M}, Y_{L,M} \right\rbrace$ are complex-valued functions of $\left\lbrace L_i, C_i, R_i, Y_s^{(j)} \right\rbrace$ ($i=1,2,3$, $j=1,2$), as given in \cite{zeng2022intelligent}.
The reflection and refraction coefficients of the IOS element can then be given by
\begin{equation}\label{reflection-refraction}
\begin{aligned}
& \Gamma^{(l)}=\frac{(A+B/Z_0)-Z_0(C+D/Z_0)}{(A+B/Z_0)+Z_0(C+D/Z_0)}\exp\left(-j2\beta d_1\right),\\
& \Gamma^{(r)}=\frac{2}{(A+B/Z_0)+Z_0(C+D/Z_0)}\exp\left(-j\beta (d_1+d_2)\right),
\end{aligned}
\end{equation}
where $Z_0$ is the characteristic impedance determined by the working frequency, $d_1$ and $d_2$ are the distances between reference planes and the element's surfaces. Since $\left\lbrace Y_F, Y_{U,M}, Y_{L,M} \right\rbrace$ take complex values, the reflection and refraction coefficients in (\ref{reflection-refraction}) may have different amplitude and phase components. In other words, \emph{the reflection and refraction responses of one IOS element can be different in both amplitude and phase domains}, depending on the element structure design.

Though the close-form relationship between the reflection/refraction coefficients and the equivalent circuit model is constructed, one is still not able to design the IOS element structure in practice only based on such information. Thus, we show in Table \ref{table-relation} how the geometric parameters (i.e., $\left\lbrace g, w_p, l_p, l_E, w_E \right\rbrace $ in Fig. \ref{structure_IOS}.) influence these coefficients according to the full-wave analysis. For convenience, we use $\uparrow$ and $\downarrow$ to depict the positive and negative correlation, respectively. Table \ref{table-relation} serves as a guideline for IOS element design to obtain different response characteristics.
\begin{table}[!t]
	\renewcommand\arraystretch{1.2}
	\begin{center}
		\caption{Relation between IOS element geometric parameters and the RLC circuit model parameters}
	\begin{tabular}{|l|l|l|l|l|}
		\hline
		$\left\lbrace C_1,C_2\right\rbrace$  & $L_1$ & $L_2$ & $C_3$ & $L_3$ \\ \hline
		$w_E \uparrow$, $w_p \uparrow$,$g \downarrow$& $l_p \uparrow$ & $l_E \uparrow$ & $w_E \uparrow$, $l_E \downarrow$ & $w_F \uparrow$ \\ \hline
	\end{tabular}\label{table-relation}
\end{center}

\end{table}
\subsection{Channel Model}
\label{channel}
The IOS introduces reconfigurable propagation paths between the Txs and Rxs. Phase and amplitude responses of the IOS with respect to both reflection and refraction should be considered to depict the wireless channel. Some research studies have contributed to the channel model construction from different perspectives, including path loss and phase response models. A general expression of the channel $h_k$ between the $k$-th transmit antenna and the $s$-th receive antenna via the $m$-th IOS element can be given by
\begin{equation}\label{h_IOS}
h^{(k,s)}_m = {\sqrt{\beta^{(k,s)}_mG_m^{(k,s)}}}g^{(k,s)}_m\Gamma_m^{(x)}e^{j\theta_m^{(x)}},
\end{equation}
where $\beta^{(k,s)}_m$ is the path loss, $g^{(k,s)}_m$ depicts the small-scale fading, $G_m^{(k,s)}$ is the normalized power radiation pattern, and $\Gamma^{(x)}_m$ is the reflection (or refraction) coefficient of IOS element $m$. When the Rx is at the refractive side of the IOS, the superscript $x=r$; otherwise, we have $x=l$.

\subsubsection{Power Radiation Pattern}\label{power radiation pattern}
\begin{figure}[!t]
	\centering
	\includegraphics[width=0.3\textwidth]{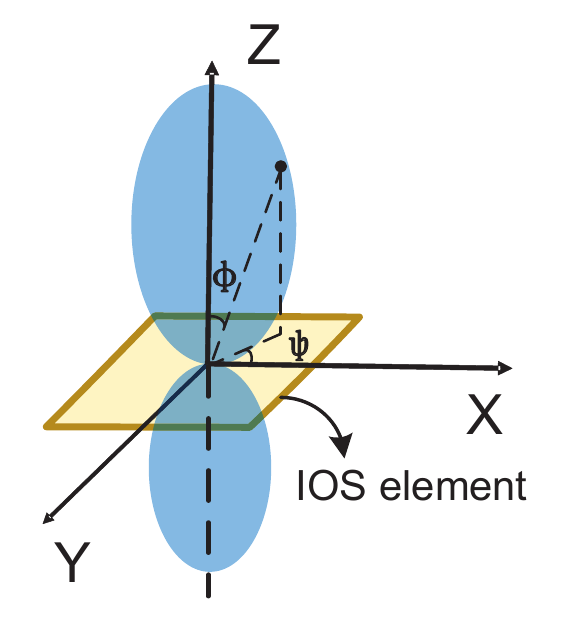}
	\caption{Normalized power radiation pattern of an IOS element.}
	\label{radi}
\end{figure}

It refers to the variation of power that arrives at the surface, usually related to the direction deviated from the surface. It depicts the maximum power in a certain direction toward the surface. According to the results in \cite{tang2020wireless}, the normalized power gain is a function of the incident angle $\phi$ in the form of, for example, $|\cos{\phi}|^3$, as shown in Fig. \ref{radi} \cite{zhang2021intelligent}. Different function forms are discussed in \cite{xu2021star},\cite[eq.(8)-(55)]{balanis2012advanced}.

\subsubsection{Phase Response Model}
\label{phase}
The phase shift response of each IOS element directly influences the quality of the Tx-IOS-Rx link. In general, the phase shifts for refraction and reflection are coupled as they are simultaneously determined by the PIN diodes implemented in the IOS. Moreover, the phase shifts are also influenced by parameters as introduced in Section \ref{modeling}, and thus can be optimized.  

A linear model is widely used in the literature. To be specific, a pair of reflection and refraction phase shifts satisfies $\theta_t(i) - \theta_r(i) = c(i)$, where the subscript $(i)$ refers to the $i$-th IOS element state and $c(i)$ is a constant. Since the number of element states is finite in practice, the number of achievable phase shifts of each element is also finite and discrete, i.e., $c(i)$ takes discrete values. In Section \ref{response}, the experimental results on phase responses with respect to two different element states will be reported for further verification.

Some recent works have further investigated the practical coupling effect. For example, in \cite{zeng2022intelligent}, the physical meaning as well as the EM mechanism behind the coupling effect were provided from the perspective of equivalent circuit theory. The work of \cite{zhu2014dynamic} pointed out that typically a lossless metasurface equipped with zero-junction-resistance diodes can achieve the reflection-refraction independence, which is non-trivial in practice. The coupling between the refraction and reflection phases and amplitudes is given in a closed form, which is applied in \cite{liu2021simultaneously,xu2022star} for outage performance analysis.

\subsubsection{Path Loss Model}\label{path_loss_part}
Inheriting the features of metasurface, the IOS shares the same path loss model with the IRS. Three key factors influencing the path loss $\beta^{(k,s)}_m$ as listed below.

\begin{itemize}
	\item \textbf{Effective surface area:} When the size of the IOS is large, not all the IOS can receive incident signals from the Tx. \emph{Effective surface area} refers to the IOS area on which incident signals from the Tx are projected. Denote the distance between the IOS and Tx as $d_1$. If $d_1 > \frac{2\eta}{\lambda}$ where $\eta$ and $\lambda$ are the border length of the IOS and the wavelength, respectively, the effective area equals the IOS area. Otherwise, the edge of the IOS may not receive the incident signal, and this area is an ellipse since the main lobe of the Tx has a conical shape according to the conic-section theory. Since the main lobe carries most transmit energy of a beam, the effective area of the IOS, denoted by ${\mathcal{S}}_e$, can be approximated by the projection of the main lobe, which is determined by the beamwidth of the incident beam and $d_1$, as detailed in \cite{ntontin2021optimal}.
	
	\quad Given the concept of effective surface area, the amplitude response of the refracted (and reflected) signal for an IOS element $m$ satisfies
	\begin{equation}
		\Gamma_{m}^{(l)} =\Gamma_m^{(r)}= 0,\mbox{if}\ \textbf{z}_m \notin {\mathcal{S}}_e \ \mbox{and}\ d_1 < \frac{2\eta}{\lambda},
	\end{equation}
	where $\textbf{z}_m$ is the position of element $m$.

	\item \textbf{Distance-related scaling law:} It differs when the metasurface acts as a focusing lens and the collection of numerous scatters \cite{tang2020wireless}. 
	
	\quad The IOS acts as a flat lens when 1) the surface size is much larger than the wavelength, 2) the reflection (and refraction) coefficients of all IOS elements are the same, i.e., all elements operate in an identical state. In this case, the IOS performs specular reflection and transmission based on the geometric optic rules. The free-space path loss of the transmission on both sides of the IOS is proportional to $(d_1 +d_2^{(l)})^2$ and $(d_1 +d_2^{(r)})^2$, respectively, where $d_1$ is the distance between the IOS and the Tx, $d_2^{(l)}$ and $d_2^{(r)}$ are the distances between the IOS and the Rx on both sides of IOS, respectively. The flat-lens mode can be utilized for near-field broadcasting toward desired directions. More details can be found in \cite{tang2020wireless}.
	
	\quad Otherwise, the IOS acts as a collection of scatters when IOS elements have different reflection (and refraction) coefficients from each other. The free-space path loss on both sides of the IOS is proportional to $(d_1 d_2^{(r)})^2$ and $(d_1d_2^{(l)})^2$, respectively. This is a widely used path-loss model in the literature \cite{zhang2020beyond} since the IOS in this case can be leveraged to generate different beams toward specific users, enabling the function of beamforming. 
	
	\item \textbf{Far-field v.s. near-field:} Similar to the IRS-aided wireless communication system, an IOS-aided one also needs to separate its near-field and far-field regions as their corresponding propagation characteristics are different. A commonly used criterion, i.e., Rayleigh distance $2L^2/\lambda$, is used to differentiate the far-field and near-field regions \cite{johnson1973determination}. Here, $L$ is the largest dimension of the IOS. Let $z$ be the distance of a particular field point to the IOS. The points where $z$ equals to the Rayleigh distance are the boundary of the far-field and near-field regions. 
	
	\quad In general, the difference between near-field and far-field regions is how the power density changes with distance. For example, assume that the energy spread over a certain area $s$ within a fixed angle $\Psi$, and we can find that the power density around the focal point is proportional to $\Psi/s$ According to the results in \cite{arun2020rfocus},  after reflection/refraction, the area in the near-field region is proportional to $\lambda^2(1 + 4(z/L)^2)$, which is a typical spherical dissipation of the signal power with the distance. In the far-field region, as $z/L$ dominates, we have $\Psi/s \approx \frac{L^2\Psi}{4\lambda z^2}$, indicating that it can be approximated as a plane wave. In other words, the channel models for far-field and near-field regions should be different, especially when the IOS is large. 
	
\end{itemize}


\subsection{Summary and Outlook}
In this section, we have introduced the refractive-refractive model for the IOS presented in Section \ref{IOS modeling} and the channel model for the IOS-aided wireless communications. From the above models, we can have the following take-home messages:
\begin{itemize}
	\item Refraction and reflection responses of an IOS can be different in both amplitude and phase shifts, which is highly related to the structure design of the IOS element. 
	
	\item For the IOS element introduced in Section \ref{IOS modeling}, once the ON/OFF states of embedded PIN diodes are given, the amplitudes and phase shifts for both refraction and reflection signals are all determined. In other words, the amplitude and phase shift characteristics should be designed jointly. As the control principle of the IOS is the same as that of the IRS, the energy split will not incur extra hardware costs. However, the hardware cost of the IOS is slightly higher than that of the IRS due to an additional layer for refraction.
	
	\item Refraction and reflection signals are highly coupled as they are determined at the same time when the states of PIN diodes are set. This leads to the essential difference between IOS-aided wireless communication networks and IRS-aided ones. Due to the coupling, the IOS cannot be regarded as a simple combination of an IRS and an RRS, which requires more sophisticated designs for such a system. 
\end{itemize} 

With the models given in this section, we can have the following two possible research lines:
\begin{itemize}
	\item \emph{Parameter optimization:} As we have introduced in this section, the response of an IOS element is influenced by many parameters, such as equivalent inductance, resistance, and capacitance. In order to improve the performance for various use cases, the amplitude and phase shifts of the IOS's response should be well designed. For example, we can use an IOS to enable the switching between reflection-only and refraction-only functions. We can select a part of the configurations to set the refraction amplitudes close to zero, while the reflection amplitudes for the other configurations are set to zero. In this way, we can switch between reflection-only and refraction-only modes by simply adjusting the states of the PIN diodes. To realize such a function, we need to carefully design these parameters to meet these requirements. In the literature, there are some initial investigations \cite{zhang2022ergodic,wang2022performance}. The authors in \cite{zhang2022ergodic} presented the impact of the number of possible phase shifts on the ergodic capacity for IOS-aided wireless communication systems. Their results showed that a small number of phase shifts are sufficient enough to realize the potential of the IOS. In \cite{wang2022performance}, the authors considered an IOS-aided non-orthogonal multiple access (NOMA) setting. Their simulation and theoretical results showed that low-precision elements with only four phase shifts can achieve performance close to that of ideal continuous phase shifts. Moreover, the average achievable rates with correlated channels and uncorrelated channels are asymptotically equivalent, which is independent of the number of IOS elements. 
	
	\item \emph{Type selection:} Different surfaces have their own use cases, but the type selection\footnote{Although it is still possible to switch among different functions, its function is determined once the surface is manufactured. We need to determine which type of the surface to produce before the deployment rather than switching among different possible modes provided by the surface. In fact, the mode selection is equivalent to the beamforming scheme to be introduced in Section \ref{fd-IOS} with the amplitudes of some states being zero.} is still an open problem in the literature. To meet the diverse performance metrics in future cellular systems, the selection criterion will be quite different even with the same user distribution, which is also valuable to be investigated. An example study on the capacity was given in \cite{zeng2021reconfigurable}. The authors derived the system capacity, and analyzed the optimal type of the surface under a specific user distribution. Their analysis reveals that the type selection is also influenced by the position of the surface. However, when the size of the surface exceeds a certain threshold, the IOS is always the best choice.
\end{itemize} 





\section{IOS enabled Full-dimensional Wireless Communications}
\label{fd-IOS}
\begin{figure*}[!t]
	\centering
	\includegraphics[width=0.75\textwidth]{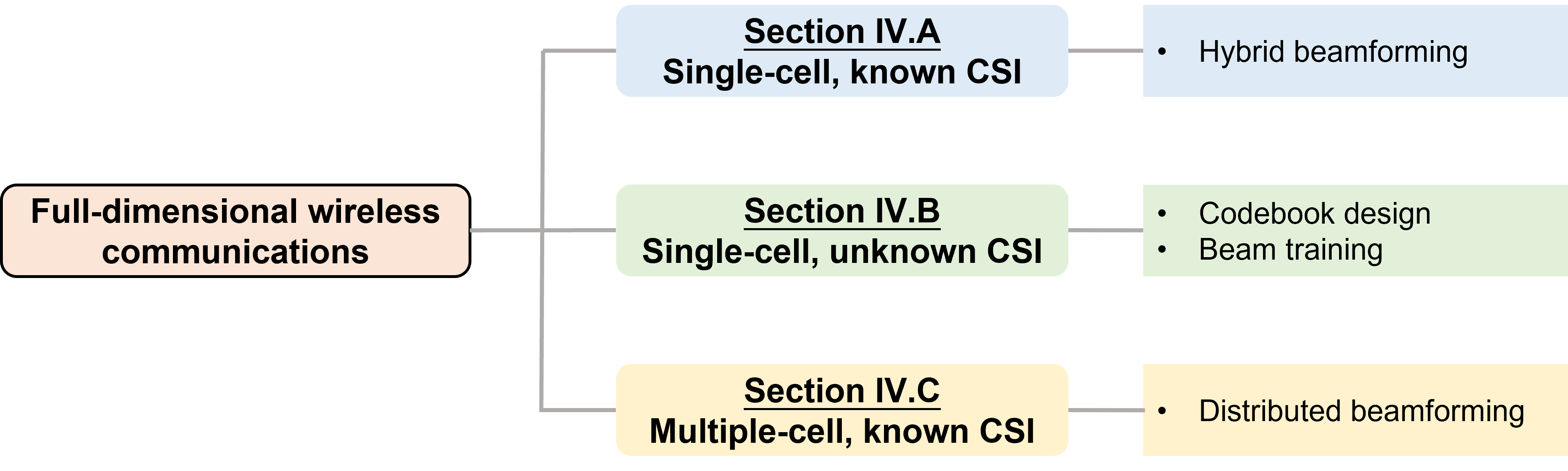}
	\caption{Classification of techniques for IOS-aided full-dimensional wireless communications.}
	\label{classification}
\end{figure*}

The IOS has shown its potential in enabling transmissions toward both sides of the surface, but the mutual influence between refraction and reflection coefficients also brings new challenges. In this section, we illustrate how the full-dimensional beamforming is achieved, assisted by the IOS, in three scenarios, as outlined in Fig. \ref{classification}. In the first two scenarios, we will consider a single-cell case, but the availability of channel state information (CSI) is different. In Section \ref{csi_aware}, we introduce the hybrid beamforming scheme with known CSI. For the CSI-unknown case, we propose a codebook-based beamforming scheme, where codebook design and beam training are introduced in Section \ref{unknown}. In Section \ref{multi-cell}, we extend the single-cell case to a multiple-cell case. Here, we present a distributed beamforming scheme with known CSI. For the unknown-CSI case, we can easily use the codebook-based scheme provided in Section \ref{unknown}. Following that, other related works are introduced and discussed. 

Before introducing the beamforming algorithms for these three scenarios, we would like to first present how the users, the BS, and the IOS are synchronized, especially when the LoS link is blocked\footnote{If the LoS link is not blocked, we can use the LoS link for the synchronization following the same process in current cellular networks, and thus we only show the synchronization procedure without the LoS link in this paper.}. Following a similar synchronization process for 5G new radio (NR) \cite{sync}, the synchronization can be achieved with four steps:
\begin{itemize}
	\item \emph{Step 1:} The user might not be able to receive the synchronization signals as the LoS link is blocked. To address this issue, the BS should have a set of predefined configurations of the IOS to ensure that each user on either side can be served at least once with these configurations. Therefore, during the discovering phase, the IOS will traverse all the configurations to make sure that the user can receive the synchronization signals \cite{hashida2022adaptive};
	
	\item \emph{Step 2:} The user obtains downlink synchronization when the user can receive the synchronization signals and decode the system information;
	
	\item \emph{Step 3:} When the downlink is synchronized, the user can access based on the channel reciprocity and obtain uplink synchronization;
	
	\item \emph{Step 4:} The BS and the user establish a dedicated connection through the IOS and the user obtains a dedicated connection identification (ID).
\end{itemize}
With the synchronization protocol, the beamforming schemes are presented under the assumption that the users, the BS, and the IOS are perfectly synchronized.

\subsection{Hybrid Full-dimensional Beamforming for the CSI-Known Case}\label{csi_aware}
In this part, we investigate how to leverage the full-dimensional transmission capability of the IOS to extend the cell coverage when the CSI is known to the BS. 

\subsubsection{Key Idea}
\begin{figure}[!t]
	\centering
	\includegraphics[width=0.45\textwidth]{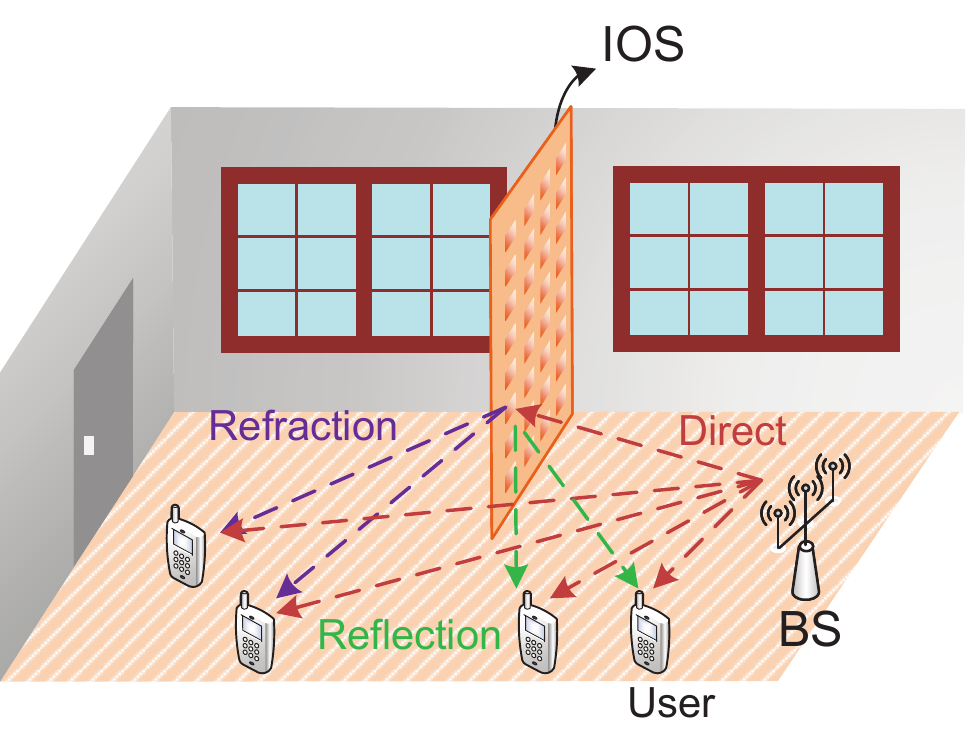}
	\caption{Scenario of the multi-user IOS-aided wireless communication.}
	\label{beamformer:system}
\end{figure}

Consider a downlink multi-user multiple-input single-output (MISO) scenario consisting of one BS and multiple users as shown in Fig. \ref{beamformer:system}. Due to the intractable scattering characteristics of propagation environments, cell-edge users may suffer severe link quality degradation. By deploying an IOS within this cell, the originally scattered signals can be collected by the IOS and intelligently reflected/refracted toward those cell-edge and blocked users \cite{zhang2020beyond}. 

\begin{figure*}[!t]
	\centering
	\includegraphics[width=0.75\textwidth]{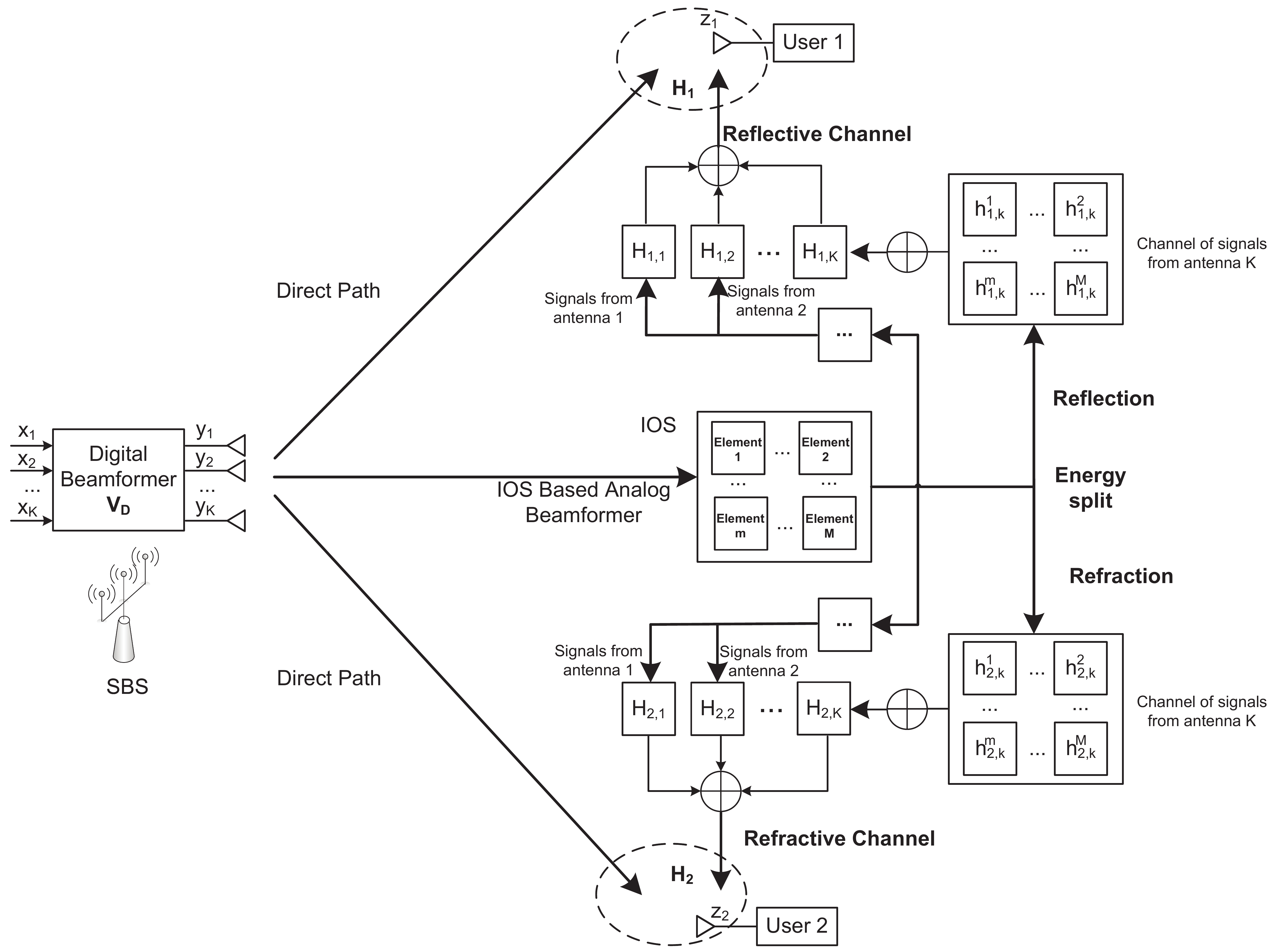}
	\caption{Block diagram of the IOS-based beamforming for 2 users.}
	\label{beamformer}
\end{figure*}

Each user receives signals from both the direct link between the BS and itself, as well as the extra reflected/refracted link introduced by the IOS. The channel between user $k$ and the $n$-th antenna of the BS can be given by
\begin{equation}\label{h_ks}
h^{(n,k)} = \underbrace{\sum_m \left( \sqrt{\frac{\kappa}{\kappa+1}}h_m^{(n,k)}  + \sqrt{\frac{1}{\kappa+1}}h_{m,NLoS}^{(n,k)}\right)}_{\text{reflected/refracted links}}  + h_D^{(n,k)},
\end{equation}
where $h_{m,NLoS}^{(n,k)}$ depicts its non-line-of-sight (NLoS) paths, $h_D^{(n,k)}$ refers to the direct link between the user and the IOS, and $h_m^{(n,k)}$ is defined in $(\ref{h_IOS})$. The BS-IOS-user link is modelled as a Rician channel with factor $\kappa$, implying that the reflection/refraction enabled by the IOS contributes to an extra LoS link between the BS and the user. Benefiting from such reflected/refracted paths, the received signal strength of the user is enhanced, and thus, the cell coverage is extended. 
\subsubsection{Hybrid Beamforming Scheme}\label{csi_aware_hybrid}
Note that the reconfigurable IOS elements have no digital processing capability, they do not serve as the RF active components as those in traditional digital phase arrays. It is thus necessary to develop a hybrid beamforming scheme where the BS performs the digital beamforming to support multiple users, and the analog beamforming is delivered by the IOS due to its phase-shifting ability.

The received signal of $K$ users, each equipped with a single antenna, can be expressed by a $K \times 1$ vector, i.e.,
\begin{equation}\label{z_all}
\bf{z} = \bf{H}\bf{V_D}\bf{x} + \bf{w},
\end{equation}
where $\bf{w}$ and $\bf{x}$ are both $N \times 1$ vectors representing the noise and the transmitted signal, respectively, $\bf{V_D}$ is a $K \times N$ digital beamforming matrix at the BS with $K$ being the number of BS transmit antennas, and $\bf{H}$ is a $N \times K$ equivalent channel matrix depicting the propagation of both direct links and IOS-aided links. Each element is given in $(\ref{h_IOS})$ and $(\ref{h_ks})$. Based on $(\ref{z_all})$, the received signal of user $n$ can be given by
\begin{equation}\label{z_k}
	z_k = \left({\bf{h}}_{D}^H + {\bf{h}}_{IU,k}^H{\bf{Q}}{\bf{H}}_{BI}\right) {\bf{V}}_D{\bf{x}} + n_k,
\end{equation}
where ${\bf{h}}_{D}$ and ${\bf{H}}_{BI}$ are the channels of the direct BS-user link and the BS-IOS link, respectively, ${\bf{h}}_{IU,k}$ denotes the channel between IOS and user $k$. All these channels contain the path-loss components as introduced in Section \ref{path_loss_part}. The notion ${\bf{Q}}$ is a $M \times M$ diagonal matrix, where $M$ is the number of IOS elements. Each element of ${\bf{Q}}$ contains the power gain and phase-related term ${\sqrt{G_m^{(n,k)}}}\Gamma_m^{(x)}e^{j\theta_m^{(x)}}$.

A block diagram for the beamforming scheme with a two-user case is shown in Fig. \ref{beamformer}, where the BS transmits two different data streams to the users separately \cite{zhang2021intelligent}. According to $(\ref{z_all})$ and $(\ref{z_k})$, the BS first encodes the transmit signal targeted at different users via a digital beamformer. After the signal arrives at the IOS, a part of the energy is reflected by the IOS and the remaining part (excluding the dielectric loss) is refracted by the IOS toward two users, respectively. Compared to the incident signal, the phases of reflected and refracted signals are first configured by the IOS, and these signals then go through the IOS-user channels to serve different users.

For the beamforming matrix design, the sum rate maximization leads to the following optimization problem:
\begin{subequations}\label{opt}
	\begin{align}
		\max\limits_{\left\lbrace {\bf{Q}},{\bf{V}}_{D}\right\rbrace } & \sum\limits_{n} R_n, \\
		\operatorname{s.t.} &~\left( \theta_{m}^{(t)} , \theta_m^{(r)}\right) \in \left\lbrace (c^{(t)}(i),c^{(r)}(i) ) \right\rbrace_{i=1}^{S}, \label{constraint_1}\\
		&~Tr({\bf{V}}_{D}{\bf{V}}_{D}^H) \le P_T,\label{constraint_2}
	\end{align}
\end{subequations}
where $R_n$ is the data rate of user $n$, which can be expressed as $R_n = \log_2(1 + \gamma_n)$. Here, $\gamma_n$ is the received Signal-to-interference-plus-noise ratio (SINR). Constraint $(\ref{constraint_1})$ depicts the coupling between discrete reflection and refraction phase shifts, and $S$ is the number of discrete reflection-refraction phase shifts. An alternate to this constraint is the linear model introduced in Section \ref{phase}, i.e., $\theta_m^{(t)} - \theta_m^{(r)} = c$, where $c$ is a constant. This model also works when continuous phase shifts are considered. We also assume that the amplitudes for different states are fixed. Constraint $(\ref{constraint_2})$ sets the upper bound of the transmit power. It is worth pointing out that unlike the hybrid beamforming problem for IRS-aided wireless communication systems, as given in \cite{ying2020gmd,gao2021reflection}, the optimization problem for an IOS-aided wireless communication system needs to consider the coupling of refraction and reflection, which necessitates the development of novel algorithms for these problems. 

Such an optimization problem can be solved by iteratively performing the digital beamforming (via zero forcing \cite{spencer2004zero} or minimum mean square error (MMSE) \cite{shi2011iteratively} methods) and the IOS-based analog beamforming (via alternative optimization \cite{di2020practical} or majorization-minimization \cite{kundu2021channel} or successive convex approximation \cite{chen2020reconfigurable}, etc.). In the following parts of this paper, we also use this hybrid beamforming scheme. Therefore, we interchange IOS phase shift designs with IOS beamforming, which generally refers to the analog beamforming at the IOS.

\subsubsection{Numerical Results}
\begin{figure}[!t]
	\centering
	\includegraphics[width=0.45\textwidth]{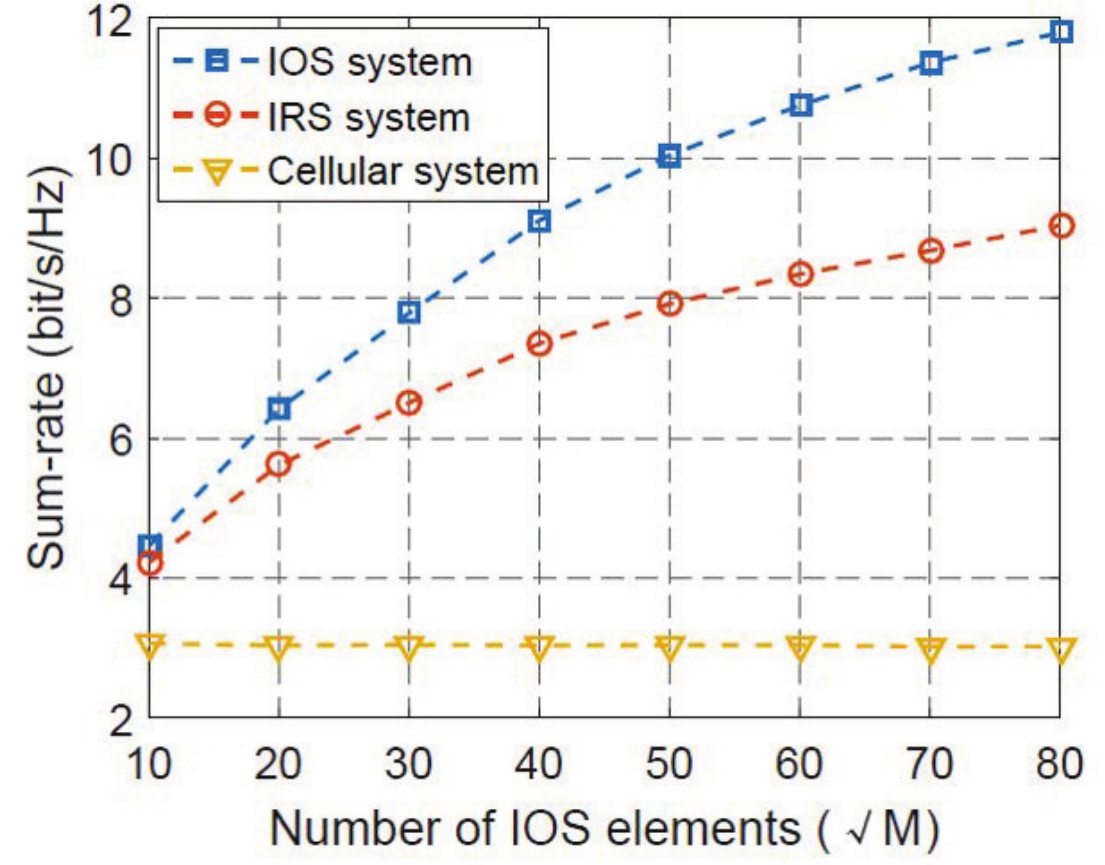}
	\caption{Sum-rate of IOS-aided communication systems \cite{zhang2021intelligent}.}
	\label{beamformer:fig1}
\end{figure}

\begin{figure*}[!t]
	\centering
	\includegraphics[width=0.75\textwidth]{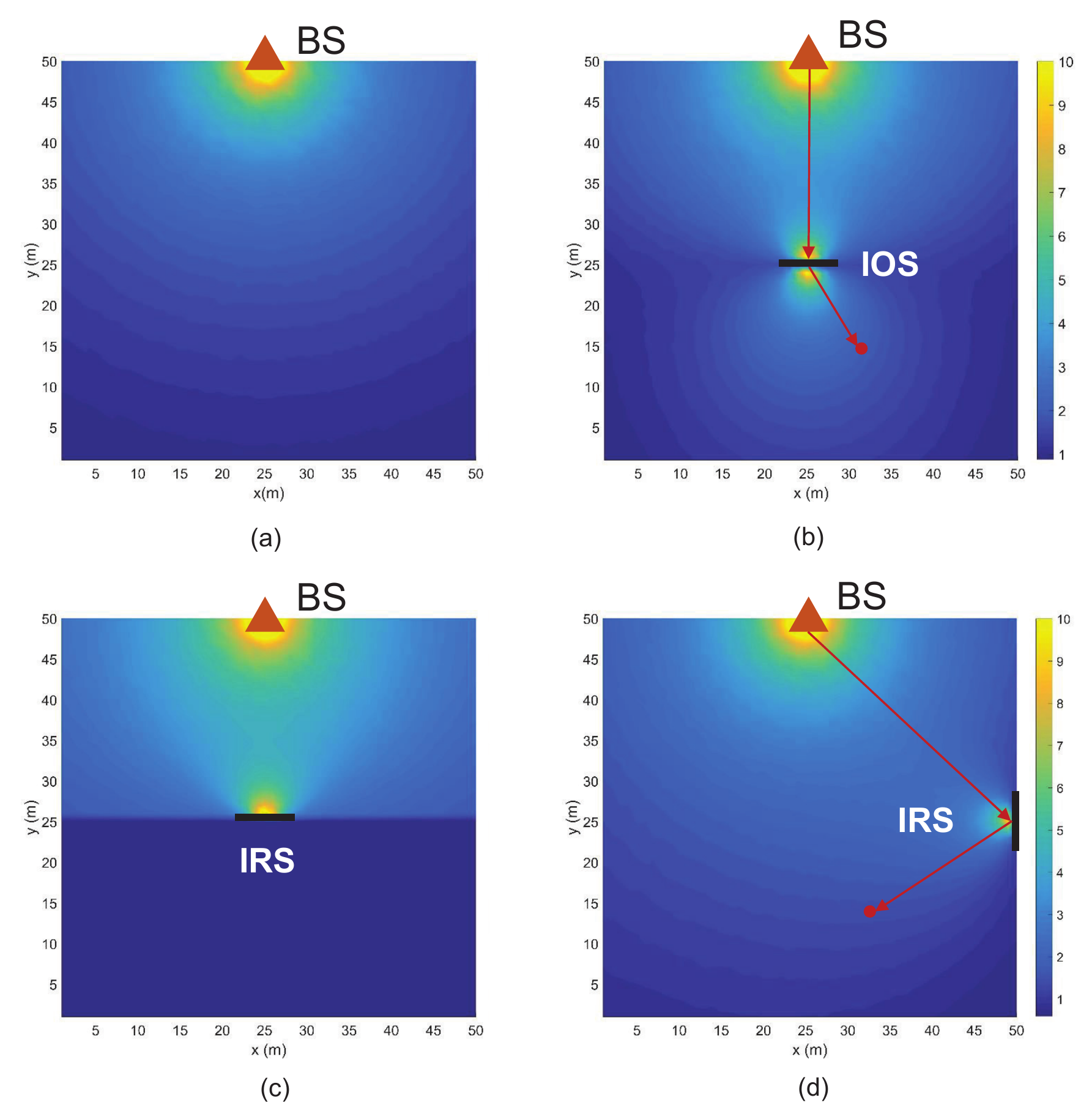}
	\caption{Data rate distribution: (a) non-IOS; (b) IOS-aided; (c) IRS-aided (in the middle); IRS-aided (on the side).}
	\label{beamformer:fig2}
\end{figure*}

The performance of a typical IOS-aided communication system is evaluated in Figs. \ref{beamformer:fig1} and \ref{beamformer:fig2} (see \cite{zhang2021intelligent} for details). As shown in Fig. \ref{beamformer:fig1}, up to a 30$\%$ data rate improvement can be achieved by the IOS case compared to the IRS-aided case, which shows the superiority of IOS in serving multiple users on both sides. Moreover, the system sum-rate will increase with the number of IOS elements, but the growth rate slows down with a larger size of the IOS. Fig. \ref{beamformer:fig2} depicts the spatial distribution of one user's data rate, i.e., the data rate of a user at different locations, under different schemes. Compared to the non-IOS case in Fig. \ref{beamformer:fig2}(a) where the data rate decreases as the user is farther away from the BS, while in the IOS-aided case as given in Fig. \ref{beamformer:fig2}(b), the data rate on both sides of the surface is improved significantly. Moreover, from Figs. \ref{beamformer:fig2}(b) and \ref{beamformer:fig2}(c) we can observe that the data rate on the lower part of the surface remains at a very low level when we replace the IOS with an IRS at the same position. These verify the capability of IOS to enhance the communication quality on both sides. Finally, we compare the results in Figs. \ref{beamformer:fig2}(b) and \ref{beamformer:fig2}(d) and we can see that at the same position (the red point), its data rate with the IOS is higher than that with the IRS. This justifies the higher capability of the IOS to extend the coverage.

\subsection{Codebook Design and Beam Training for the CSI-Unknown Case}\label{unknown}
In this subsection, we discuss how to serve multiple users in the IOS-aided systems when the CSI is not known in advance by the BS. 
\begin{figure*}[t]
	\centering
	\includegraphics[width=0.95\textwidth]{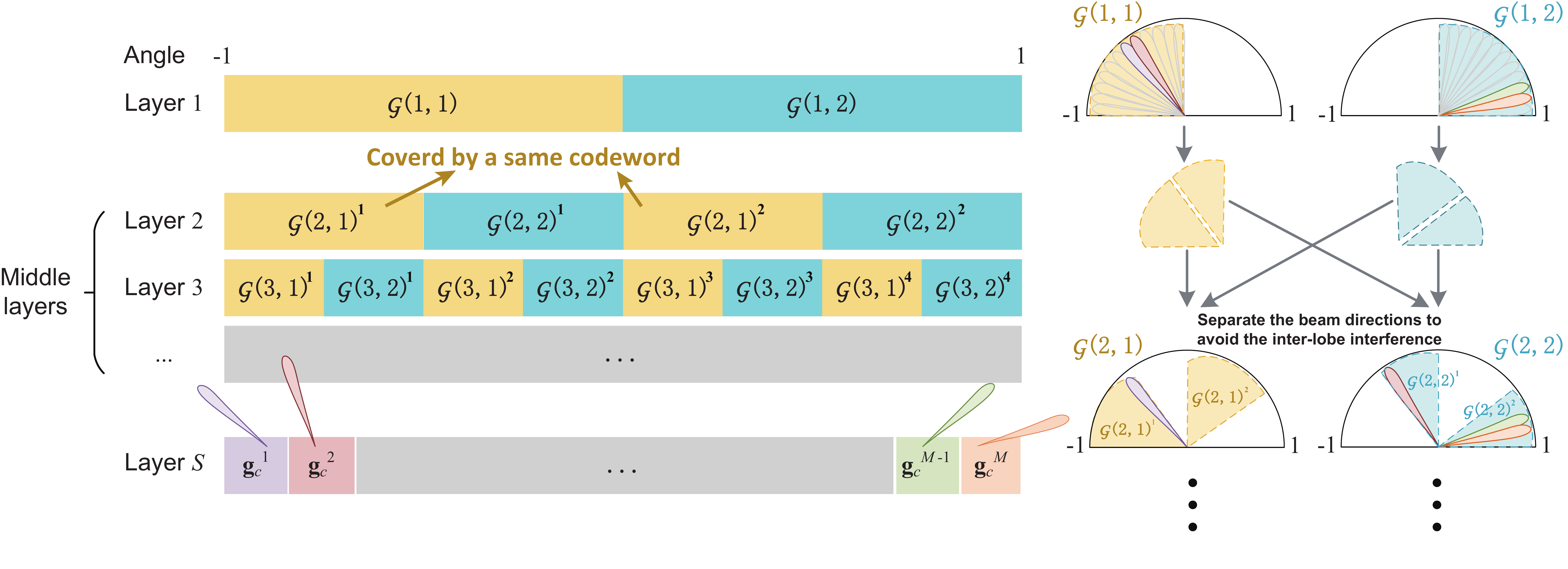}
	\caption{Illustration of the codebook design.}
	\label{Fig:codebook}
\end{figure*}

\subsubsection{Key Idea}
Consider an IOS-aided system where the BS equipped with $K$ antennas serves $N$ users, each equipped with $N_r$ antennas, via an IOS of $M$ elements. When the IOS size $M$ is rather large, the traditional channel estimation methods may suffer a huge complexity since the number of channels is proportional to $KMNN_r$, which leads to tens of thousands of channels to be estimated \cite{wang2020compressed}. To avoid such a huge complexity, the codebook-based beam training method that does not rely on the acquisition of real-time CSI is a promising solution. Each codeword refers to a directional beam, and the users send a feedback on the received signal strengths with respect to different codewords. In this way, the BS selects suitable codewords to generate the beamforming matrix for serving multiple users. 

However, the traditional beam training methods do not fit the IOS-aided systems. As mentioned in Section \ref{power radiation pattern}, the power radiation pattern of the IOS depends on the angle of arrival from the BS, which then influences the user's received signal. Given such a coupling between the BS and the IOS, the existing beam training methods where the codebook is designed solely for the BS or the metasurface do not work well in this case.

To depict the coupling between the BS and IOS, a reflective-refractive beam training method is proposed by optimizing the BS and IOS codebooks as well as codewords selection jointly such that the selected IOS codewords are adaptive to the incident signal of the BS. Below the joint codebook based beamforming scheme, the codebook design and beam training methods are presented, respectively. Compared to the IRS-aided systems, which only consider the reflection region, the optimal codebook for reflection might not work for the refraction region in the IOS-aided systems. Therefore, it is necessary to find a trade-off between the performance for reflection and that for refraction.


\subsubsection{Codebook Based Beamforming Scheme}

Recall that the received signal at user $k$ is given in $(\ref{z_k})$. The beamformer of the BS $\bf{V}$ can be written as $\bf{V} = {\bf{V}}_A{\bf{V}}_D$, where ${\bf{V}}_A$ and ${\bf{V}}_D$ are analog and digital precoders, respectively. The codebook based beamforming scheme aims to obtain the optimal $\bf{V}$, $\bf{Q}$, and ${\bf{w}}_k$ when  ${\bf{H}}_{IU,k}$ and ${\bf{H}}_{BI}$ are unknown. The following procedures are performed \cite{zhang2022dual}.

\textbf{Codebook-based analog precoding at the BS}: a codebook employed at the BS contains optimal codewords ${\bf{v}}_c^*(k)$ for each user $k$ from a candidate set $\mathcal{V}$ to form the analog precoder ${\bf{V}}_A = \left[{\bf{v}}_c^*(1),\cdots,{\bf{v}}_c^*(K)\right] $. Divide the area between the BS and the IOS into $N_b$ sections, each of which is depicted by a codeword in $\mathcal{V}$. That is, the $i$-th codeword enables the beam to be projected toward the center point $(x_i,y_i)$ of the $i$-th section. At the analog precoding stage, the BS sets a random IOS phase shift matrix $\bf{Q}$ and sequentially selects different codewords to send signals to the users. Each user sets an omni-directional combiner ${\bf{w}}_k$ and reports the received signal strengths corresponding to different codewords to the BS. Based on the feedback from users, the BS then selects $K$ orthogonal codewords from $\mathcal{V}$ that maximize users' signal strengths to form ${\bf{V}}_A$.

\textbf{Codebook based IOS beamforming}: we first rewrite ${\bf{H}}_{BI}$ as ${\bf{H}}_{BI} = {\bf{H}}_{I}{\bf{H}}_{B}$, where ${\bf{H}}_{I} \in {\mathbb{C}}^{M \times K}$ and ${\bf{H}}_{B} \in {\mathbb{C}}^{K \times N}$ are steering matrices of the IOS and the BS, respectively. That is, the channel between the BS and the IOS consists of $K$ propagation paths, referring to $K$ sections selected by the BS in procedure (a). The ${\bf{Q}}{\bf{H}}_{BI}$ term in $(\ref{z_k})$ can then be rewritten as ${\bf{Q}}{\bf{H}}_{BI} = {\bf{Q}}{\bf{H}}_{I}{\bf{H}}_{B} = {\bf{G}}{\bf{H}}_{B}$. A codebook is deployed at the IOS such that an optimal codeword $g_c^*(k)$ is selected for each user $k$ from a candidate set $\mathcal{G}$ to form $\bf{G} = \left[g_c^*(1),\cdots,g_c^*(K) \right] $. Denote the number of codewords in $\mathcal{G}$ as $N_G$. The $p$-th codeword in $\mathcal{G}$ enables a refracted/reflected beam in the direction of $\phi_I(p) = -1 + \frac{2p-1}{N_G}$ with a beam coverage of ${\mathcal{C}}=\left[-1+\frac{2p-2}{N_G},-1+ \frac{2p}{N_G}\right]$. As such, the codeword $g_c^*(k)$ is selected to refract/reflect the incident signal toward user $k$ to maximize the received signal strength, i.e.,
\begin{equation}\label{ios_code_max}
{\textbf{g}}_c^*(k) = \arg\max\limits_{{\textbf{g}}_c(k) \in \mathcal{G}} \gamma_k.
\end{equation}
Note that the influence of the incident signal on the IOS phase shifts is reflected by the matrix ${\bf{G}}$. Instead of training an optimal ${\bf{Q}}$, we target at the optimal ${\bf{G}}$, which is then used to construct the optimal ${\bf{Q}}$, as shown below.

\textbf{Analog combining at users}
In addition to the IOS-based beamforming, each user $k$ performs the analog combining by selecting an optimal codeword from the candidate set $\mathcal{W}$ in order to achieve the maximum received signal strength:
\begin{equation}
	{\textbf{w}}_k = {\textbf{w}}_c^*(k) = \arg\max \gamma_k, {\textbf{w}}_c^*(k) \in \mathcal{W}.
\end{equation}

\textbf{IOS phase shift matrix construction}: based on the beam training results $\bf{G}$ and ${\bf{V}}_A$, the IOS phase matrix $\bf{Q}$ can then be constructed.
\begin{equation}
\begin{split}
\bf{Q} = &\arg\min_{\bf{Q}}\left| {\bf{G}} - {\bf{Q}}{\hat{\bf{H}}}_I\right|^2 \\
=& \arg\min_{\bf{Q}}\sum_{k=1}^{K}\left| {\bf{g}}_c^*(k) - {\bf{Q}}{\hat{\bf{h}}}_{I,k}\right|^2,
\end{split}
\end{equation}
where ${\hat{\bf{H}}}_I$ is the estimated receive steering matrix. Each column vector ${\hat{\bf{h}}}_{I,k}$ can be generated based on ${\hat{\bf{h}}}_{I,k} = \left[e^{\frac{2\pi}{\lambda}d_1},\cdots, e^{\frac{2\pi}{\lambda}d_n}\cdots \right]^T $, where $d_n$ is the distance between the $n$-th BS antenna and the center of section $k$, i.e., $(x_k,y_k)$, corresponding to the $k$-th selected codeword ${\bf{v}}_c^{*}(k)$. This is because the $k$-th propagation path passes a scatter located at $(x_k,y_k)$.

\textbf{Digital precoding at the BS}: based on the selected codewords for the BS, users, and the IOS, we can then construct the equivalent channel matrix between the BS and users. Therefore, the digital precoding can be performed by the BS via traditional methods such as zero-forcing or MMSE.

\subsubsection{Codebook Design and Beam Training}
Since the BS and each user only target projecting toward the IOS, the conventional single-lobe hierarchical codebook \cite{alkhateeb2014channel} is employed. To search for the optimal codewords for multiple users simultaneously, we consider a multi-lobe beam training method for the IOS. 

The first step is to generate a multi-layer codebook based on the basic single-lobe candidate set $\mathcal{G}$. For a $S$-layer codebook as shown in Fig. \ref{Fig:codebook}, the bottom $S$-th layer consists of $N_G$ codewords, and each covers a single lobe \cite{zhang2022dual}. From the first to the $S-1$-th layer, each layer consists of two codewords (colored by yellow and blue in each layer). For example, on the second layer, each codeword consists of two lobes, which can be generated by the weighted sum of multiple basic codewords in $\mathcal{G}$.

The beam training process is performed layer by layer. By collecting the feedback of each user on these multi-lobe codewords, the BS can then infer the optimal single-lobe codeword ${\bf{g}}_c^{*}(k)$ for each user $k$ according to $(\ref{ios_code_max})$.

In the literature, there are two other codebook-based methods:
\begin{itemize}
	\item \emph{Independent codebook \cite{zheng2019intelligent}:} In this method, the codebook for the BS and the IOS will be designed separately. Although its complexity is lower, it neglects the coupling effect of the BS and the IOS, e.g., the effect of the incident angle as introduced in Section \ref{measure}, thus leading to performance degradation, which can be observed in Fig.~\ref{codebook:fig1}. 
	
	\item \emph{Impedance codebook \cite{kim2022learning}:} In this codebook, the codebook is realized by changing the impedance of each IOS element, i.e., in a hardware manner. This would pose some extra hardware design challenges. 
\end{itemize}
These facts motivate the proposed scheme.

\subsubsection{Numerical Results}
\begin{figure}[!t]
	\centering
	\includegraphics[width=0.45\textwidth]{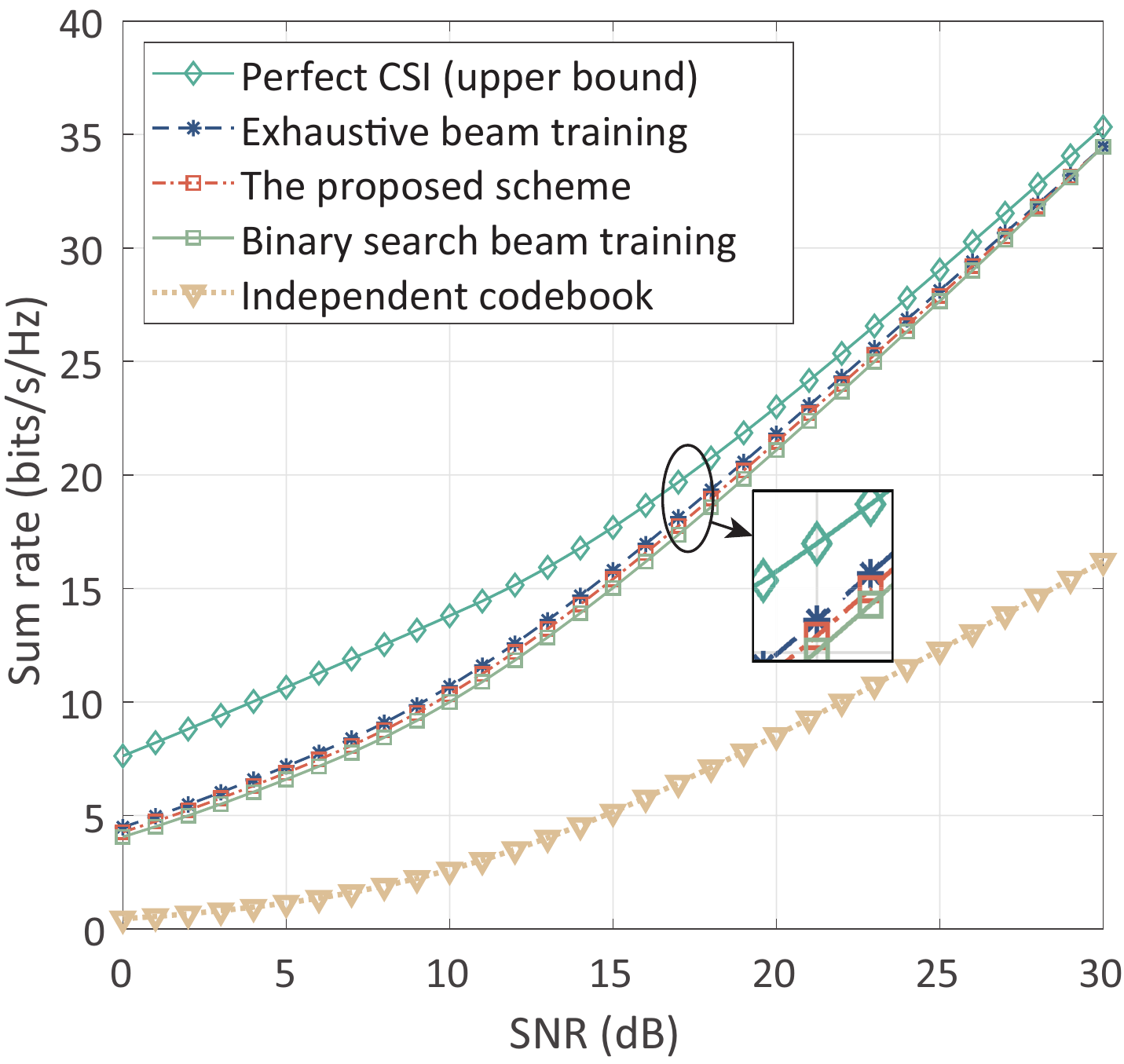}
	\caption{Sum-rate of the codebook-based scheme \cite{zhang2022dual}.}
	\label{codebook:fig1}
\end{figure}

\begin{figure}[!t]
	\centering
	\includegraphics[width=0.45\textwidth]{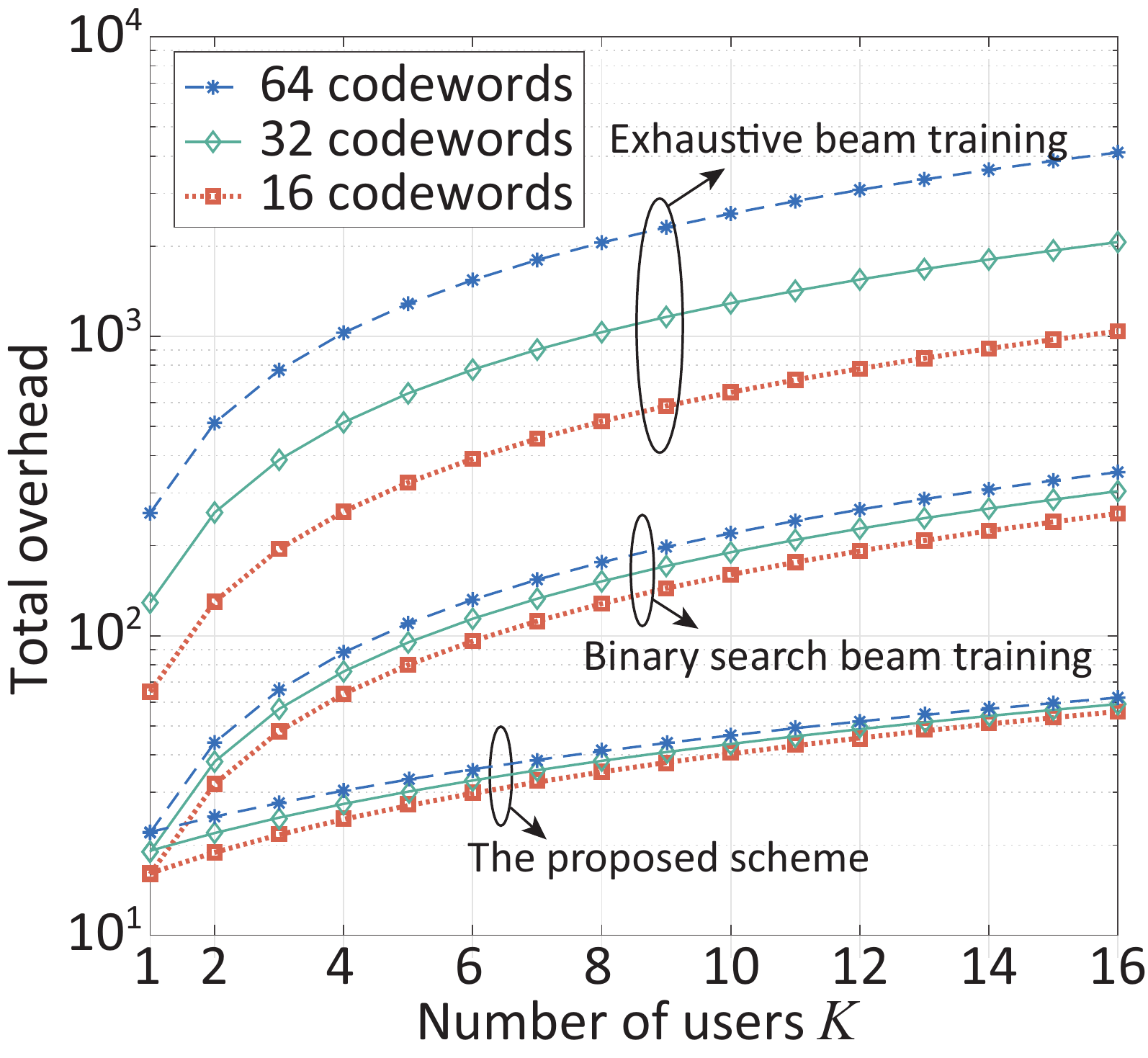}
	\caption{Training overhead of the codebook-based scheme \cite{zhang2022dual}.}
	\label{codebook:fig2}
\end{figure}

The performance of the proposed joint BS-IOS codebook based beam training method is evaluated in Figs. \ref{codebook:fig1} and \ref{codebook:fig2} in terms of the sum rate and training overhead (see \cite{zhang2022dual} for simulation settings). In Fig. \ref{codebook:fig1}, we can observe that the sum rate of different schemes grows with the SNR. Moreover, by comparing with the independent IOS codebook based scheme and the binary search beam training, the proposed scheme achieves a much higher sum rate, indicating the effectiveness of such a joint BS-IOS codebook based beamforming scheme. We can also observe that the sum rate of the proposed scheme is guaranteed to be very close to the upper bound obtained in the CSI-aware case while the training overhead is significantly lower, as shown in Fig. \ref{codebook:fig2}.

\subsubsection{Discussions: Hybrid Near-/far-field Codebook}
It is worth pointing out that when the size of the IOS is large, the Rayleigh distance (as introduced in Section \ref{path_loss_part}) will expand. As a result, users are very likely to be located in both near-field and far-field regions \cite{wei2021codebook}. Unlike the far-field region, where EM waves can be regarded as plane waves, EM waves transmitted in the near-field area are spherical waves, and thus the codebook designed for users in the far-field area is not applicable to those in the near-field area. This requires a hybrid near-far codebook design for an IOS with a large size.  

The basic idea of the hybrid near-far codebook design is to consider different steering vectors for users in near-field and far-field areas. To be specific, for far-field users, the channel gain from an IOS element to the user can be regarded as the same and the phase is linear with the index of the IOS element, while for near-field users, the channel gain and the phase are both a function of the distance between the IOS element and the user.

\subsection{Distributed Beamforming for Multi-Cell Communications}\label{multi-cell}
In this subsection, we focus on a multi-cell indoor communication system where the IOS is deployed in the overlapping coverage areas of adjacent cells such that inter-cell interference can be alleviated by controlling the phase shifts of the IOS. The concept of ``intelligent wall" is introduced in the sense that an intelligent component of the indoor environment is achieved by embedding the IOS into the real wall. A distributed IOS-based beamforming scheme is then discussed for such a multi-cell case.


\begin{figure}[t]
	\centering
	\includegraphics[width=0.45\textwidth]{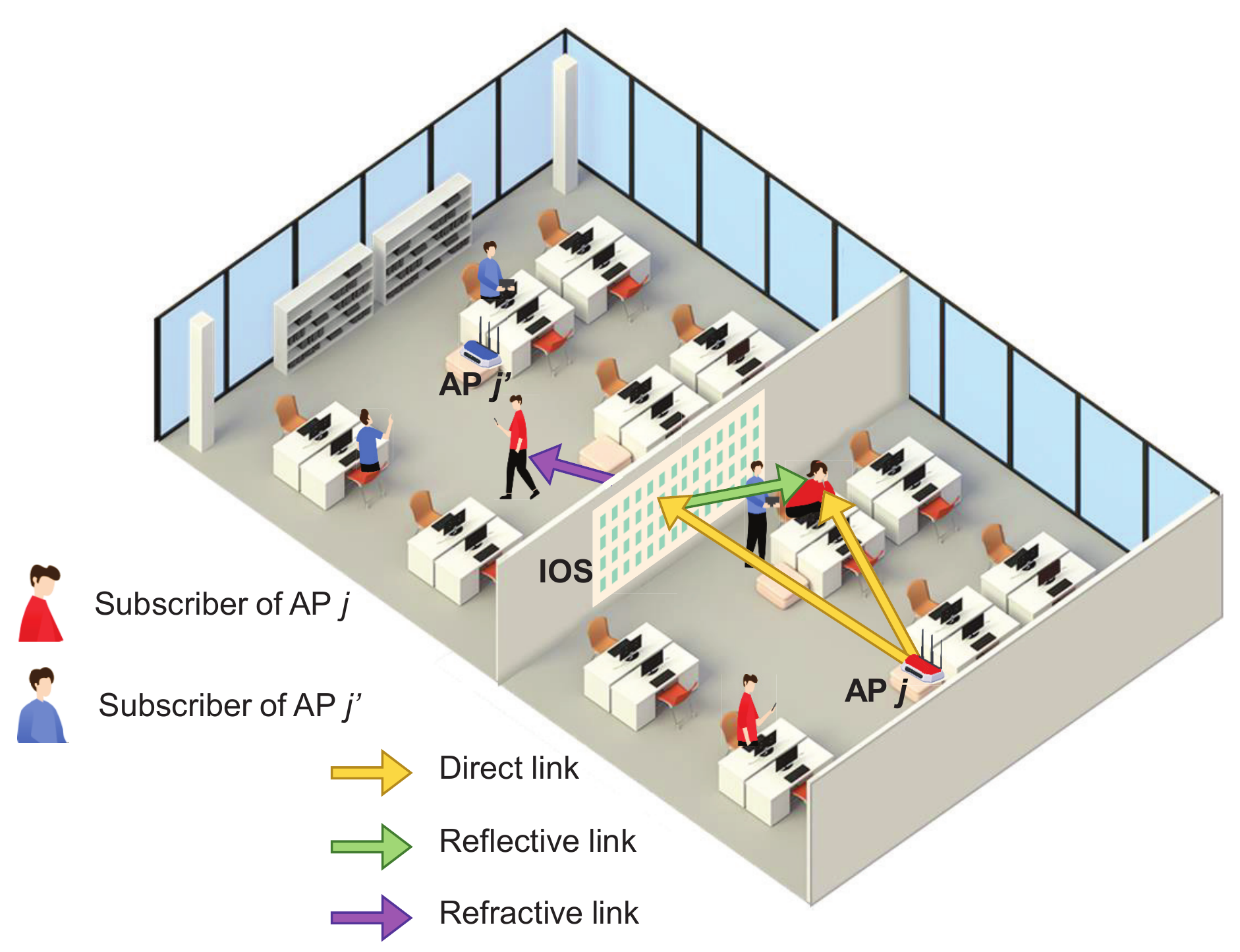}
	\caption{Two-cell case for the IOS-aided wireless communication.}
	\label{Fig:distributed}
\end{figure}

\subsubsection{Key Idea}
For ease of illustration, a two-cell case is shown in Fig. \ref{Fig:distributed}, where two APs serve their own subscribing users, forming two overlapping cells. For each cell, the serving users are distributed in both rooms. Due to the coverage overlap, each user inevitably suffers interference from the AP that it does not subscribe to. By embedding the IOS into the wall between these two rooms, each AP can control the IOS phase shifts to constructively add favourable reflected/refracted signals to the serving users and to decrease received signals for unintended users. In this way, the inter-cell interference can be mitigated and the desired signals can be strengthened simultaneously. It is worth pointing out that the inter-cell interference management capability is introduced by the refraction links of the IOS. To be specific, if we use two IRSs instead of the IOS, each IRS can only serve one cell. In other words, for one cell, only the subscribed users in the reflection regions can be served, those in the neighboring cell cannot enjoy the beamforming gains or the received signal strength will even be reduced due to the blockage of IRSs.

\subsubsection{IOS-aided Beamforming for Inter-cell Interference Cancellation} 
Following a traditional centralized method, the inter-cell interference mitigation issue can be tackled when two APs are willing to cooperate with each other on the IOS configuration and exchange the CSI of their own serving users. However, it is not the case in practice where the central controller is not always available or desirable. Due to the independence of different APs, it is difficult to exchange CSI between them, rendering the coordination of multiple APs non-trivial. A distributed IOS-based beamforming scheme is thus necessary so that a consensus on the IOS phase shifts can be achieved via negotiation between different APs.

The goal is to design a distributed mechanism to maximize the sum rate of all users by coordinating different APs to agree on the IOS phase shifts. Without a centralized node, each AP can only control its own beamformer and the phase shifts at the IOS, and thus each AP has its own optimization problem to solve. An example of the optimization problem for each AP can be shown as below \cite{zhang2022meta}: 
\begin{subequations}\label{opt_dis}
	\begin{align}
		\max\limits_{\left\lbrace {\mathbf{\Theta}}_j,{\bf{V}}_{D,j}\right\rbrace}  & \sum\limits_{n} R_{j,n}, \\
	\bf{s.t.} &~Tr({\bf{V}}_{D,j}{\bf{V}}_{D,j}^{H}) \le P_T,\label{constraint_5}\\
	&~\theta^{(l)}_{j,m} - \theta^{(r)}_{j,m} = c, \forall m,\label{constraint_3}\\
	&~\theta^{(l)}_{j,m} = \frac{s\pi}{2^{S-1}}, s=0,1,\cdots,2^S-1,\forall m, \label{constraint_4}\\
	&~{\mathbf{\Theta}}_{j} = {\mathbf{\Theta}}_{j'}, \forall j \ne j',\label{constraint_6}
\end{align}
\end{subequations}
where $R_{j,n}$ is the data rate of user $n$ at cell $j$, ${\bf{V}}_{D,j}$ is the digital beamformer of AP $j$, ${\mathbf{\Theta}}_j$ is the phase shift matrix of IOS proposed by AP $j$, in which the reflected and refracted phase shifts of $m$-th IOS element are denoted by $\theta^{(l)}_{j,m}$ and $\theta^{(r)}_{j,m}$. The coupling between the reflected and refracted phase shifts of each element is given in $(\ref{constraint_3})$, which makes the optimization problem different from that in IRS-aided systems, and the available phase shifts are limited and discrete for each element, as shown in $(\ref{constraint_4})$. Unlike traditional centralized problems, we introduce a consensus constraint $(\ref{constraint_6})$, implying that the proposed IOS phase shift configuration by all the APs should agree with each other \cite{di2021sharing}.

Based on the Lagrangian dual transform, the above problem can be decoupled to two local problems, each for an AP. The objective function of each AP contains both the sum rates of all the serving users and a penalty term which is proportional to $\left\| {\mathbf{\Theta}}_{j} - {\mathbf{\Theta}}_{j'} \right\| ^2$. Each AP aims to maximize its own sum rate while achieving a consensus with other APs on the IOS phase shifts. 

A negotiation scheme consisting of multiple rounds of information exchange between APs can be developed to solve such a distributed problem. In iteration $t$, each AP $j$ proposes an IOS phase shift matrix ${\mathbf{\Theta}}_{j}^{(t)}$ and optimizes its own digital beamformer ${\bf{V}}_{D,j}^{(t)}$. It broadcasts such information to its neighboring APs and receives the proposals from them simultaneously. Each AP then adopts the ADMM method \cite{ye2019mobility}, where the inter-cell interference is estimated based on other APs' locations and their reported digital beamformers, to update its own proposal for the next iteration, i.e., ${\mathbf{\Theta}}_{j}^{(t+1)}$ and ${\bf{V}}_{D,j}^{(t+1)}$. Note that each AP is only aware of other APs' locations and digital beamformers, but not the CSI of their serving users. Therefore, the distributed mechanism is performed to reach a consensus on the IOS phase shifts without revealing the user information of each cell.

\subsubsection{Numerical Results}

\begin{figure}[t]
	\centering
	\includegraphics[width=0.45\textwidth]{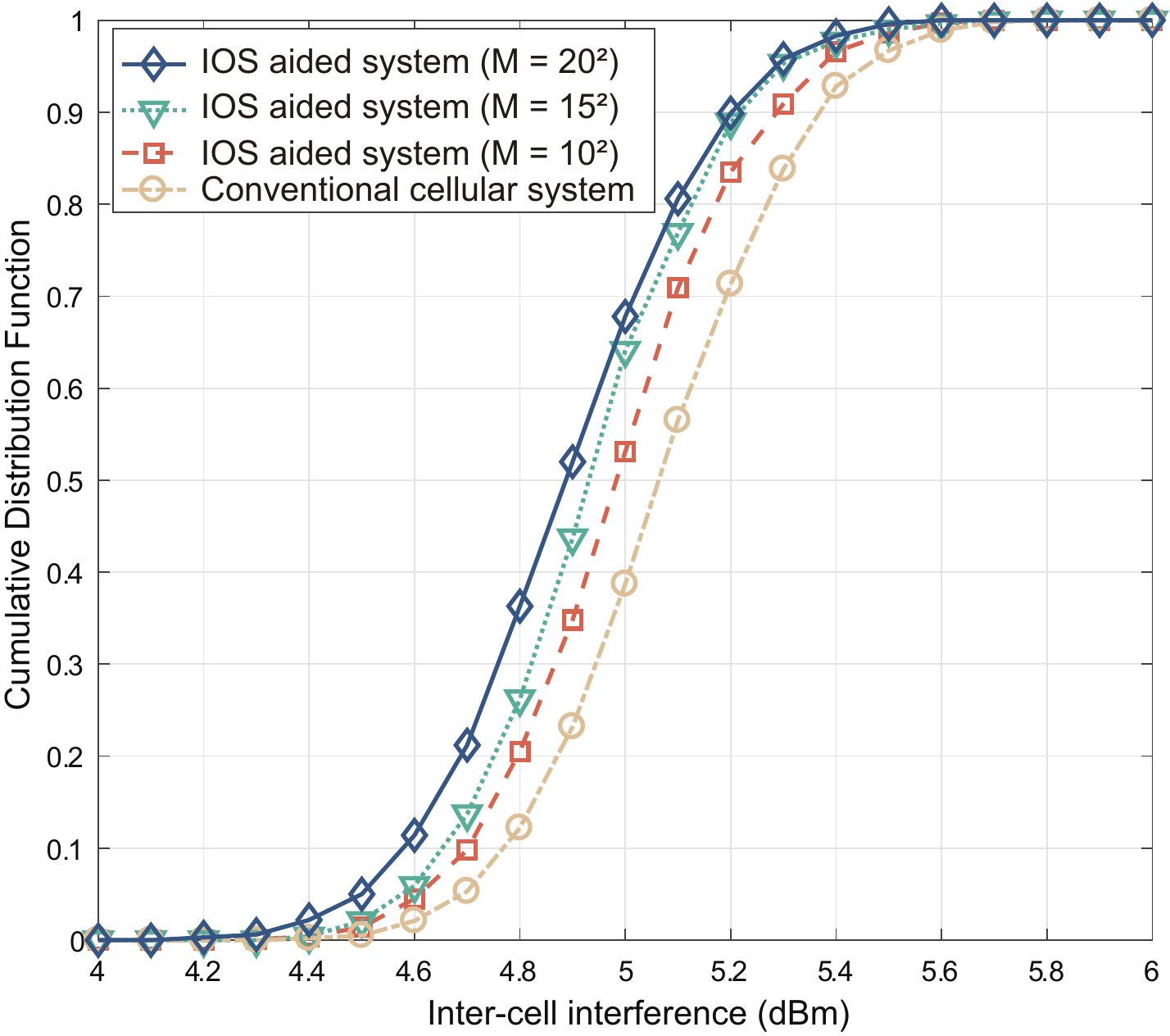}
	\caption{CDF of the inter-cell interference \cite{zhang2022meta}.}
	\label{Fig:cdf}
\end{figure}

\begin{figure}[t]
	\centering
	\includegraphics[width=0.45\textwidth]{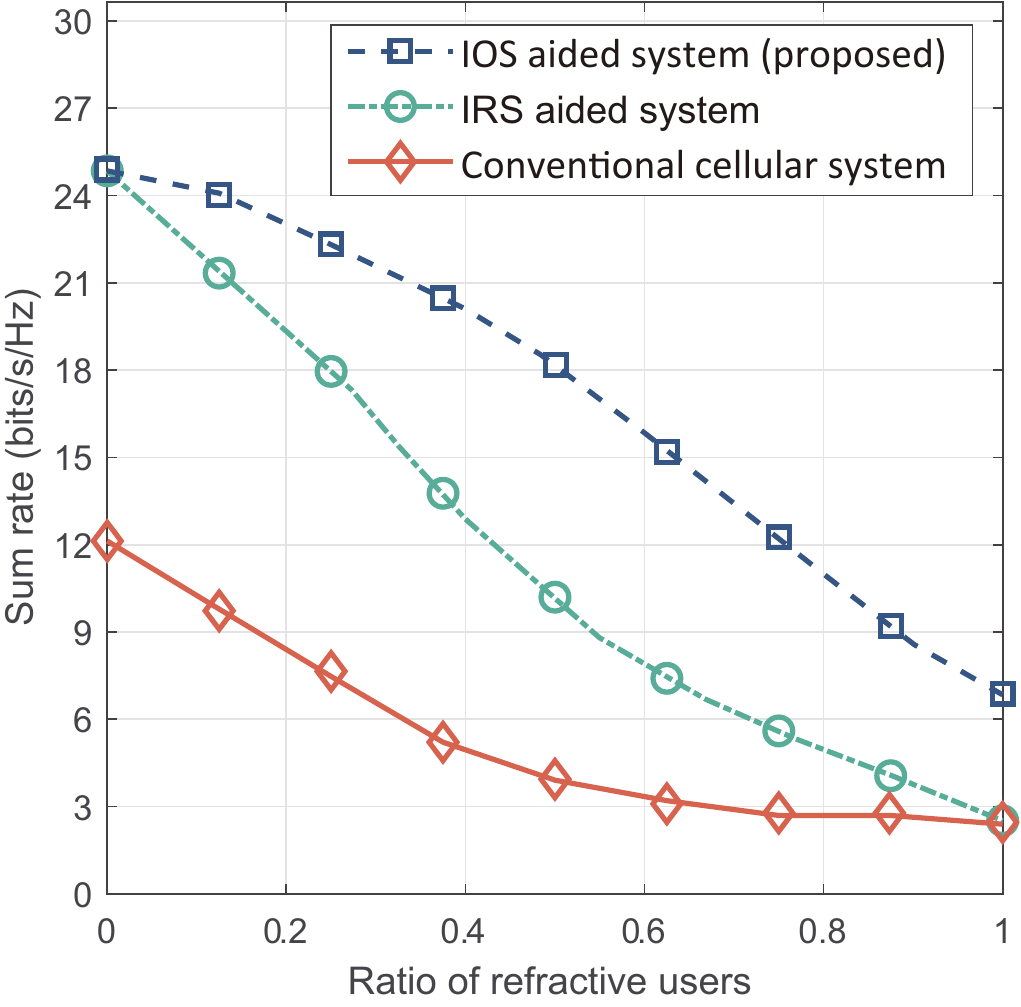}
	\caption{Sum rate vs. the ratio of users in the refractive area \cite{zhang2022meta}.}
	\label{Fig:tratio}
\end{figure}

The performance of the proposed IOS-aided multi-cell distributed beamforming scheme is evaluated in Figs. \ref{Fig:cdf} and \ref{Fig:tratio} (see \cite{zhang2022meta} for details). Fig. \ref{Fig:cdf} compares the IOS-aided system with the traditional non-IOS case in terms of the cumulative distribution function (CDF) of the inter-cell interference. The IOS-aided system achieves a higher value of CDF and converges to 1 faster. This implies that the deployment of IOS helps alleviate inter-cell interference efficiently. With a larger IOS size, such an interference mitigation effect becomes better. In Fig. \ref{Fig:tratio}, the sum rate of IOS-aided system compared with traditional non-IOS system and the IRS-aided systems\footnote{In the IRS-aided systems, two IRSs are used and each IRS serves users in each room separately.} is evaluated. The IOS-aided system shows  better performance than the traditional case and IRS-aided systems. It also performs close to the centralized scheme, which serves as an upper bound of the proposed distributed scheme.

\subsection{Other Related Works}
\label{related}
In this part, we will overview some existing works including beamforming schemes and performance analysis for IOS-aided wireless networks. 

\subsubsection{Beamforming Schemes}
As discussed in Section \ref{IOS modeling}, the state of each IOS element will have an impact on reflected and refracted signals simultaneously. In other words, the phase shifts for reflected and refracted signals cannot be adjusted independently. In \cite{wu2022bios}, the IOS was connected with another RRS to facilitate the separate control of the phase shifts for reflection and refraction, where the states of IOS elements were adjusted for the reflected signals while the states of RRS elements were adjusted for the refracted signals. The authors jointly optimized the transmit beamformer at the BS, the states of the IOS and RRS, and the reflection-to-refraction ratio of the IOS to minimize the total transmit power, subject to the SINR requirement for each user. The authors in \cite{cai2021joint} considered a two-user scenario, where the phase shifts and beamformers for both reflection and refraction were optimized, and the energy split ratio of the IOS was optimized independently as well. However, we would like to point out that the design given in Section \ref{principle} cannot achieve the independent control of amplitudes and phase shifts as the amplitudes and phase shifts are all controlled by the states of embedded PIN diodes. New architecture designs of IOSs are necessary to realize the independent control of amplitudes and phase shifts.

The authors in \cite{mu2021simultaneously} discussed a more ideal version of the IOS, where each element can be switched among reflective-only, refractive-only, and reflective-refractive modes. Moreover, they assumed separate control of phase shifts for reflective and refractive signals. Under such a setting, they proposed three protocols: energy splitting (enabling energy splitting between refractive and refractive signals), time switching (a portion of time is for reflection and the other portion of time is for refraction), and mode switching (switching between reflective-only and refractive-only modes according to users' requirements). They formulated the transmit power minimization problem by optimizing the beamformer at the BS and the phase shifts of the IOS based on these protocols. Following this line of work, the authors in \cite{niu2021weighted} maximized the weighted sum-rate of IOS-assisted MIMO systems under these three protocols. Their results showed that for unicast communication, the time switching protocol outperforms the energy splitting and mode switching protocols, while for broadcast communication, the energy splitting protocol outperforms the time switching and mode switching protocols. In \cite{ndjiongue2022double}, the authors studied the beamforming design for IOS-aided optical wireless communication systems. Their results show that there is an optimal number of IOS elements where the maximum rate is achieved. However, it should be noted that such a variant of the IOS is still under theoretical study but has not been implemented yet.

\subsubsection{Performance Analysis} 
In \cite{xu2021star}, the authors showed the diversity gain obtained by the IOS.  For comparison, the authors also presented a composite surface where one part is an IRS and the other part is an RRS. Through the derivation of the outage probability, they concluded that the diversity order obtained by the IOS is almost twice of that obtained by the composite surface. Further results are reported in \cite{xu2022star}. In this paper, the authors studied three configuration strategies for the IOS: 1) primary-secondary phase shift configuration strategy, where the IOS is optimized for the channel gain improvement for one primary user while simultaneously serving other secondary users; 2) diversity preserving phase shift configuration strategy where full diversity order for a user in the reflection region and a user in the refraction region; and 3) T/R-group phase shift configuration strategy where a group of elements is full refractive while the others are full reflective. Their results showed a trade-off between the power of desired signals and that of interference, with no matter which configuration strategy chosen. Moreover, their results indicated that Strategy 2 could achieve the best performance in terms of the received power.

\subsection{Summary and Outlook}

In this section, we have presented problem formulations and beamforming schemes for three scenarios to achieve full-dimensional wireless communications. Due to the high coupling of refraction and reflection, the methods proposed for IRS-aided wireless communications cannot be applied to IOS-aided wireless communications directly, thus requiring novel beamformer designs.

In addition to the beamforming techniques discussed above, the critical issues listed below are also valuable to be investigated before implementing the IOS-enabled full-dimensional wireless communications.

\begin{itemize}
	\item \emph{Channel Estimation:} To fully exploit the performance gain brought by the IOS, it is crucial to acquire accurate CSI. However, in consideration of 1) the passive nature of the IOS; 2) the cascaded effect of channels; and 3) a possible large number of IOS elements, channel estimation is challenging in practice \cite{wang2020channel}. Different from the channel estimation scheme for IRS-aided wireless communications \cite{zheng2019intelligent,shtaiwi2021channel,alwazani2020intelligent}, the channels for reflection and refraction signals in IOS-aided wireless communication systems are highly related, which makes the channel estimation more difficult. To be specific, the channels for reflection and refraction share a common channel from the Tx to the IOS, and thus the channels for these two regions cannot be estimated separately. An initial attempt at channel estimation in IOS-aided wireless communications is documented in the literature \cite{wu2021channel}. They used the MMSE metric for channel estimation by jointly designing the pilot sequences of the users, energy splitting ratio, and training pattern matrices under a coupled phase shift model for reflection and refraction. 
	
	\item \emph{System Synchronization:} Although we have introduced an example protocol for synchronization and user discovery in this section, this topic is still in its infancy. In contrast to the IRS-aided protocol \cite{cao2021reconfigurablemac}, it is not easy to find the configurations to enable the full coverage for both reflection and refraction regions in the IOS case. Moreover, there is a balance between the discovery latency caused by using more configurations and the coverage to be exploited, especially for ultra-reliable low-latency communications (URLLC). Furthermore, in practice, the synchronization could not be perfect. Therefore, it is also important to investigate how many synchronization errors are acceptable in order to put this technique into commercial use.
	
	\item \emph{Frequency-selective Beamforming:} Current beamforming designs for IOSs either focus on narrowband communications or ignore their frequency-selective profile of each element \cite{zheng2019intelligent,zhang2021spatial}, which provides increased flexibility for the design of wideband operation by matching the spectral behavior of each element to the equivalent wideband channel \cite{shlezinger2021dynamic}. As a result, this feature for wideband communications gives rise to research issues regarding the fundamental limits of IOS-aided wireless communications, including the achievable data rate \cite{zhang2020reconfigurable} and the impact of the number of elements on the system data rate \cite{zhang2021howmany}. Moreover, the design of frequency-selective beamformers is another important question in order to achieve the maximum data rate \cite{li2021intelligent}. Compared to an IRS, the frequency-selective characteristics for reflection and refraction of an IOS could be totally different, which makes performance analysis and beamformer design more challenging. 
	
	\item \emph{IOS Placement:} According to the model given in Section \ref{channel}, the incident angle has a significant impact on the radiation pattern. This indicates that the orientation of the IOS influences the link quality, which requires the optimization of the IOS's orientation in order to improve the system performance. In addition to the orientation, the position of the IOS should also be considered since it directly changes the propagation environment between the Tx and the Rx. 
	
	\quad Although there are some existing works on the IRS placement \cite{zeng2020reconfigurable,hashida2020intelligent,kang2021irs}, IOS deployment remains challenging for the following reasons. First, users on both sides of the IOS should be jointly considered in this case as the reflection and refraction are coupled. It is not trivial since the rate performance in the refraction and reflection zones usually do not agree with each other for the same IOS phase shifts. It is still unclear how to achieve the trade-off between the performances in the above two zones by carefully designing both the placement and element states of the IOS. Second, as we have discussed before, both the orientation and location of the IOS will jointly influence the coverage. However, as the impact of the incident angle on the reflection and refraction signals might not be symmetric, it is difficult to decouple the orientation and location as suggested in \cite{zeng2020reconfigurable}. Therefore, it is necessary to develop a new tractable optimization framework to study the effects of orientation and location. Moreover, how to deploy the IOS under the cell-free framework \cite{ohyama2022quantum} is also interesting to be investigated.
\end{itemize}

\section{Implementation and Experiments}
\label{implementation}

\begin{figure}[!t]
	\centering
	\includegraphics[width=0.48\textwidth]{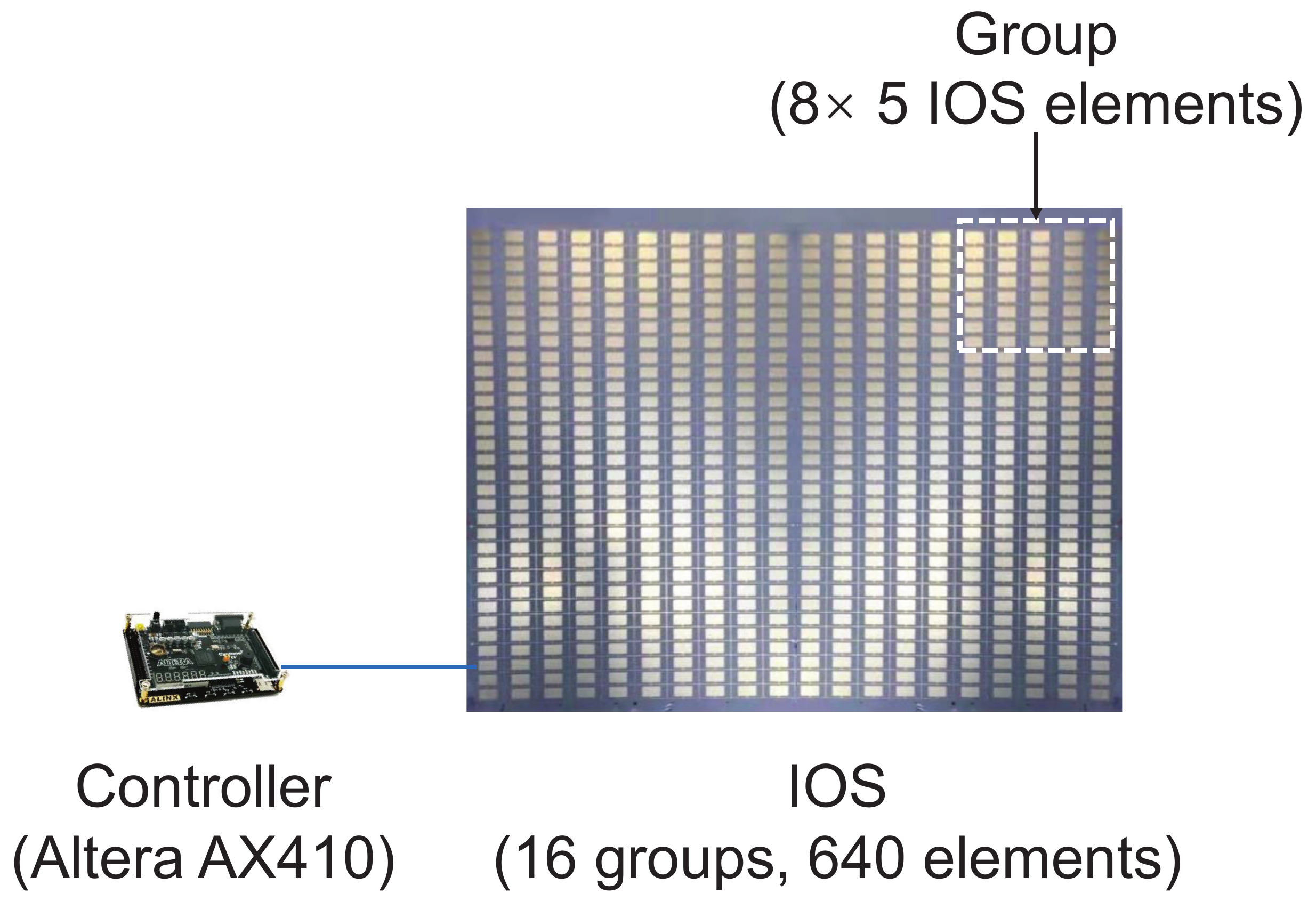}
	\caption{An implementation of the IOS \cite{zhang2022intelligent}.} \label{implement}
\end{figure}

\begin{table*}[!t]
	\renewcommand\arraystretch{1.2}
	\caption{Reflection-refraction response of each IOS element \cite{zhang2022intelligent}.}
	\centering
	\scriptsize
	\begin{tabular}{| c | c | c |  p{1.0cm}<{\centering}| p{1.2cm}<{\centering} | p{1.2cm}<{\centering}| p{1.0cm}<{\centering}|}
		\Xhline{1.pt}
		\multirow{2}*{\textbf{State}} &
		\multirow{2}*{\textbf{PIN Diode-1}}  & \multirow{2}*{\textbf{PIN Diode-2}} &
		\multicolumn{2}{c|}{\textbf{Reflection Response}} &
		\multicolumn{2}{c|}{\textbf{Refraction Response}}\\ 
		\cline{4-7} 
		& & & \textbf{Phase} & \textbf{Amplitude} &\textbf{Phase} & \textbf{Amplitude} \\
		\hline
		$1$ & OFF & OFF & $20^\circ$ & 0.46 & $300^\circ$ & 0.58\\ \hline
		$2$ & ON & ON & $215^\circ$ & 0.55 & $123^\circ$ & 0.81 \\
		\Xhline{1.pt}
	\end{tabular}
	\label{state}
\end{table*}

\begin{figure*}[!t]
	\centering
	\includegraphics[width=0.75\textwidth]{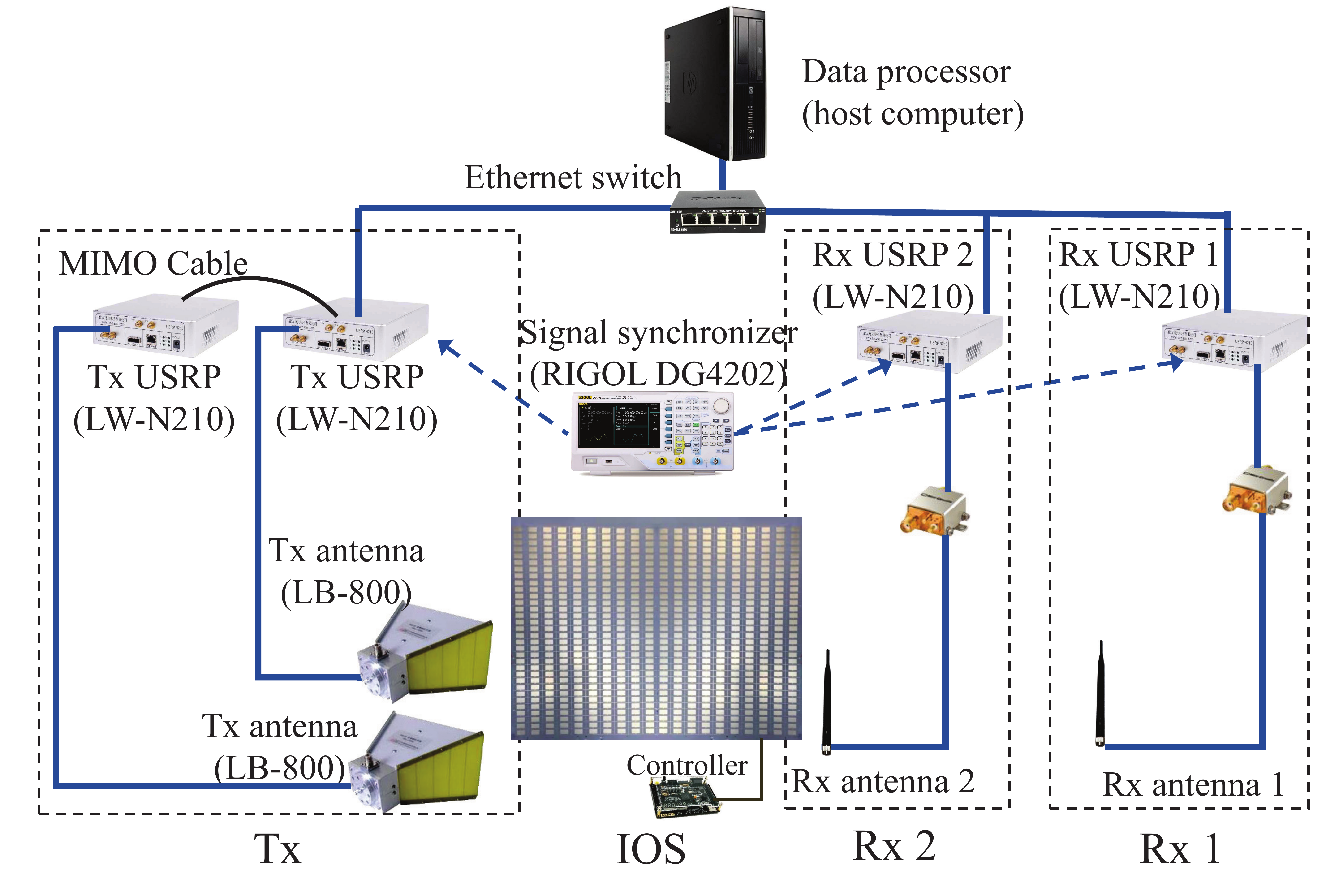}
	\caption{Hardware components of the implemented IOS-aided wireless communication prototype \cite{zhang2022intelligent}.} \label{prototype}
\end{figure*}

In this section, the implementation of an IOS and an IOS-aided point-to-point communication prototype are presented, along with experimental results. In the prototype, we use the hybrid beamforming scheme introduced in Section \ref{csi_aware} as an example.

\subsection{IOS Implementation}
\label{response}

Based on the IOS element structure introduced in Section \ref{IOS-epsilon}, an IOS consisting of 640 elements is designed and implemented, with the center working frequency being 3.6 GHz, as shown in Fig. \ref{implement} \cite{zhang2022intelligent}. Each IOS element of size 2.87 $\times$ 1.42 $\times$ 0.71 cm$^3$ consists of two symmetric layers. A PIN diode is deployed on each layer, enabling two element states tuned by the diode switches (ON, ON) and (OFF, OFF). A field-programmable gate array (FPGA) with a pre-loaded program serves as the IOS controller to apply different voltages to each PIN diode such that the state of each IOS element can be independently tuned to manipulate the incident wave. The FPGA is a Cyclone IV EP4CE10F17C8 platform that is manufactured by Intel Altera corporation. The development board that contains the FPGA and various interfaces is manufactured by ALINX For the ease of control and to locally fulfill the assumption of periodic boundary conditions, we divide the IOS into groups, each consisting of $5 \times 8$ elements. In each group, the states of elements are set as the same.

We utilize a rectangular waveguide system connected to a vector network analyzer to measure the reflection and refraction coefficients of the IOS element, with periodic boundary conditions for the IOS element and normal illumination, as shown in Table \ref{state}. We observe that the phase shift difference of the reflected/refracted signal in ON and OFF states is around 180$^\circ$ (i.e., $\left| 215^\circ - 20^\circ \right| = 195^\circ$, $\left|123^\circ - 300^\circ\right| = 177^\circ $). Meanwhile, the amplitude responses of different states are as close as possible. This satisfies the design principle introduced in Section \ref{IOS-epsilon}. 

The coupling effect between reflection and refraction coefficients is also verified in Table \ref{state}. That is, for both ON and OFF states, the phase shift of the reflected signal is not independent of the refracted signal. The difference in phase shifts between the reflected and refracted signals for the same state is approximately 100$^\circ$ (i.e., $\mod(\left|20^\circ -300^\circ \right|, 180^\circ) = 100^\circ$, $\mod(\left|215^\circ -123^\circ \right|, 180^\circ) = 92^\circ$), as depicted in $(\ref{constraint_3})$ and $(\ref{constraint_4})$.

\subsection{IOS-aided Wireless Communication Prototype}

The prototype of the IOS-based communication system is shown in Fig. \ref{prototype}, consisting of one two-antenna Tx, two single-antenna Rxs, an IOS, and other hardware for signal processing and control \cite{zeng2022intelligent}. 
\begin{itemize}
	\item \textbf{Tx}: A universal software radio peripheral (USRP) is implemented as the Tx (LW N210 with an SBX-LW120 RF daughterboard). The USRP has the functions of RF modulation/demodulation and baseband signal processing using a GNU Radio software development kit in Python \cite{ettus2015universal}. Its output port is connected to a ZX60-43-S+ low-noise amplifier (LNA), which amplifies the transmitted signal. A directional double-ridged horn antenna is employed (part number LB-880 manufactured by A-INFO Corporation). The maximum transmit power of each USRP is set as 1 dBm, the gain of the LNA is 14.3 dB, and the Tx antenna gain brought by the horn antenna is 10 dB.
	
	\item \textbf{Rx}: It is a USRP as well, whose input port is connected to an LNA, and directional double-ridged horn antennas are utilized. An external clock (10MHz OCXO) is used to provide a precise clock signal to the Tx and Rx. The Rx antenna gain is the same as that for Tx, which is also 10 dB.
	 
	\item \textbf{Signal synchronizer}: A RIGOL DG4202 signal source generator is used for generating the reference clock signal and the pulse-per-second signal for the modulation and demodulation at the Tx and Rx.
	
	\item \textbf{Data processor}: It is a host computer that serves as the signal processing module of the Tx and Rx through a software program written in Python.
	
	\item \textbf{Ethernet switch}: It connects the Tx, Rx, and the host computer for transmit and receive signal acquisition, and its bandwidth is 1 GHz.
\end{itemize}

The different modules of the IOS prototype are coordinated through a program written in Python on a host computer. Specifically, the host computer configures the USRP through the GNU Radio software development kit. The FPGA, in turn, transforms the input from the host computer into a control signal for the IOS with the aid of a program written in Verilog.

\subsection{Measurement and Results}
\label{measure}
\begin{figure}[!t]
	\centering
	\includegraphics[width=0.48\textwidth]{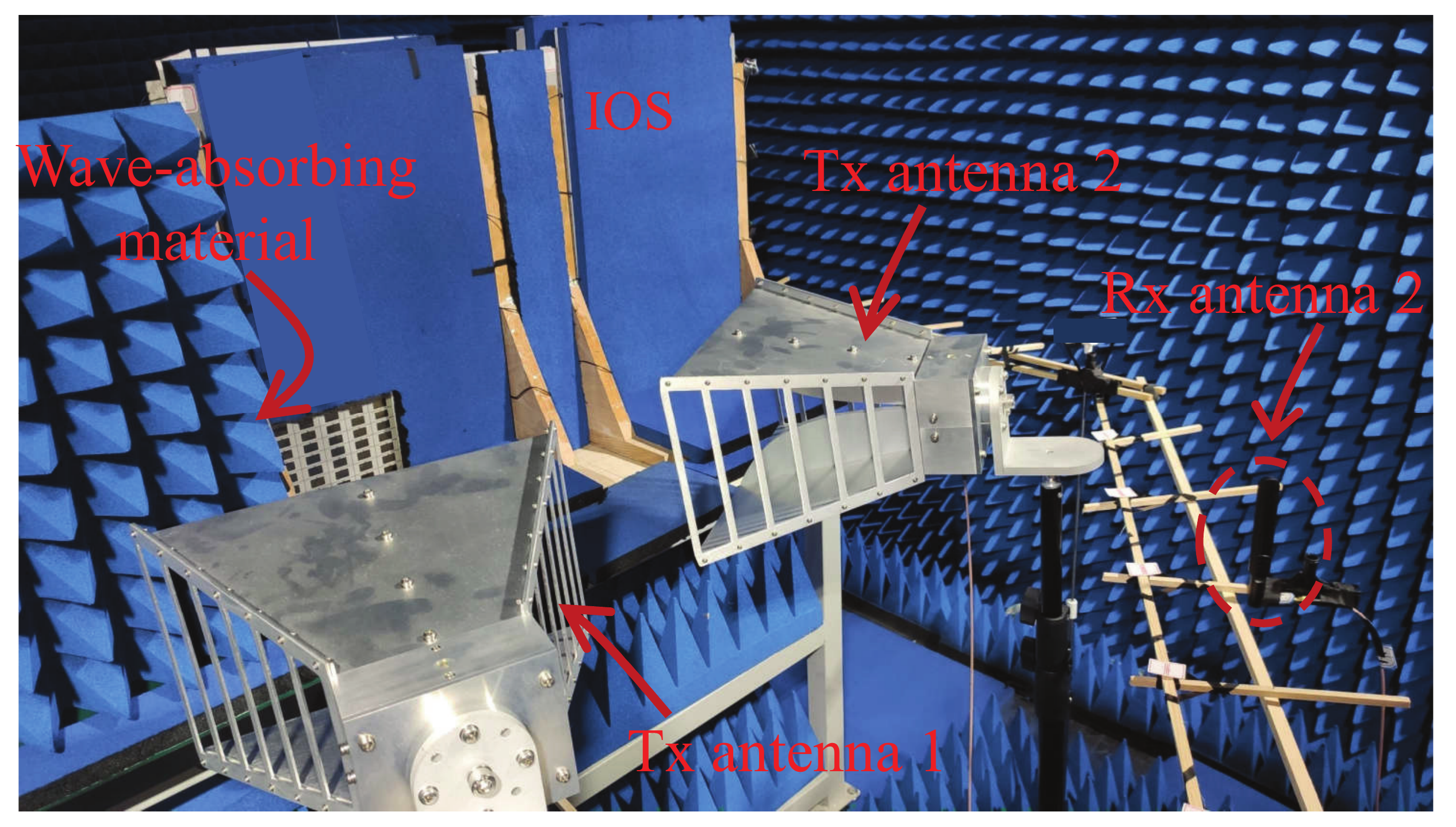}
	\caption{Experiment snapshot \cite{zeng2022intelligent}.} \label{setting}
\end{figure}

The measurement environment is designed to evaluate the performance of the IOS system in terms of the beam steering ability and the data rate. As shown in Fig.~\ref{setting}, the prototype is deployed in an empty room with aluminium walls covered by wave-absorbing material to block the EM interference in the propagation environment. More details as well as the measurement results are presented below \cite{zeng2022intelligent}.

\subsubsection{Beam Pattern}
\label{beam_pattern}
It is important to evaluate the beam steering capability of the IOS, which can be achieved by measuring the beam pattern, i.e.,
\begin{equation}
F(\psi, \phi) = \left|\sum_k E_k(\psi, \phi)v_{D_{k}} \right|^2,
\end{equation}
where $v_{D_{k}}$ is element $k$ of the digital beamformer, and $E_k(\psi, \phi)$ is the electric field excited by the IOS at direction $(\psi, \phi)$ when the $k$-th antenna of the Tx sends the incident signal toward the IOS. 

\begin{figure}[!t]
	\centering
	\includegraphics[width=0.45\textwidth]{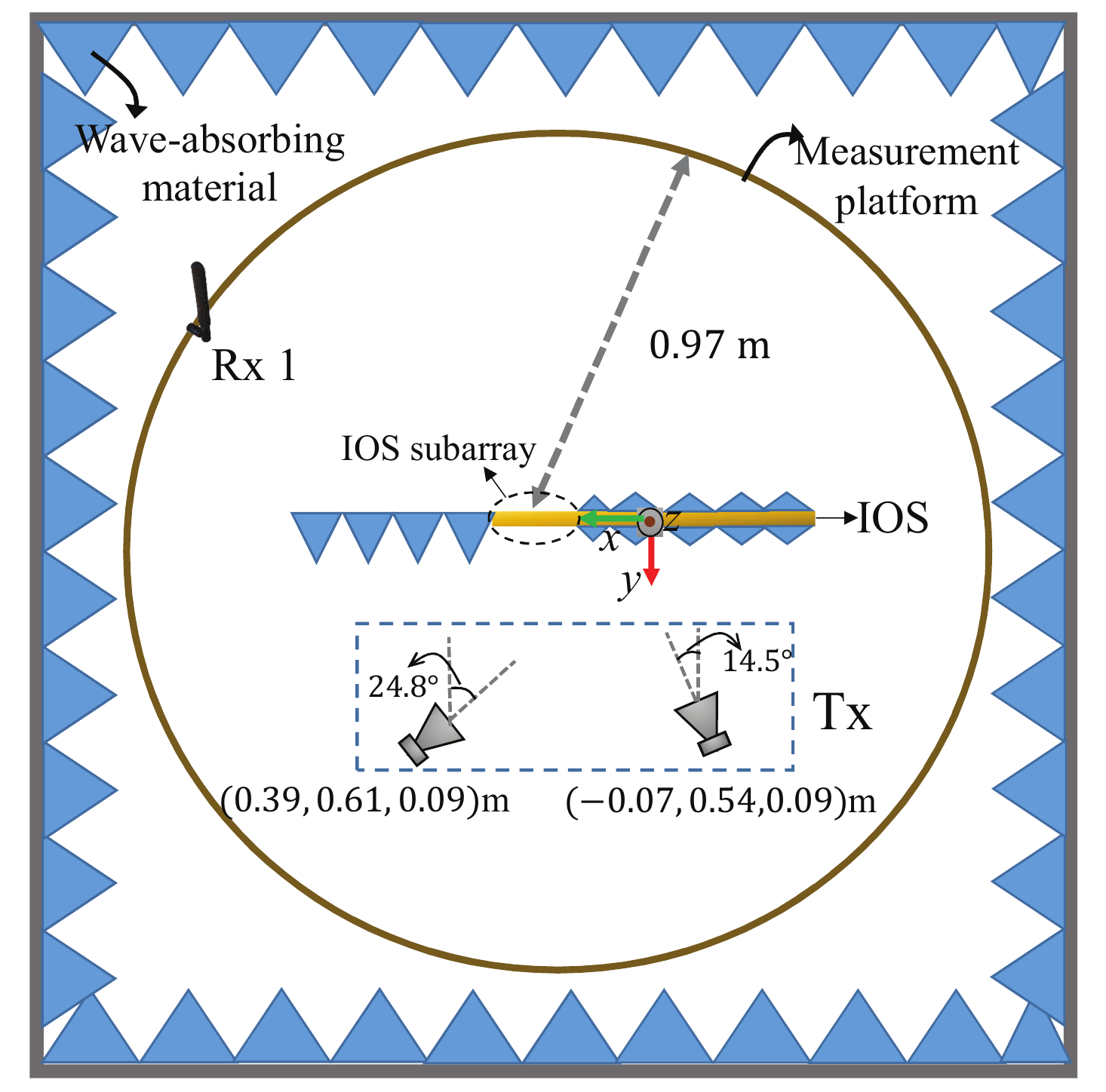}
	\caption{Layout of the experiment for beam pattern measurement \cite{zeng2022intelligent}.} \label{layout1}
\end{figure}

A circular measurement platform is constructed parallel to ground such that $E_k(\psi, \phi)$ can be measured when a Rx is deployed at different locations on this platform. The layout of the experiment is shown in Fig.~\ref{layout1}. To avoid unwanted scattering, we only use eight rows at the middle of the IOS, and the rest is covered with wave-absorbing material. The IOS-based system is evaluated in terms of the following aspects.

\begin{figure}[!t]
	\centering
	\includegraphics[width=0.48\textwidth]{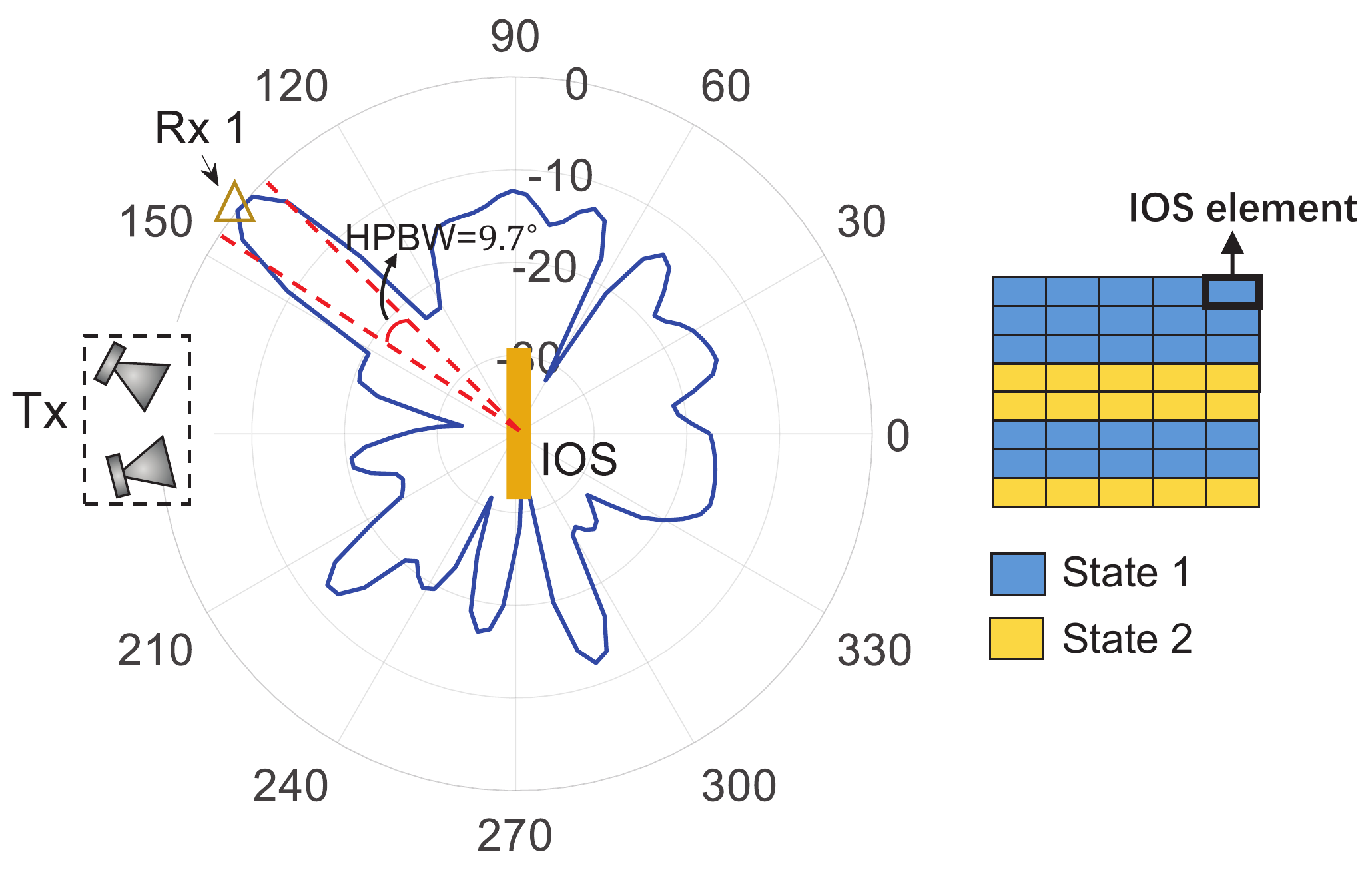}
	\caption{The scattered beam when the IOS is configured to point toward 141$^\circ$ in the reflection region \cite{zeng2022intelligent}.} \label{HPBW_reflection}
\end{figure}

\begin{figure}[!t]
	\centering
	\includegraphics[width=0.48\textwidth]{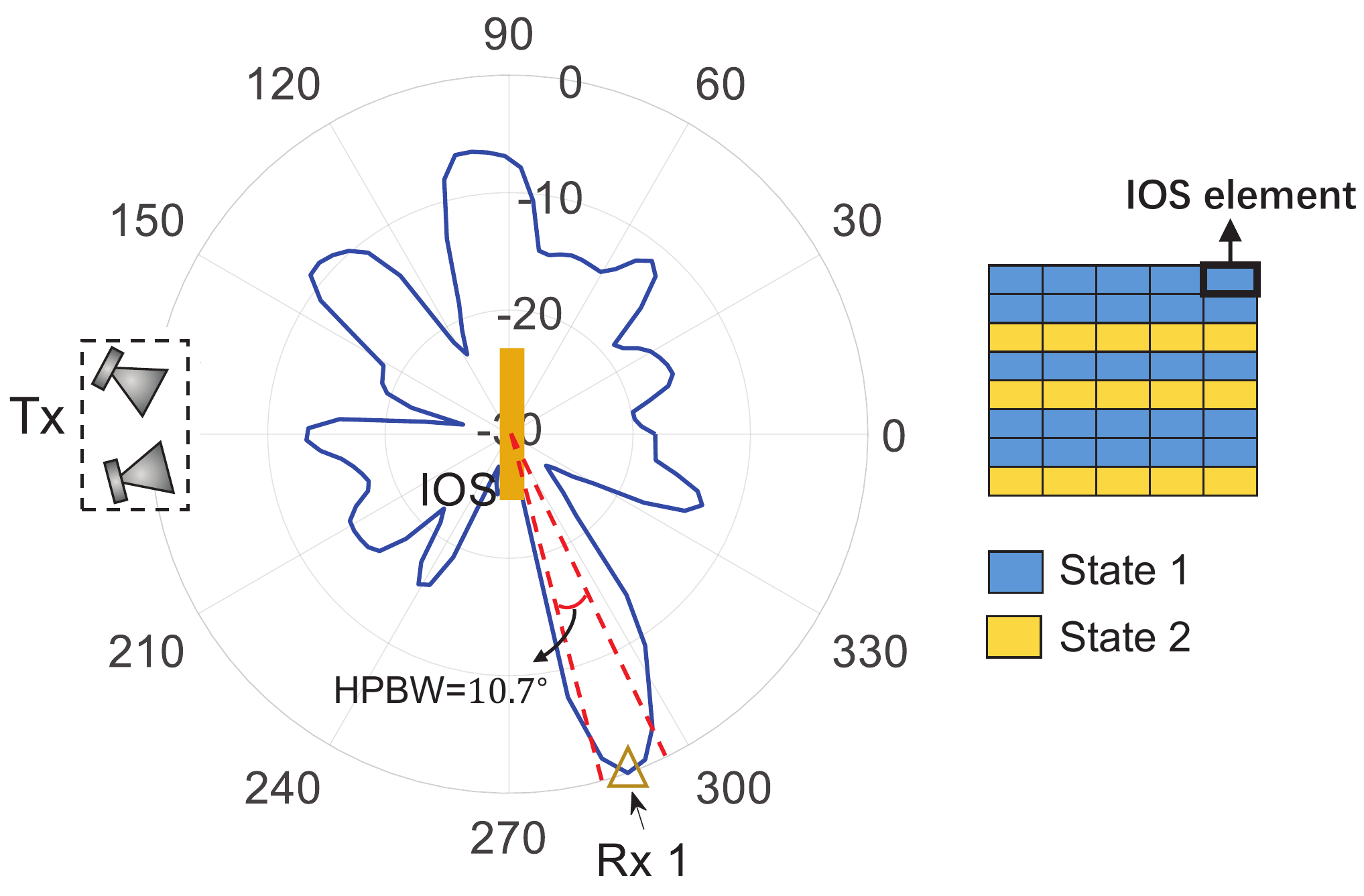}
	\caption{The scattered beam when the IOS is configured to point toward 289$^\circ$ in the refraction region \cite{zeng2022intelligent}.} \label{HPBW_refraction}
\end{figure}

\textbf{Pattern configuration}: we evaluate the beam steering ability of the IOS by reconfiguring the element states to serve different users, which shows the capability of the IOS to change the wireless channels. As shown in Figs. \ref{HPBW_reflection} and \ref{HPBW_refraction}, two radiation patterns are configured by the IOS with different main lobes projecting toward the refraction-zone and reflection-zone users, respectively.

\textbf{Half-power beamwidth (HPBW)}: it refers to the range of the main lobe within which the radiated power does not fall below half of the peak radiation direction, which can indicate the spatial resolution of the IOS. We evaluate the HPBW of both reflected and refracted beams in Fig. \ref{HPBW_reflection} and \ref{HPBW_refraction}, respectively. A narrow HPBW of 10$^\circ$ is observed at the main lobe projected toward the Rx direction. Given that the HPBW of the incident wave of the Tx is 37.57$^\circ$, the IOS shows a strong ability of energy concentration to narrow the beamwidth of the main lobe.

\begin{figure}[!t]
	\centering
	\includegraphics[width=0.48\textwidth]{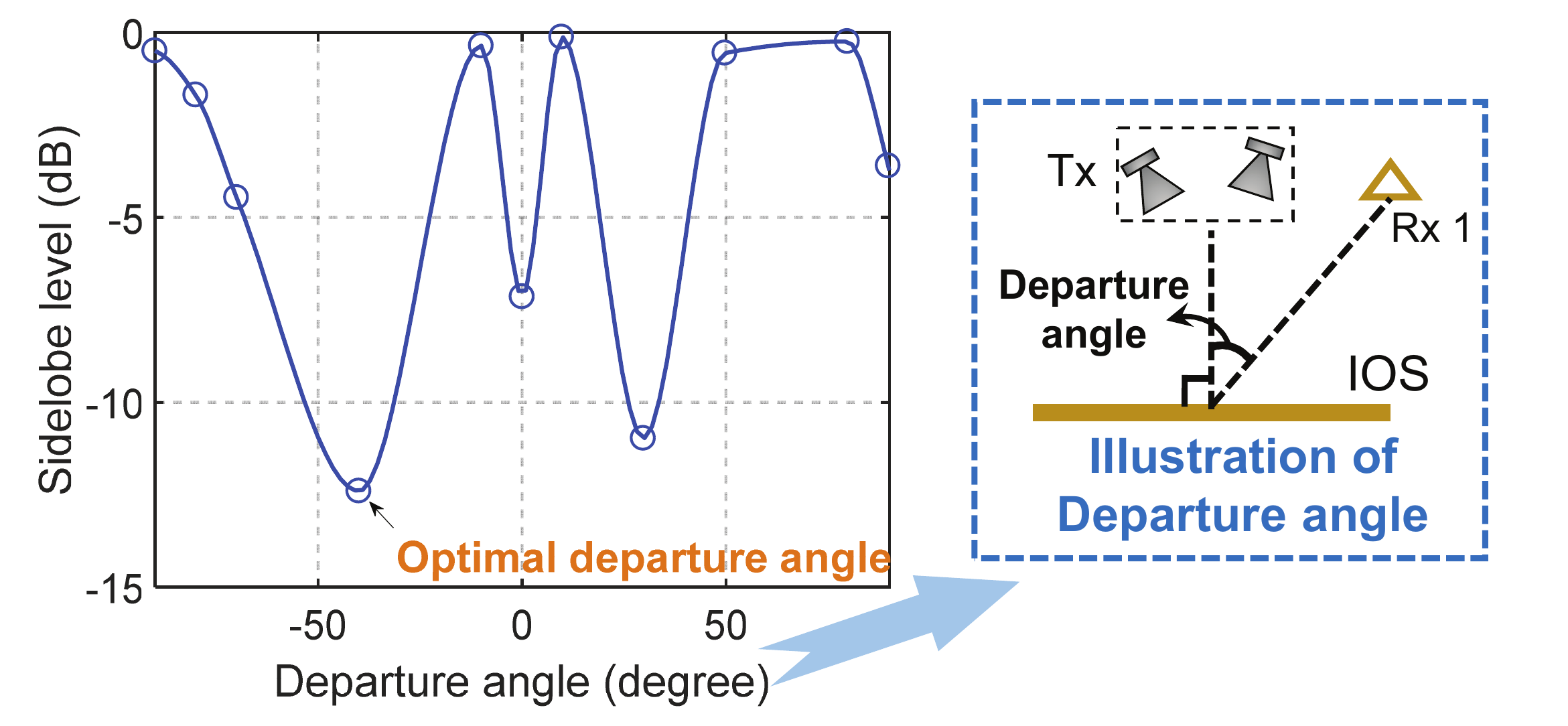}
	\caption{Sidelobe level vs. the angle of departure from the IOS, when the scattered beam points toward the desired Rx in the reflection region (Rx~1) \cite{zeng2022intelligent}.} \label{SLL}
\end{figure}

\textbf{Influence of the departure angle on the sidelobe level}: the sidelobe level (SLL) is depicted by the power gap between the largest sidelobe and the main lobe. It serves as a representative metric to measure the peak inter-user interference. A low SLL indicates a low peak interference. Fig. \ref{SLL} shows the SLL versus the departure angle when the Rx is located at the reflection zone\footnote{Without loss of generality, we omit the case of refraction zone here.} of the IOS. The orientation of the IOS is adjusted to achieve different departure angles. For different departure angles, the IOS phase shifts are configured in a way that the main-lobe beam always points toward the Rx. We observe that the SLL varies with the departure angle and the departure angle achieving the lowest SLL is within $\left[40^\circ, 50^\circ \right] $. This implies that there exists an optimal IOS orientation such that the SLL can be minimized.

\begin{figure}[!t]
	\centering
	\includegraphics[width=0.48\textwidth]{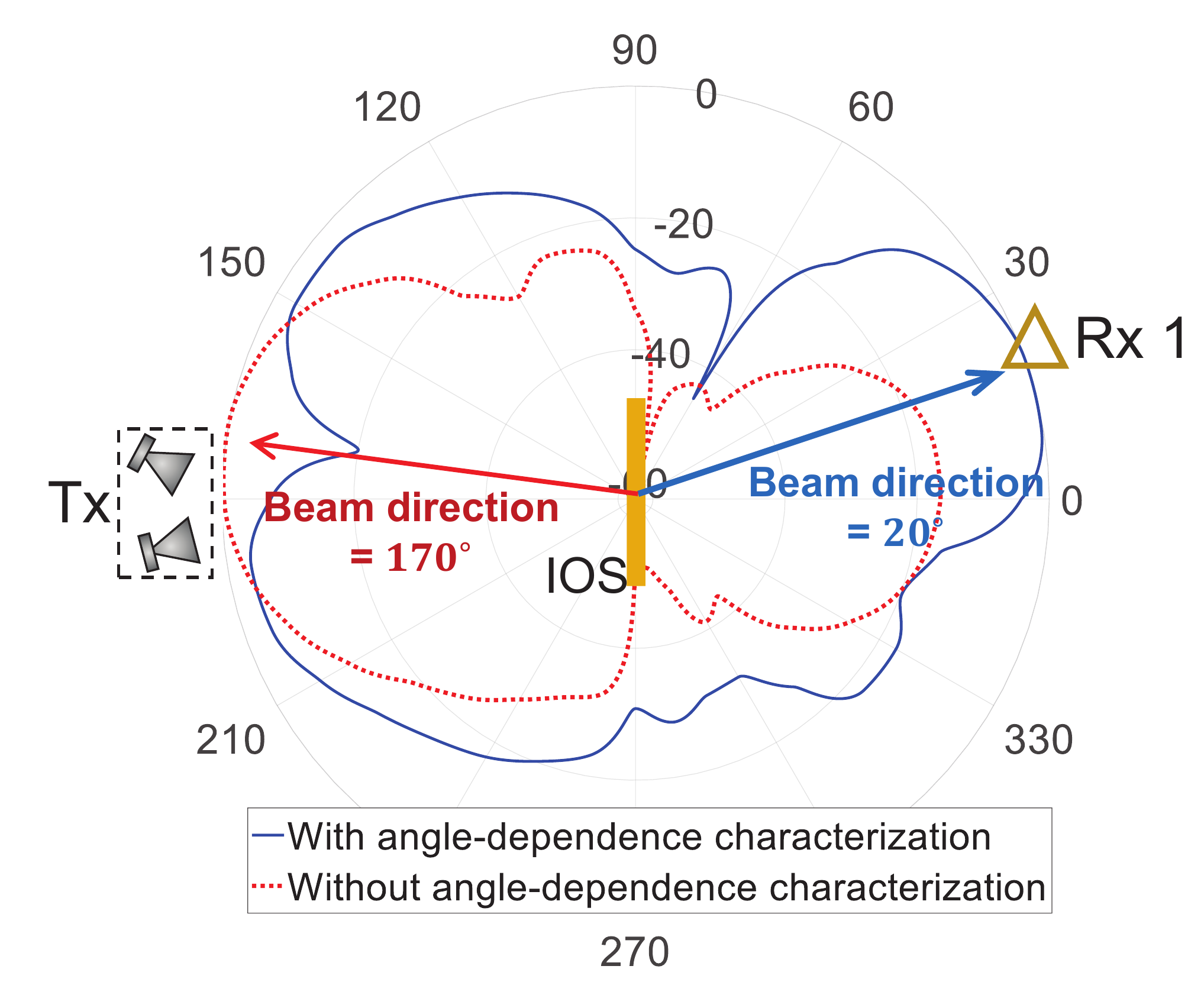}
	\caption{Comparison of the scattered beams when the angle-dependence response of the IOS is or is not considered \cite{zeng2022intelligent}.}
	\label{cmp_angle_dependence}
\end{figure}

\textbf{Influence of the incident angle on the beamforming scheme}: to explore the influence of the incident angle, we consider two different beamforming schemes. For the first scheme without angle-dependence characterization, we simply assume that the refraction and reflection coefficients of the IOS only relate to the state of each IOS element and have no relation to the departure angle. That is, the coefficients are always constants as long as the IOS configuration is fixed, and the beamforming scheme is performed given such constants. For the beamforming scheme with angle-dependence characterization, we take into account different refraction and reflection coefficients with respect to different departure angles. The radiation patterns given two schemes are shown in Fig. \ref{cmp_angle_dependence}. We observe that the angle-dependent scheme can generate a beam precisely pointing toward the desired direction, i.e., the direction of Rx 1. In contrast, the beam direction in the angle-independent scheme deviates from the target direction by 150$^\circ$. This verifies the influence of the incident angle, which should be considered in the beamforming scheme introduced in Section \ref{fd-IOS}. 

\subsubsection{Data Rate}

\begin{figure}[!t]
	\centering
	\includegraphics[width=0.45\textwidth]{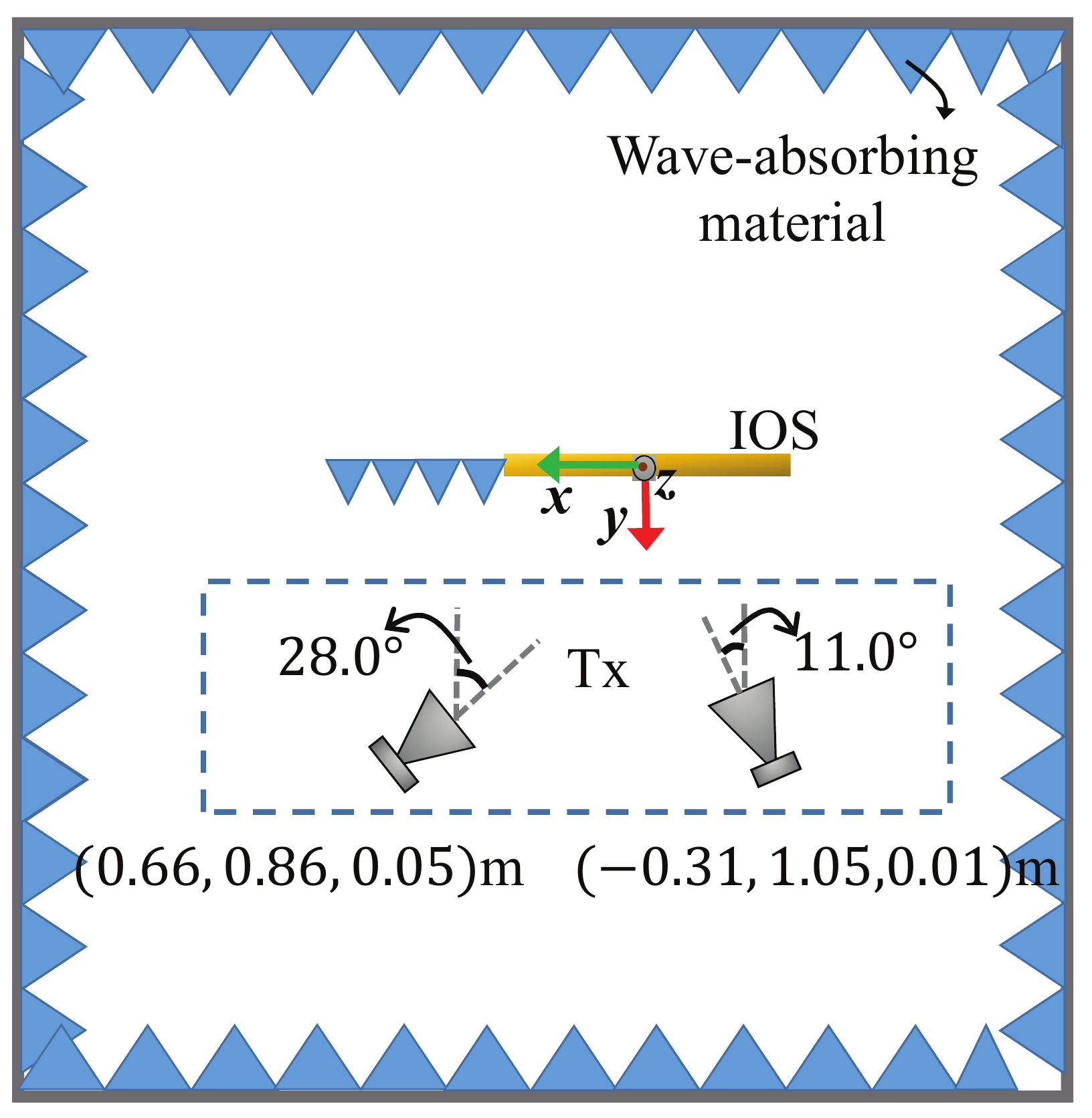}
	\caption{Layout of the experiment for data rate measurement \cite{zeng2022intelligent}. }
	\label{layout_data}
\end{figure}

\begin{figure*}[!t]
	\centering
	\subfigure[]{
		\begin{minipage}[b]{0.32\textwidth}
			\centering
			\includegraphics[width=1\textwidth]{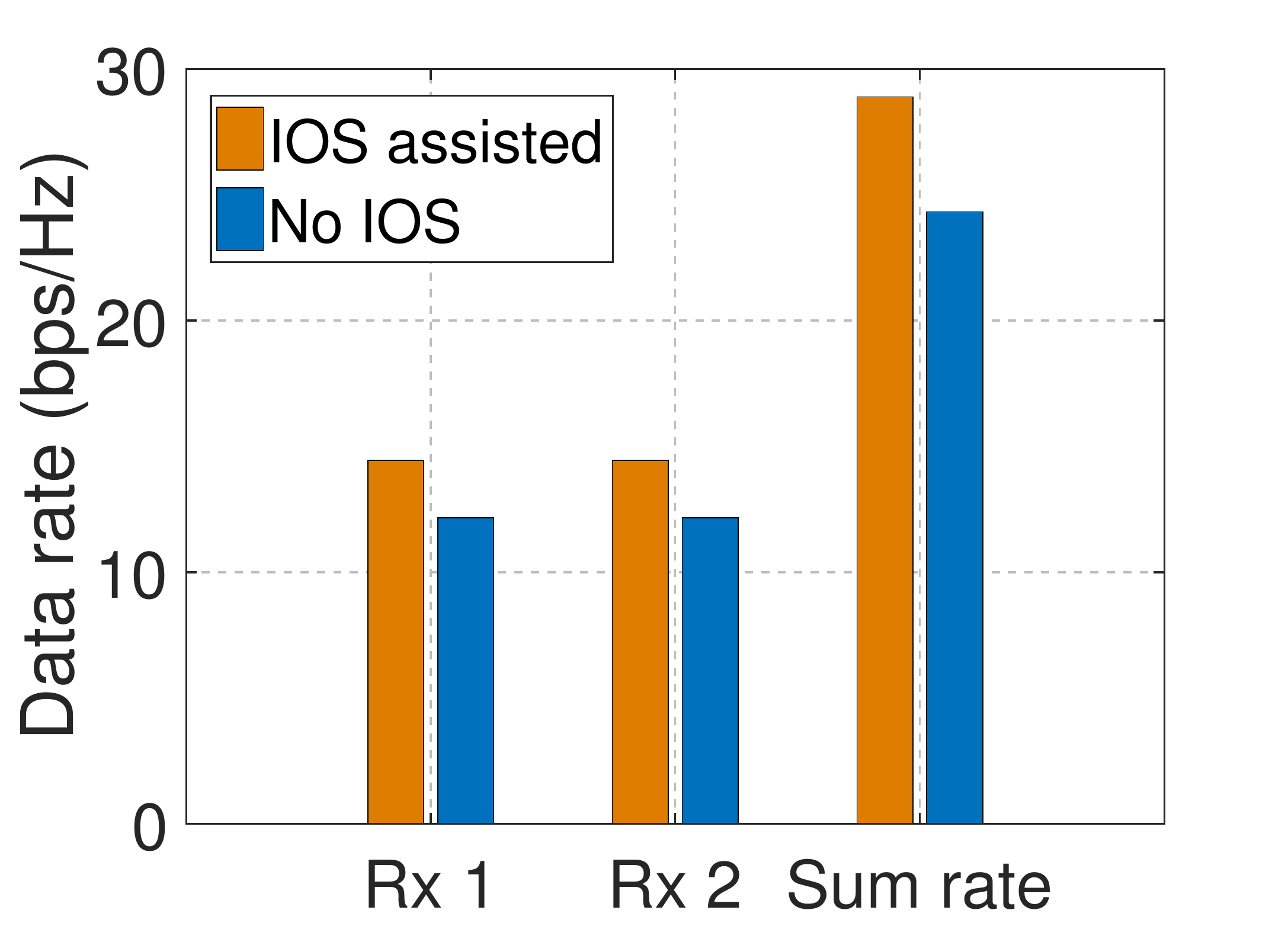}
			\label{data_rate_rr}
	\end{minipage}}
	\subfigure[]{
		\begin{minipage}[b]{0.32\textwidth}
			\centering
			\includegraphics[width=1\textwidth]{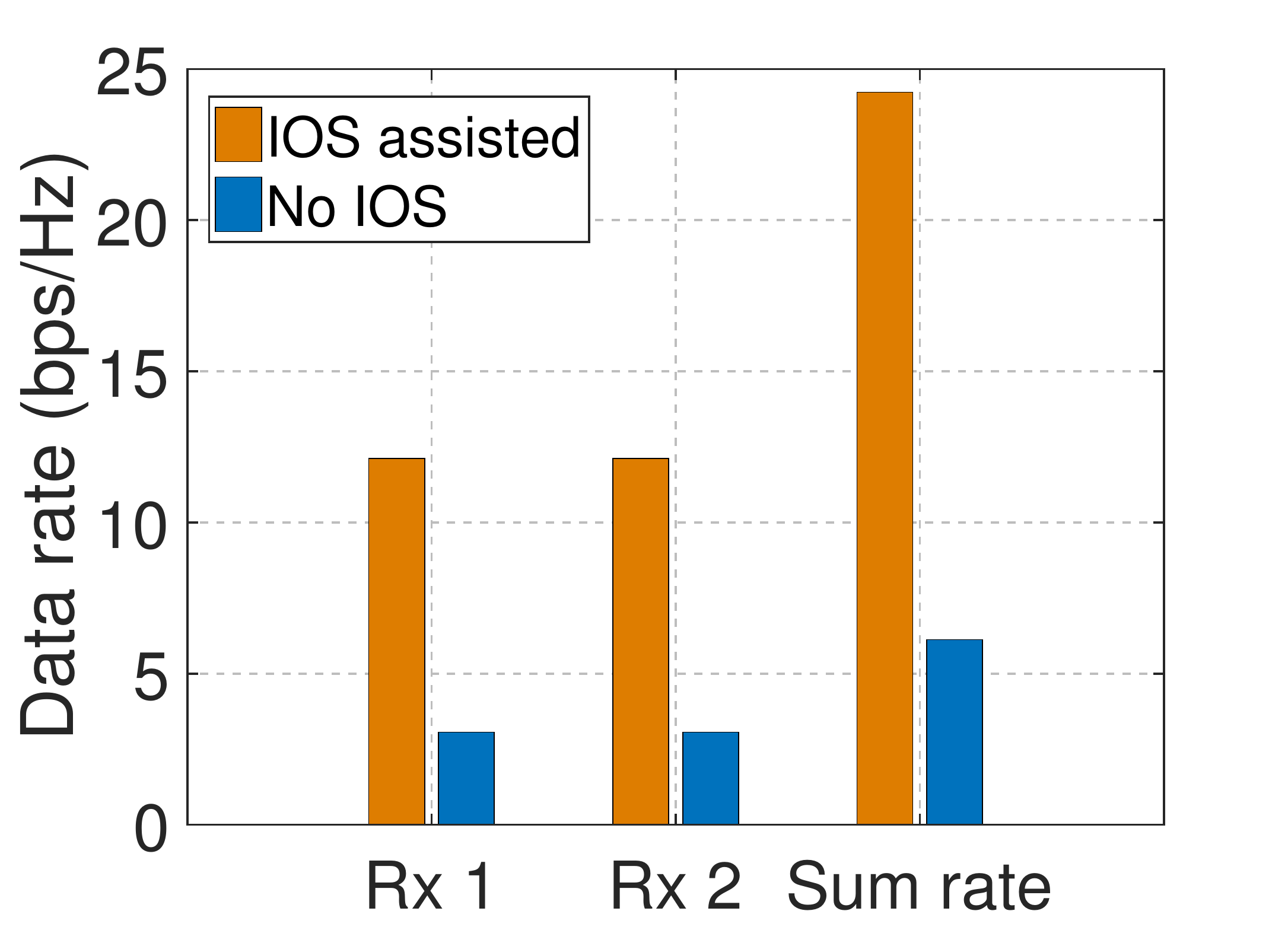}
			\label{data_rate_tt}
	\end{minipage}}		
	\subfigure[]{
		\begin{minipage}[b]{0.32\textwidth}
			\centering
			\includegraphics[width=1\textwidth]{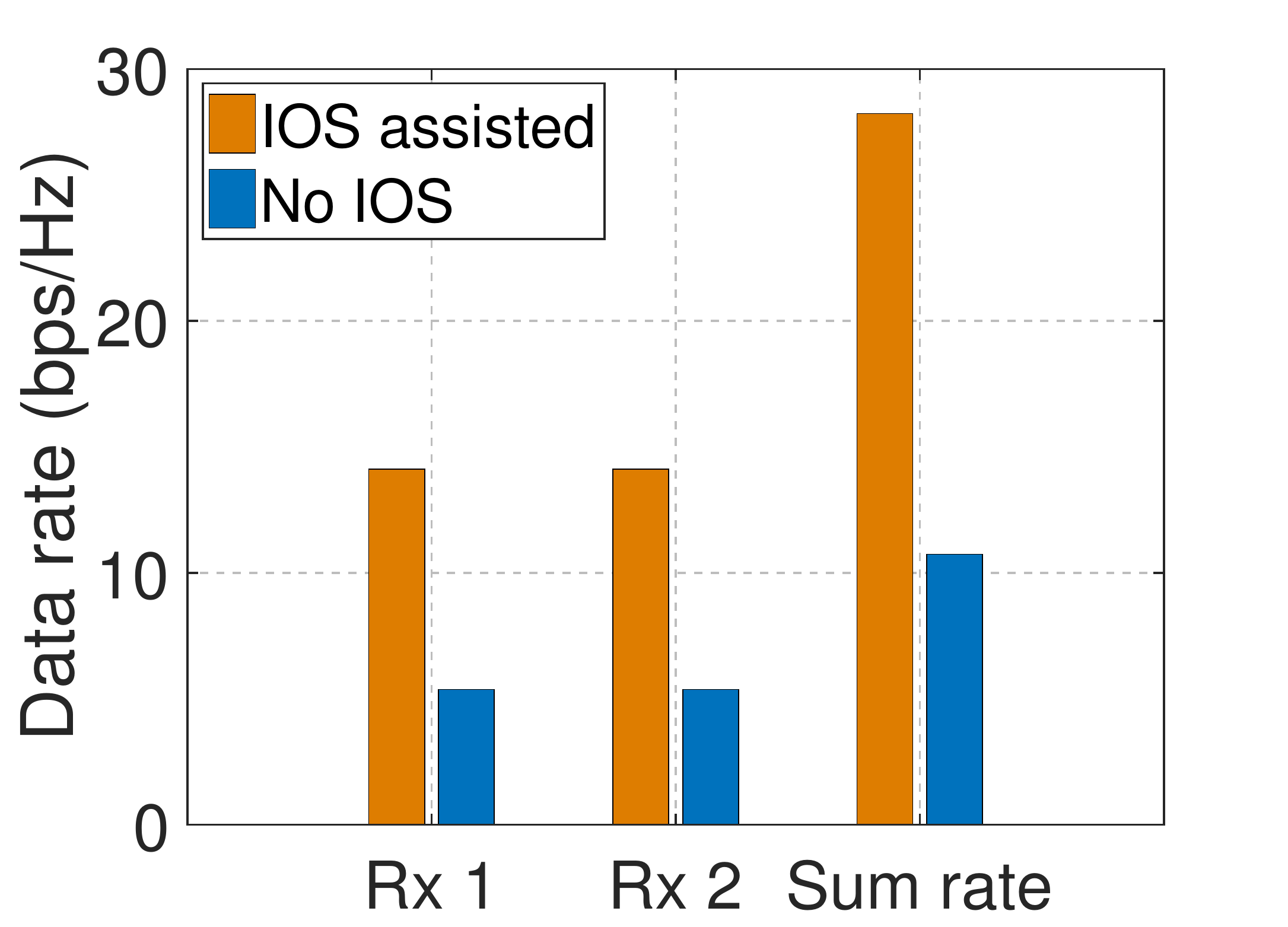}
			\label{data_rate_rt}
	\end{minipage}}
	\caption{Data rate of the prototype, where (a) both Rxs are in the reflection region, (b) both Rxs are in the refraction region, and (c) one Rx is located in the reflection region and one Rx is located in the refraction region \cite{zeng2022intelligent}.}
	\label{data_rate_fig}
\end{figure*}

To evaluate the data rate, three different Rx deployments are considered in  i.e., the refection-zone-only, refraction-zone-only, and refraction-reflection-zone cases. The layout of the experiment is shown in Fig. \ref{layout_data}, where the whole IOS is utilized. The beamforming scheme is performed taking into account the departure angles of users to maximize the minimum rate of the Rxs. 

A practical channel estimation method is utilized below. It is noted that the equivalent channel between the BS and user $k$, ${\widetilde{\textbf{h}}}_{k} = {\bf{h}}_{IU,k}^H{\bf{Q}}{\bf{H}}_{BI}$, is jointly influenced by the BS-IOS link, the IOS refraction and reflection coefficients for different element states, and the IOS-user link. It can be modeled as
\begin{equation}\label{channel_estimate}
\begin{split}
{\widetilde{h}}_{k,n}(s_1,\cdots,s_M) = & {\widetilde{h}}_{k,n}(s_1=0,\cdots,s_M=0) +\\
& \sum_{m}s_m\Delta{\widetilde{h}}_{k,n}^{(m)},
 \end{split}
\end{equation}
where $s_m \in \left\lbrace 0,1 \right\rbrace$ denotes the OFF and ON states of IOS element $m$, $\Delta{\widetilde{h}}_{k,n}^{(m)}$ is the change of when $s_m$ varies from state 0 to state 1 while the other elements remain state 0, i.e.,
\begin{equation}
	\begin{split}
\Delta{\widetilde{h}}_{k,n}^{(m)} =& {\widetilde{h}}_{k,n}(s_1=0,\cdots,s_m=1,\cdots,s_M=0) -\\
&{\widetilde{h}}_{k,n}(s_1=0,\cdots,s_m=0,\cdots,s_M=0).
\end{split} 
\end{equation}

In practice, to reduce the complexity of the channel estimation, we divide the IOS elements into $G < M$ groups and the IOS elements in the same group share the same state. In this way, $M$ in (\ref{channel_estimate}) is replaced by $G$, and each group of IOS elements can be viewed as a single equivalent ``element".

Given the channel information, the beamforming scheme is then performed to maximize the sum rate of two Rxs. As shown in Fig. \ref{data_rate_fig}, both Rxs can always receive from the Tx regardless of their positions in the IOS-assisted system, verifying the full-dimensional transmission of the IOS-based system. The sum rate of two Rxs is higher than that obtained in the no-IOS case, indicating the effectiveness of the IOS. It is worth noting that Rxs in the refraction region obtain a more pronounced improvement in the data rate with respect to the no-IOS case due to a stronger refraction coefficient.

\subsection{Other Prototypes}
The verification of the proposed IOS has also attracted broad interest from the research community. In the following part, we will introduce three other prototypes built using different materials, and their differences are summarized in Table \ref{comparison2}.

\begin{table*}[!t]
	\renewcommand\arraystretch{1.2}
	\begin{center}
		\caption{Comparison among different materials}
		\begin{tabular}{|c|m{5.5cm} |m{5.5cm} |}
			\hline
			{\textbf{Materials}}& \makecell[c]{\textbf{Advantages}}& \makecell[c]{\textbf{Limitations}} \\ \hline
			Glass & Transparent for visible light frequency range & Require mechanical control, which is less reliable and flexible\\ \hline
			Liquid Crystal & Transparent for visible light frequency range; Wider working range compared to semiconductor components & Longer switching time between two states compared to semiconductor components  \\ \hline
			Graphene & Beneficial EM wave characteristics & Immature fabrication process \\ \hline
		\end{tabular}\label{comparison2}
	\end{center}
\end{table*}

\subsubsection{Glass}
\begin{figure}[!t]
	\centering
	\includegraphics[width=0.48\textwidth]{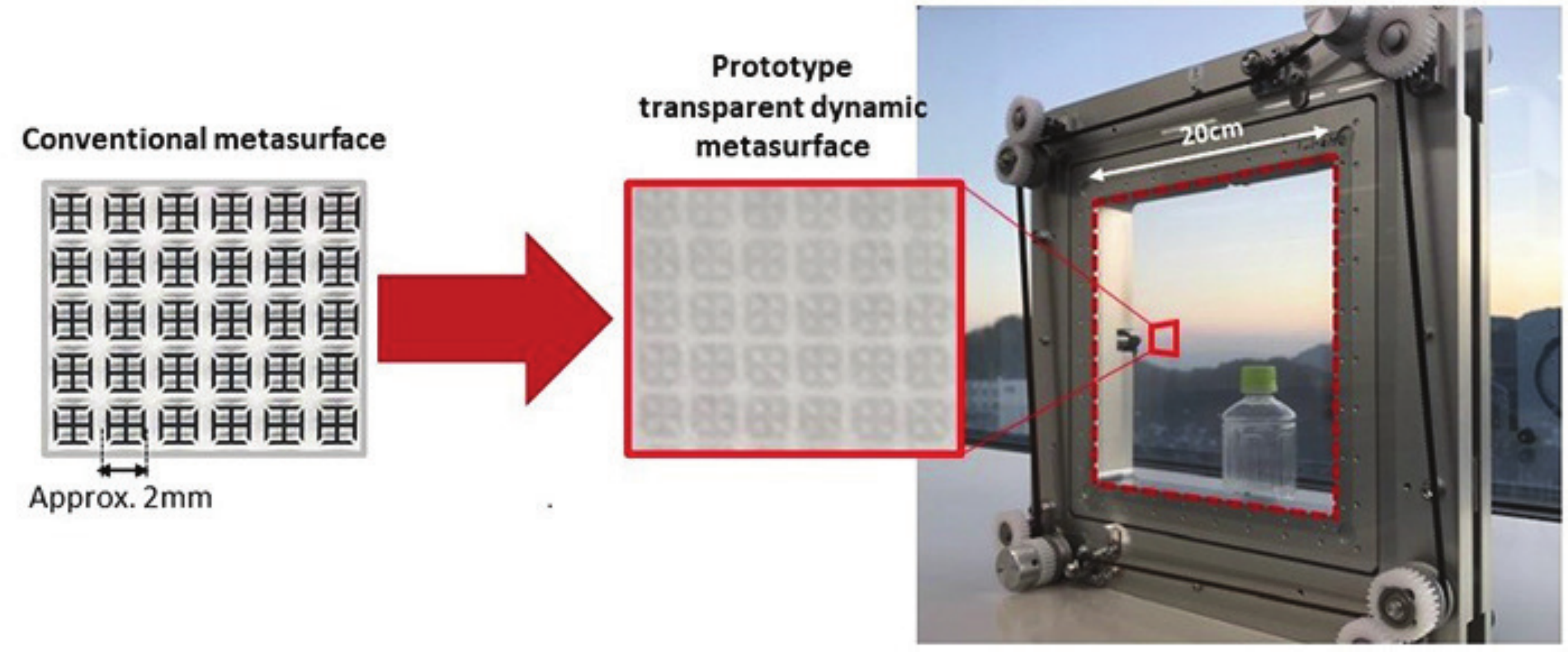}
	\caption{Prototype of Smart Glass in DOCOMO \cite{docomo2020docomo}. }
	\label{docomo}
\end{figure}

In collaboration with the global glass manufacturer AGC Inc., NTT DOCOMO, Japan built a prototype of IOS based on a transparent dynamic metasurface \cite{docomo2020docomo}, as shown in Fig. \ref{docomo}. This metasurface comprises a large number of sub-wavelength unit cells placed in a periodic arrangement on a two-dimensional surface covered with a glass substrate. It is reported that the metasurface works at a frequency of 28 GHz, which is the radio band for 5G. It allows dynamic control of its reflection and refraction. When deployed in a window, the transparency of the window can also be guaranteed. With a different width of the dielectric material and its distance between substrates, the smart glass can have different modes leading to different responses to incident signals, such as full penetration (refraction), full reflection, and partial reflection. In their trials, radio waves were beamed perpendicularly to the metasurface, and they measured the strength of refracted signals in full refraction and full reflection modes. Test results showed that radio waves passed through the substrate in the full refraction mode and were blocked in the full reflection mode. 

The appealing advantage of using glass for prototype implementation is its transparency in the visible light frequency range, and thus the IOS can be easily integrated into buildings. However, this implementation requires very accurate control of the distance between substrates. Such a mechanical control will be less reliable and flexible compared to the electrical control enabled by PIN diodes as shown in Section \ref{response}.

\subsubsection{Liquid Crystal}
\begin{figure}[!t]
	\centering
	\includegraphics[width=0.48\textwidth]{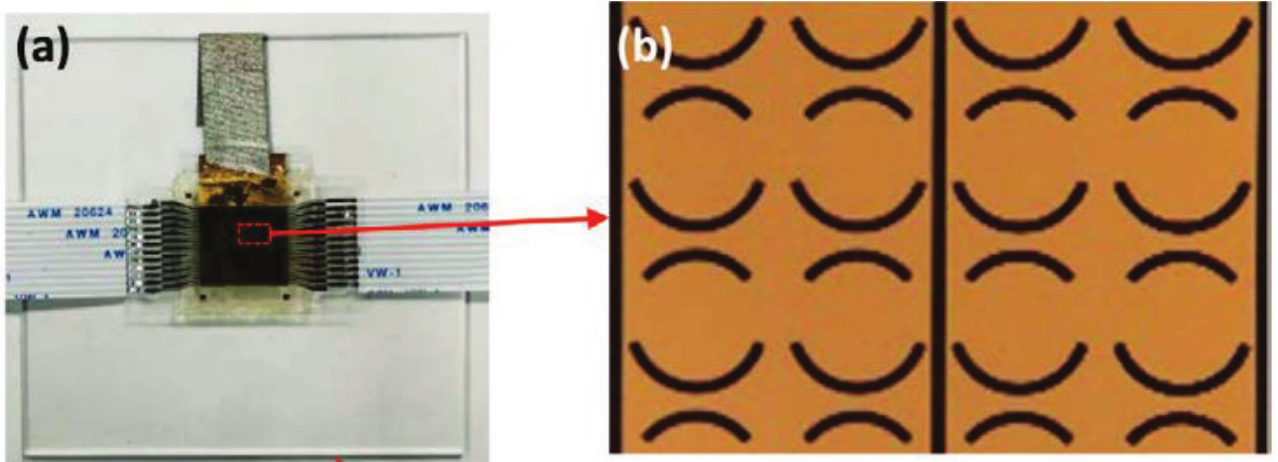}
	\caption{(a) Fabricated  prototype. (b) Microscopic  image  of  unit  cells. \cite{liu2021programmable}. }
	\label{liquid}
\end{figure}

Infusing a metasurface with liquid crystals and changing its EM characteristics with external controls, such as electric field and magnetic field, is an alternative implementation for the IOS. A prototype has been reported in \cite{liu2021programmable}, as shown in Fig. \ref{liquid}. The orientations of embedded liquid crystals are determined by a polyimide alignment layer when the elements are under unbiased conditions, while their orientations are changed by an external electrical field for the biased case, and thus the reconfigurability is achieved.

The liquid crystals inherit the feature from glass that they are transparent for visible light frequency range. Moreover, compared to the implementation introduced in Section \ref{response}, they have shown their potential for a wider working range. For example, they can work in the THz region while the application of semiconductor components, e.g., PIN diodes, is limited in the frequency, while their switching time between two states is longer \cite{vasic2019tunable,li2019machine}. However, its tuning range will be limited compared to PIN diode based implementation.

\subsubsection{Graphene}

\begin{figure}[!t]
	\centering
	\includegraphics[width=0.48\textwidth]{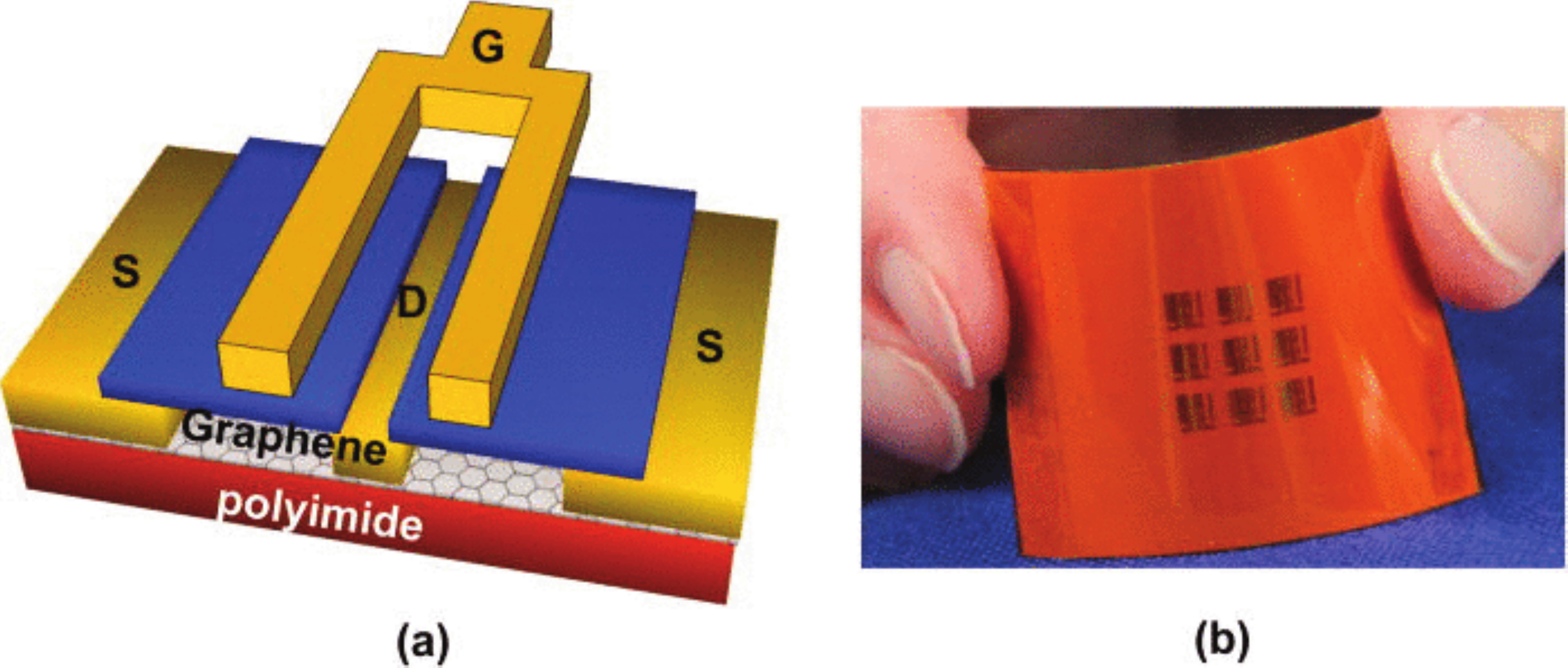}
	\caption{(a) Schematic of a graphene device. (b) Image of a graphene device array. \cite{zhu2013graphene}. }
	\label{graphene}
\end{figure}

It has been widely proven that a single graphene layer has beneficial EM wave characteristics. This feature may also be used for building IOSs \cite{zhang2020graphene}. Indeed, experimental graphene-based RF devices have already existed, as shown in Fig. \ref{graphene} \cite{zhu2013graphene}. To achieve reconfigurability, a single layer of graphene has tunable reflection and refraction coefficients by adjusting its conductivity. 

Owing to graphene's extraordinary properties, we can envision that it might provide infinite probabilities for the implementation of smart radio, such as wearable skin-like surfaces. Unfortunately, its fabrication process is not mature at the current state compared to the PIN diode empowered implementation shown in Section \ref{response}, and thus it is not economical for wide deployment.

\subsection{Summary and Outlook}
In this section, we have introduced how to implement an IOS and how to build an IOS-aided communication system. From the measurement results, we can conclude the following insights:
\begin{itemize}
	\item The coupling effect between reflection and refraction coefficients is confirmed. We also show the gap between phase shifts for reflection and refraction is almost a constant, which verifies the linear model we introduced in Section \ref{phase}.
	
	\item Experimental results confirm that IOS can generate narrow beams and steer the reflected and refracted beams over 360$^\circ$.
	
	\item The incident angle has a significant impact on the reflection and refraction coefficients.
\end{itemize}

In addition to the example prototype and the measurement results obtained from it, we also discuss different prototypes in the literature that are made of other materials. In the future work, the following issues can be further exploited for the practical implementation of the IOS:
\begin{itemize}
	\item \emph{Outdoor Measurement:} In the given example prototype, these measurement results are obtained in an empty room with aluminium walls covered by absorbing materials. Under such a setting, the EM interference from the environment is blocked and cannot reflect the exact propagation conditions that the system might face. For the commercial use of the IOS, efforts are needed to make some measurements for typical outdoor scenarios \cite{trichopoulos2022design}, and develop unique techniques to deal with the interference EM from the environment.  
	
	\item \emph{Higher Frequency:} In future wireless communication systems, mmWave \cite{dai2020reconfigurable} and THz \cite{venkatesh2020high} have attracted much attention as a higher working frequency can provide a wider bandwidth. However, a shorter wavelength leads to a smaller size of the IOS elements, which is typically harder to implement and manufacture. Moreover, some unique characteristics of higher-frequency bands also require signal processing techniques specific to these	bands \cite{sarieddeen2021overview}. Therefore, it is necessary to exploit new designs of IOS elements and develop novel signal processing algorithms for these high-frequency systems.
\end{itemize}

\section{Future Research Directions and Challenges}
\label{future direction}
In this section, we provide a discussion on the future research directions and potential challenges for IOS-aided full-dimensional communications, as summarized in Table \ref{future}.

\begin{table*}[!t]
	\renewcommand\arraystretch{1.2}
	\begin{center}
		\caption{List of future research directions}
		\begin{tabular}{|m{5.5cm} |m{10.5cm}|}
			\hline
		     \makecell[c]{\textbf{Topics}}& \makecell[c]{\textbf{Future Research Directions}} \\ \hline
			\makecell[l]{IOS-aided Secure Communications} & \makecell[l]{1. New beamforming schemes for secrecy rate improvement\\ 2. IOS phase shift optimization without CSI information\\ 3. IOS sharing for multi-user cases with fairness constraint} \\ \hline
			\makecell[l]{IOS-aided Wireless Sensing and Localization} & \makecell[l]{1. Joint optimization of IOS phase shifts and decision function to reduce estimation errors\\ 2. Integrated sensing and communication designs}\\ \hline
			\makecell[l]{IOS-aided Next Generation Multiple \\ Access (NGMA)} & \makecell[l]{1. Performance gain analysis for IOS-aide NGMA\\2. Joint designs of NGMA and IOS} \\ \hline
			\makecell[l]{IOS-aided Aerial Access Networks (AANs)} & \makecell[l]{1. IOSs for performance improvement in AANs \\ 2. Aerial vehicles for a flexible IOS deployment}\\ \hline
			\makecell[l]{IOS-aided Simultaneous Wireless Information\\and Power Transfer (SWIPT)} & \makecell[l]{1. Over-the-air energy splitting \\2. SWIPT for multiple users with one Rx}\\ \hline
		\end{tabular}\label{future}
	\end{center}
\end{table*}

\subsection{IOS-aided Secure Communications}
Due to the broadcast nature of wireless transmission, private data can be leaked to the propagation environment. PLS is an effective technique to solve this issue without the high-complexity encryption \cite{mukherjee2014principles}. It exploits the inherent noise as well as the channel fading in the wireless environment to reduce the amount of private information leaked to those unauthorized users \cite{yang2020intelligent}. Since the IOS is capable of reflecting and refracting the incident waves toward any direction in full space, it provides a novel method to enhance the quality of the legitimate communication link and to destructively add the received signal to the eavesdroppers. Thus, secure communication can be enhanced, helped by the IOS. 

Compared to the IRS-aided PLS improvement \cite{yang2020secrecy,almohamad2020smart,yang2020deep,han2021reconfigurable}, new research challenges have arisen due to such unique characteristics. First, the reflection and refraction coefficients of the IOS vary with different departure angles, and this feature should be considered when modeling the secure capacity of users on both sides of the IOS. To alleviate the signal leakage on the side of the IOS opposite to the Tx, a new beamforming scheme for secrecy rate improvement is required, especially for the uplink case where incident signals from both sides of the IOS experience both refraction and reflection. Second, since the IOS is a device without any signal processing ability, it is difficult to obtain the channel information of the eavesdroppers, leading to the open issue of how to optimize the IOS phase shifts in such a CSI-unknown case. Third, for the general multi-user case where all the user-BS links share the same IOS, it is not trivial to support secure transmission while guaranteeing the user fairness.

Some initial attempts to address these challenges can be found in \cite{niu2021simultaneous,fang2022intelligent,han2022artificial,zhang2021secrecy}. In \cite{niu2021simultaneous}, the authors exploited the use of an IOS to improve the security in a MISO network, and maximized the weighted sum secrecy rate by jointly designing the beamformer at the BS and the reflection and refraction phase shifts at the IOS. Their results showed that the IOS outperforms the IRS. In \cite{fang2022intelligent}, the authors proposed to use an IOS to improve the secrecy performance of a legitimate Rx in the presence of a multi-antenna eavesdropper, and artificial noise (AN) was used to provide additional security. They jointly optimized the beamformer, the AN, and the phase shifts for both reflection and refraction to maximize the secrecy rate. In \cite{han2022artificial}, they extended the AN-based scheme to a multi-user case with NOMA. Their findings showed that more IOS elements can reduce the AN power, but this effect shrinks when the size of the IOS is sufficiently large. In \cite{zhang2021secrecy}, secure communication in an IOS-aided uplink NOMA system was considered, where legitimate users send confidential signals to the BS by exploiting the IOS to customize the propagation environment proactively. Full-CSI and statistical-CSI cases of an eavesdropper were both discussed. For the full-CSI case, an adaptive-rate wiretap code scheme was used to maximize the minimum secrecy capacity subject to the successive interference cancellation (SIC) decoding order constraints. For the statistical-CSI case, a constant-rate wiretap code scheme was employed to minimize the maximum secrecy outage probability subject to the quality-of-service (QoS) requirements of legitimate users. Their results revealed that it was better to deploy the IOS around the users or the BS under the adaptive-rate wiretap code setting, while it was preferable to deploy the IOS far away from them for cascaded channel degradation under the constant-rate wiretap code setting.


\subsection{IOS-aided Wireless Sensing and Localization}
It has attracted much attention to utilizing wireless signals for sensing and localization services due to the low cost and privacy preservation \cite{wang2018device}. The principle of wireless sensing and localization is to recognize the changes of wireless fingerprints (i.e., certain characteristics of wireless channels) caused by the environmental dynamics \cite{xu2015device}. It should be noted that the sensing and localization precision will be higher if the received fingerprints at two arbitrary positions are easily distinguished. In this context, the IOS is a useful tool for wireless since it can actively customize the wireless channels and enlarge their differences. Although some papers have proposed using the IRS to modify the channels to improve the sensing and localization precision \cite{hu2022metasketch,zhang2020towards,zhang2020metaradar}, the coverage extension capability brought by the IOS can effectively reduce the blind zones for wireless sensing and localization systems, which can further improve the systems' reliability.

However, practical designs of IOS-assisted sensing and localization systems are still challenging. One of the challenges is the phase shift optimization of the IOS to reduce the sensing and localization errors. As the optimizations of the decision function for the received signals and the phase shifts at the IOS are highly coupled, they should be optimized jointly. Different from IRS-aided systems, which only focus on the users/objects in the reflection region, the IOS needs to balance the performance in both the reflection and refraction regions, and thus the methods for IRS-aided systems cannot be directly applied. Since the mapping from the received signals to locations/existence of objects is complicated, deep reinforcement learning might be a solution for such a problem \cite{hu2020reconfigurable,hu2021metasensing,yang2020artificial,xiong2019deep}. Moreover, as suggested in \cite{zhang2021metalocalization}, the system can have a higher sensing or localization precision with more configurations\footnote{A configuration refers to states of all the IOS elements.} of the IOS as we can obtain more fingerprints, while it will cost more training time which not satisfy the latency requirements for practical applications. In this regard, we need to carefully optimize the number of configurations to use for the precision-latency trade-off.  

In addition to the aforementioned sensing and localization applications, the IOS can also find its own role in integrated sensing and communication (ISAC), which is originally proposed to save bandwidth for integrating sensing functions in communication systems \cite{liu2022integrated}. Compared to the IRS \cite{zhang2022holographic,shtaiwi2021sum,jiang2021intelligent}, the IOS is expected to benefit the ISAC more as it can double the coverage. To fully reap these benefits, it is necessary to properly deploy the IOS and optimize their real-time reflection and refraction responses by taking into account the new co-channel sensing-communication interference. On the other hand, ISAC might also be useful for improving IOS-aided wireless communications in return. To be specific, with the sensing function, it will be possible for the users to estimate some communication parameters (e.g., angle-of-arrival or angle-of-departure), thus facilitating the CSI acquisition, which is a challenging task in IOS-aided wireless communication systems as we have discussed before. 

\subsection{IOS-aided Next Generation Multiple Access}

%

For future wireless communication systems, next generation multiple access (NGMA) has been developed to support massive numbers of users in a more resource- and complexity-efficient manner than existing multiple access schemes \cite{liu2022evolution}. In particular, NOMA has been widely considered as a promising solution and has drawn significant research interest in the past few years. NOMA enables multiple users to transmit over the same channel simultaneously, but it also introduces significant inter-user interference \cite{di2016sub}. To deal with the inter-user interference, power domain multiplexing at the Tx and SIC at the Rx are applied. However, the successful decoding probability of SIC heavily relies on the channel quality gap between different users, which may not be guaranteed, especially when the user channels are highly correlated. Owing to the EM regulating ability of the IOS on both sides, the full-dimensional propagation environments can be reshaped such that the SIC decoding orders can be scheduled flexibly based on the user priorities to enhance the successful decoding probability as well as user fairness. 

In the literature, there have been several proposals to integrate the IOS into NOMA. Existing research can be divided into two categories. The first category is to exploit the performance gain from the IOS. For example, the authors in \cite{zhang2022star} derived closed-form analytical expressions of outage probabilities for the paired NOMA users using the central limit model and the curve fitting model, and discussed the performance gains under three protocols as introduced in Section \ref{related}. They have shown that the time-switching protocol has the best performance but requests more time blocks than other protocols, and the energy-splitting protocol can achieve higher diversity gains compared to the mode-switching protocol. \cite{wu2021coverage} investigated IOS applications for NOMA, and reported that the gain obtained by the IOS for NOMA is greater than that obtained by the IOS for orthogonal multiple access (OMA). In \cite{wang2021performance}, the authors investigate the performance of an IOS-assisted downlink NOMA network with phase quantization errors and channel estimation errors, where the channels related to the IOS are spatially correlated. It was shown that the average achievable rate converges to the same value with a large size of IOS, no matter whether the channels are correlated or not.

The second category is to develop efficient joint designs for NOMA and IOS. For example, the authors in \cite{zuo2021joint} proposed efficient algorithms to solve a joint optimization problem of decoding order, channel assignment, power allocation, and reflective-refractive beamformer at the IOS for downlink NOMA with the objective of maximizing the system sum-rate, while in \cite{zuo2021uplink}, the authors jointly optimized the transmit-power of users, receive-combining vectors at the BS, beamforming vectors at the IOS, and time slots for uplink NOMA. In \cite{hou2021joint}, an IOS was proposed to be deployed for a NOMA enhanced coordinated multi-point transmission (CoMP) network, where inter-cell interference and desired signals can be simultaneously eliminated and boosted via the IOS. A joint beamformer and phase shift optimization scheme was developed to achieve this. The authors in \cite{aldababsa2021simultaneous} considered the mode switching protocol for an IOS-aided NOMA system, and proposed an algorithm to partition the IOS among the available users, aiming to determine the proper number of reflective or refractive elements for the system sum-rate maximization while guaranteeing QoS requirements for individual users. The work in \cite{ni2021star} integrated NOMA and over-the-air federated learning into a unified framework with the assistance of an IOS for coverage extension. The transmit power at users and the configuration of the IOS are optimized jointly to accelerate the convergence while satisfying QoS requirements for individual users.

\subsection{IOS-aided Aerial Access Networks}
Providing ``connectivity from the sky" is an innovative trend for future wireless communication systems. In particular, aerial vehicles, such as satellites, high and low altitude platforms, drones, etc., are being considered as candidates for deploying wireless communications complementing the terrestrial communication infrastructure, forming aerial access networks (AANs) \cite{zhang2020unmanned}. In AANs, an IOS can be employed as an EM lens to complement the path loss caused by the long transmission distance. On the other hand, AANs can in return facilitate a flexible deployment of the IOS thanks to their high mobility \cite{ye2022non}. As a result, the interplay between IOSs and AANs can be an interesting topic.

In general, there are two research lines for the integration of IOSs and AANs. The first line proposes to deploy the IOS for the performance improvement in AANs \cite{li2020reconfigurable,xu2021intelligent,cao2021reconfigurable,xu2021envisioning}. Compared to the IRS, the IOS can significantly reduce the signaling burden. For example, when a UAV flies from the reflection region to the refraction region, the IRS will be out of coverage and handover to another IRS while the IOS can still serve the aerial user through refractive beams. One critical issue for this line of research is how to dynamically adjust the phase shifts of the IOS to fit the high mobility of aerial vehicles. To deal with the highly dynamic environment, reinforcement learning can be a powerful tool \cite{hu2020reinforcement}. In particular, we can learn the channel information from the historical data to achieve beamforming rather than estimate the channel directly, which might cause a significant delay, and thus it can work well in high-speed scenarios.

The second line is to attach the IOS to the aerial vehicles, e.g., drones and high/low altitude platforms, to enable the flexible deployment of the IOS \cite{jeon2022energy,abdalla2020uavs}. Under such a setting, the placement/trajectory of aerial vehicles needs to be optimized together with the phase shift design at the IOS. However, the aerial vehicles are prone to being influenced by the airflow, and the perturbation of aerial vehicles will inevitably decrease the system QoS \cite{liu2022deployment}, which requires robust designs for these systems where the position errors should be taken into account for beamforming. It worth pointing out that the effects of the perturbation on users in reflection and refraction regions will be totally different, and thus designs for IRS-aided aerial systems cannot be applied directly.

Some attempts have been made to address these challenges. For example, the authors in \cite{wang2022intelligent} suggested the integration of multi-access edge computing with the IOS as the full-dimensional coverage provided by the IOS can support the computing offloading of more users. In this paper, they considered an IOS-aided aerial secure offloading system in the presence of multiple ground eavesdroppers, where an IOS was deployed to prevent information leakage, improve legitimate reception quality, and increase the security range of UAVs. They optimized computing frequency, offloading strategy, transmit power, phase shifts, and UAV locations in order to maximize the secret energy efficiency of the considered system. Their results show that compared with IRSs, IOSs can fully utilize the flexibility of UAVs to expand their secure offloading space. In \cite{duo2021full}, the authors investigated the achievable rate maximization problem of a downlink UAV-enabled communication system aided by an IOS by jointly optimizing the IOS’s phase shift and the UAV trajectory. Their simulation results show that the IOS-aided system can achieve a significant gain compared to an IRS-aided system.

\subsection{IOS-aided Simultaneous Wireless Information and Power Transfer}
SWIPT is a potential technique for future IoT networks, where IoT devices can opportunistically harvest energy from ambient EM sources or from sources that intentionally transmit EM waves for energy harvesting purposes \cite{krikidis2014simultaneous}. However, the energy harvesting efficiency is typically low, which is a critical issue for the practical implementation of SWIPT systems \cite{zhu2022aerial}. To address this issue, deploying an IOS for the SWIPT system can be a promising solution. To be specific, the IOS can focus the EM waves, thus improving the energy harvesting efficiency.

Compared to the IRS-aided SWIPT \cite{pan2020intelligent,xu2022optimal,wu2019weighted}, the IOS-aided one could have the following two novel functions: 
\begin{itemize}
	\item  \emph{Over-the-air energy splitting:} In traditional SWIPT, the receiver will split the energy into two streams: one stream is for energy harvesting, and the other is converted to baseband for information decoding \cite{perera2017simultaneous}. However, in the IOS-aided SWIPT system, the power splitting can be done at the IOS instead. As a result, more flexible scheduling of these IoT devices can be achieved. To be specific, in traditional systems, information and power need to be transferred to the same IoT devices. In an IoT device with differential requirements, some IoT devices might need more energy to complete their own tasks, e.g., sensing or transmitting data \cite{qiu2018can}. With the help of the IOS, we can enable information and energy transmission for different IoT devices with a single Rx, and thus heterogeneous IoT networks can be well supported. However, such a function also brings new challenges: 1) We need to design an IOS with different power splitting levels to meet the requirements for heterogeneous IoT devices, but how to determine these splitting levels is still a challenge; 2) Innovative signal processing techniques are necessary for the real-time control of the IOS as a different performance measurement is adopted here.
	
	\item \emph{SWIPT for multiple users with one Rx:} In the IRS-aided SWIPT systems, only one user can be served with one Rx. Owing to the energy splitting capability introduced by the IOS, the signals can be divided into two streams, each being transmitted to one user. In this way, multiple users can be served when the system is equipped with multiple IOSs and even one Rx. However, such a vision is difficult to achieve. \emph{First}, different users might have different information to transmit. Therefore, it is necessary to develop sophisticated coding techniques to combine the information for different users into one data stream. \emph{Second}, how to coordinate these users through proper control of these IOSs is another critical issue.
\end{itemize}

\section{Conclusions}
\label{conclusion}
In this paper, we have given a comprehensive tutorial on the emerging IOS technology for full-dimensional wireless communications, which can provide greater coverage compared to the existing IRS technique. We first introduced the design principles of the IOS elements, as well as its reflective-refractive model. To facilitate the full-dimensional wireless communications, we have elaborated on the beamforming techniques. In what follows, we have demonstrated the implementation of an IOS and an IOS-aided wireless communication prototype, and we have reported the measurement results to verify the feasibility of the IOS for full-dimensional wireless communications. Finally, we have also highlighted several potential research directions for IOS-aided wireless communications, including IOS placement, channel estimation, IOS-aided secure transmission, as well as IOS-aided wireless sensing and localization. We hope that this paper can be a useful resource for future research on IOS-aided full-dimensional wireless communications, unlocking its full ecosystems in future cellular networks.

\bibliographystyle{IEEEtran}%
\bibliography{ref}

\end{document}